\documentclass[aps,prd,superscriptaddress,nofootinbib,10pt,english,tightenlines,floats,showpacs,
showkeys,showpacs,preprintnumbers,amsmath,amssymb]{revtex4}
\usepackage{euscript}\usepackage{epsfig}
\usepackage{graphicx}% Include figure files
\usepackage{dcolumn}% Align table columns on decimal point
\usepackage{bm}% bold math
\usepackage[unicode=true,pdfusetitle,
 bookmarks=true,bookmarksnumbered=false,bookmarksopen=false,
 breaklinks=false,pdfborder={0 0 1},backref=false,colorlinks=false]
 {hyperref}
\usepackage{color}
\usepackage{amssymb}
\usepackage{fancyhdr}

\def\nn{\nonumber\\}
\newcommand{\f}[2]{\frac{#1}{#2}}
\def\be{\begin{equation}}
\def\ee{\end{equation}}
\def\bea{\begin{eqnarray}}
\def\eea{\end{eqnarray}}

\def\bwt{\begin{widetext}}
\def\ewt{\end{widetext}}

\begin{document}

%\preprint{APS/123-QED}

\title{Non-singular Brans-Dicke collapse in deformed phase space}

\author{S. M. M. Rasouli}

\email{mrasouli@ubi.pt}

\affiliation{Departamento de F\'{i}sica, Universidade da Beira Interior, Rua Marqu\^{e}s d'Avila
e Bolama, 6200 Covilh\~{a}, Portugal}

\affiliation{Centro de Matem\'{a}tica e Aplica\c{c}\~{o}es (CMA - UBI),
Universidade da Beira Interior, Rua Marqu\^{e}s d'Avila
e Bolama, 6200 Covilh\~{a}, Portugal}

%\affiliation{Physics Group, Qazvin Branch, Islamic Azad
%University, Qazvin, Iran}

\author{A. H. Ziaie}
\email{ah_ziaie@sbu.ac.ir}
\affiliation{Department of Physics, Shahid Beheshti University, G. C., Evin, 19839
Tehran, Iran}
\affiliation{Department of Physics, Shahid Bahonar University, PO Box 76175, Kerman, Iran}
\author{S. Jalalzadeh}
\email{shahram.jalalzadeh@unila.edu.br}
\affiliation{Federal University of Latin-American Integration, Technological
\\ Park of Itaipu PO box 2123, Foz do Igua\c cu-PR, 85867-670, Brazil}

\author{P. V. Moniz}
\email{pmoniz@ubi.pt}

\affiliation{Departamento de F\'{i}sica, Universidade da Beira Interior, Rua Marqu\^{e}s d'Avila
e Bolama, 6200 Covilh\~{a}, Portugal}

\affiliation{Centro de Matem\'{a}tica e Aplica\c{c}\~{o}es (CMA - UBI),
Universidade da Beira Interior, Rua Marqu\^{e}s d'Avila
e Bolama, 6200 Covilh\~{a}, Portugal}

\date{\today}% It is always \today, today,

\begin{abstract}
We study the collapse process of a homogeneous perfect fluid (in FLRW background) with a
barotropic equation of state in Brans-Dicke (BD) theory in the presence of phase space deformation effects. Such a deformation is introduced as a particular type of non-commutativity between phase space coordinates. For the commutative case, it has been shown in the literature $\left[{\rm M\,A\,Scheel, S\,L\,Shapiro\,\&\,S\,A\,Teukolsky,
Phys\,Rev\,D\,\textbf{51}\,4236\,(1995)}\right]$, that the dust collapse in BD theory leads to the formation
of a spacetime singularity which is covered by an event horizon. In comparison to general relativity (GR), the authors concluded that the final state of black holes in BD theory is identical to the GR case but differs from GR during the dynamical evolution of the collapse process. However, the presence of non-commutative effects influences the dynamics of the collapse scenario and consequently a non-singular evolution is developed in the sense that a bounce emerges at a minimum radius, after which
an expanding phase begins. Such a behavior is observed for positive values of the BD
coupling parameter. For large positive values of the BD coupling parameter, when non-commutative effects are present, the dynamics of collapse process differs from the GR case. Finally, we show that for negative values of the BD coupling parameter, the singularity is replaced by an oscillatory bounce occurring at a finite time, with the frequency of oscillation and amplitude being damped at late times.
\end{abstract}

\maketitle
\section{Introduction}
\indent
\label{Intro.}
The general theory of relativity proposed by Albert Einstein provides a comprehensive and coherent
description of gravity at the level of large-scale interactions. It is a geometrical
theory which is formulated in such a way that space and time are not the absolute
entities of classical mechanics, but rather, dynamical quantities determined
together with the distribution and motion of matter and energy.
%The theory is also successful in describing the present day observed universe.However,
Notwithstanding the successes and experimental validations, in the last thirty years
several shortcomings of GR were nevertheless found leaving
the scientists with the idea that there is no reason to believe that GR is the
only fundamental theory of gravitation~\cite{vfsct}.
%Many alternative theories of gravity have been proposed after
%the formulation of GR and many of those have been discarded on grounds of physical validity.
Among the alternative theories to GR, the BD theory~\cite{BD61} is
one of the simplest and well studied generalizations of GR with the aim to fully
incorporate Mach\rq{}s principle into the theory. Though the cornerstone of GR is
based on Mach\rq{}s ideas, it admits solutions that are explicitly
anti-Machian such as G{\rm$\ddot{o}$}del universe \cite{godel} and pp-waves~\cite{ppw}. In
BD theory the gravitational (coupling) constant $G$ is no longer a constant and constitutes instead
a field that varies in the  spacetime, namely, by the inverse of a
dynamical scalar field, named the BD scalar field. Thus, within the
framework of this theory the gravitational effects are described by two
fundamental non-matter fields, i.e., the metric tensor field $g_{\mu\nu}$
and the BD scalar field $\phi$. This field, in the Jordan representation \cite{Jordan-rep},
couples to gravity with an adjustable parameter, $\omega$, therefore
acting as a mediator between matter fields and spacetime geometry.
%The collapse of a dust cloud was firstly
%investigated in Ref.~\cite{OS39}, in which they have considered
%the interior and exterior regions as a Friedmann-like
%and a static Schwartzchild line elements, respectively.
%By employing a numerical analysis, the dust collapse
%has been investigated in the BD theory~\cite{SST95}.

Although the BD theory must agree with GR in the weak field regime,
(from the results of solar system experiments \cite{SSEX}) the theory
predicts remarkable deviations from GR in the presence of strong gravitational
fields, e.g., superdense regimes of extreme gravity that occur during the dynamical evolution of the collapse process. The study of the collapse scenario in the framework of BD theory and its comparison to GR has attracted significant interest over the past decades. Gravitational collapse of an ideal gas in BD theory has been investigated in \cite{nuhyidgas} using numerical simulations of relativistic hydrodynamics. Using numerical techniques, the authors in \cite{SHNANA} studied the collapse process of a spherically symmetric dust fluid and investigated the waveform and amplitude of scalar-type  gravitational waves in the context of BD theory. The numerical results of both works suggest that the end product of a collapsed object in BD theory is the Schwarzschild black hole, being in agreement with Hawking\rq{}s theorem \cite{H72} which states that stationary black holes as the final state of gravitational collapse in BD theory are no different than GR\footnote{The extension of this proof to general scalar tensor and $f(R)$ gravity theories has been investigated in \cite{sofa-1109.6324}.}. Oppenheimer-Snyder collapse in BD theory has been studied numerically in \cite{SST95} and the authors have concluded that the black holes produced as the final outcome of the collapse are in general similar to those in
GR in final equilibrium, while they behave
differently in comparison to GR, during the dynamical evolution. It has also been demonstrated that both the area and the apparent horizon theorems associated to dust collapse do not always hold for all
values of the BD coupling parameter \cite{SST95,K96}. Theoretical as well as astrophysical aspects of collapse scenario and formation of black holes in scalar tensor theories have been widely investigated in~\cite{thorndyk1971,J99,KKMC86,CL93,AC94,crit-bdcollapse,4jap-bdcol-97,NBA98,K99,bhbdcosmol-2000,SB01,HGC02,sarkarstbh2005,GW07,NS09,HY10,F10,nsbdexistence,dilnearns,RBD14,bhstkorea,shman-ltbbd,jnovak98}.

The idea of non-commutativity between spacetime coordinates was first
proposed by Snyder \cite{Snyder1947} in an attempt to introduce a short
distance cut-off (the non-commutative parameter) in a Lorentz covariant
way in order to cure the renormalizability features of relativistic quantum
field theory. Since then,
there has been a great deal of interest in this research area
(see e.g. \cite{reviewnoncom}, \cite{reviewnoncom1}).
The main motivation was triggered by
works establishing the connection between non-commutativity
and string and M theory \cite{Pbook98}. Several investigations
have been carried out to study properties of non-commutative
theories, such as IR/UV mixing and nonlocality~\cite{MRS00},
Lorentz symmetry violation~\cite{LV}, new physics at very short
distances \cite{reviewnoncom1} and non-commutative classical mechanics~\cite{RV03}.
Non-commutative extensions of models concerning quantum mechanics such
as the harmonic oscillator \cite{noncomhosc} and the spectrum of Hydrogen
atom~\cite{HYatomnc} have also been probed in order to seek for theoretical
values of the non-commutative parameter. Additionally, non-commutative settings
are investigated to describe some physical effects such as quantum Hall effect \cite{qHall} and Landau problem \cite{Landaupr}.

Soon after non-commutative field theory appeared in the
literature \cite{SW99,reviewnoncom1}, the interest in this arena
has made its way slowly but steadily into the realm of gravitational theories, from
which several applications to non-commutative gravity \cite{noncomgrav} have been
proposed. However, different formulations of gravity theory in non-commutative
spacetime have common being highly non-linear, so that the non-commutative
equations of motion are too complicated to be solved. In addition, efforts have been devoted to verify the possible role of non-commutativity within Newtonian cosmology \cite{NCNC111}, cosmological perturbation theory and inflationary cosmology \cite{NONCOMINF1400}, quantum cosmology \cite{NONCOMQC12,BP0412}
and non-commutativity based on generalized uncertainty principle \cite{GUP}. Cosmological scenarios within the framework of non-commutative geometry provide us the formulation
of semiclassical approximations of quantum gravity allowing to deal with the cosmological constant problem \cite{CCPNONCOM,CCPNONCOM1}. More interestingly, non-commutativity can provide a reasonable groundwork for non-singular cosmological scenarios where big-bang/crunch singularities are dissolved \cite{bbbcsinnc}.
%(see e.g.\cite{NCGreview120} for reviews on different approaches to non-commutativity within gravity theories).
In the context of Kantowski-Sachs cosmological model, non-commutativity has been introduced  into the classical phase space and classical non-commutative equations of motion have established \cite{BP04}. For scalar field cosmology, in \cite{SFCOSDEFPH} the classical minisuperspace is deformed and a scalar field is used as the matter component of the universe and the cosmological constant problem and removability of initial curvature singularity is studied in \cite{GUP}. In \cite{GUZSASO}, the study is focused on the consequences that the non-commutative deformation has on the slow-roll parameter, when an exponential potential for the scalar field is considered. In particular the non-commutative deformation gives a mechanism that ends inflation.  The compactification and stabilization of internal extra dimensions in multidimensional cosmology at the presence of non-commutativity are studied in \cite{compacnjs}. The main idea of the above cosmological models in the framework of classical non-commutativity is based on the assumption that modifying the Poisson brackets of the classical theory gives the non-commutative equations of motion which leads to some non-trivial phenomena such as UV/IR mixing.

Beside the cosmological models, of particular
interest are non-commutative black hole solutions.
In the herein paper, we have chosen (for practical reasons) a
particular choice for the non-commutativity setting. Of course,
a wide discussion on BD non-commutativity and gravitational
collapse will need to peruse on different choices and non-commutativity ingredients.
Non-commutativity is a vast subject, see, e.g.,~\cite{MN97,BBDP09,NONCOMINF1400,S16} and references therein.
Nevertheless, we think that without (much)
loss of generality, our investigation and subsequent
research show that when non-commutativity is present,
it conveys important deviations in terms of effective
density and pressure terms, which imply a clear modification of
the gravitational collapse and of singularity formation and possible avoidance.
Noncommutativity of
the space time can indeed be relevant in the context of black hole physics~\cite{BBDP09} and, moreover,  noncommutativy
 in a BD theory allows noncommutative parameter to couple to the variables which
are absent in the GR. Therefore, the range of
solutions and possible scenarios is much wider indeed.
It was this broad scope of possibilities we investigated herein, regarding a collapse scenario in a
BD noncommutative setting.
%Physicists have believed that a quantum theory of
%gravity may resolve the black hole singularity
%problem. However, as they have not reached such a setting, yet, an appropriate insight can be
%provided by a quantum cosmology approach based on the minisuperspace approximation.
%Indeed, it has been believed that the nonocommutativity of the space time can be relevant to the context of black hole physics~\cite{BBDP09}.}
 In this respect, a great amount of work has been done in search of non-singular neutral and charged black holes as the exact solutions of Einstein\rq{}s equations in non-commutative framework \cite{Nicollinipapers}. More recently, the collapse scenario of a homogeneous minimally coupled scalar field has been studied in the context of classical non-commutativity~\cite{RZMM14}. It was shown that contrary to the commutative case (in
which the collapse scenario ends in a spacetime singularity), introducing a non-commutativity between the momenta
associated to the scale factor and the scalar field, the singularity can be either removed or instead attained faster.

As discussed above, the process of gravitational collapse and singularity (naked or covered) formation has been a long standing issue in scalar-tensor theories of gravity. For a spatially flat~Friedmann-Lema\^{\i}tre-Robertson-Walker~(FLRW) metric in the absence of non-commutative effects, the solutions to the field equations of BD theory are shown to exhibit spacetime singularities, both in cosmological \cite{BDSINCOS} and gravitational collapse scenarios \cite{BDSINCOL}. From considering the non-singular cosmological models within non-commutative setting, we are motivated to get closer to the idea of curing the formation of spacetime singularities as the collapse end state in a non-commutative framework. Assuming a spatially flat FLRW metric, our aim here is to investigate the effects of classical non-commutativity on the collapse of a perfect fluid in BD theory. We argue that introducing non-commutativity (with constant non-commutative parameter) within the phase space, causes the behavior of classical trajectories of the collapse to be completely different in comntrast to the one obtained from the standard BD theory (commutative case).
We analyze in detail the solutions associated to some special cases, namely, when there is only a pressure-less matter and/or when the BD coupling parameter takes large values.
\par
Our paper is then organized as follows. In Sec.~\ref{NC-BDT}, we will derive the Hamiltonian
equations of motion for a concrete choice of deformation in the BD setting.
%(containing a scalar potential) for when the geometry of the
%interior spacetime is taken to be the flat FLRW line-element.
In Sec.~\ref{NC-SR}, assuming a vanishing scalar potential, we investigate
numerically the collapse of barotropic matter.
%in the BD theory in the presence of the noncommutative parameter.
%We further analyze the behavior of the quantities related to the collapse
%dynamics and
We show that non-commutative effects could remove the spacetime singularity occurring in
the standard BD theory. Finally, in Sec.~\ref{Concl.},
we summarize and discuss our results together with complementary discussions.
In appendix~\ref{APPBD}, by employing the Taylor series about the
bounce time, we present approximate analytic solutions
for two special cases, the collapse of dust and stiff fluids within non-commutative BD theory.

\section{Noncommutative Setting in Brans-Dicke Theory}
\indent
\label{NC-BDT}
Our aim in this section is to find the modified field equations in the context of
BD theory when a special type of non-commutativity is present. We consider a
spherically symmetric homogeneous perfect fluid undergoing gravitational collapse
in BD theory. We parametrize the interior spacetime of the
collapsing volume with a spatially flat FLRW line element given by
\begin{equation}\label{metric}
ds^{2}=-N^2(t)dt^2+e^{2\alpha(t)}\left(dx^2+dy^2+dz^2\right),
\end{equation}
where $a(t)=e^{\alpha(t)}$ is the scale factor and $N(t)$ is a lapse function.
For this line element, the collapse scenario amounts to a cloud that begins to collapse
from rest at an infinite initial radius \cite{McVittieMTW}. As we have discussed earlier, in the context of BD theory, the spacetime representing the above metric admits both cosmological \cite{BDSINCOS} as well as astrophysical singularities~\cite{BDSINCOL}. We are therefore motivated to investigate the existence of non-singular solutions by introducing non-commutativity between phase space coordinates. We shall see that such a deformation can be the agent for removing the spacetime singularity that occurs in standard collapsing scenarios in BD theory.

Let us start with the action functional of BD
theory\footnote{Note that in the original BD theory there is no scalar potential.
Nevertheless,
%in order to obtain late time acceleration
it can be added by hand (see, e.g.,~\cite{SS01}) or
%Although, it has recently been demonstrated that,
%instead of an {\it ad hoc} assumption, such a scalar potential
 it can be geometrically induced in the context of
a modified BD theory (MBDT)~\cite{RFS11-R14}.
In some special cases, the MBDT reduces to its concrete setting in GR, see, e.g.,~\cite{stm99-DRJ09-RJ10}}
in the Jordan frame~\cite{BD61,Jordan-rep}
\begin{equation}\label{lag1}
S=\int\sqrt{-g}\left[\phi {\mathcal R}-
\frac{\omega}{\phi}g^{\mu\nu}\phi_{,\mu}\phi_{,\nu}-V(\phi)\right]d^4x+\int\sqrt{-g}{\cal L_{\rm matt}}d^4x,
\end{equation}
where the greek indices run from zero to 3, $g$ is the determinant of the metric $g_{\mu\nu}$
associated to the four dimensional spacetime and
%$S_{\rm matter}$ denotes
%the (ordinary) matter part of the action and
% It should be Noted that the BD scalar
%field $\phi$ indeed couples only indirectly to the matter fields by coupling to the
%Ricci scalar ${\mathcal R}$ and therefore to the  geometry, which in turn
%interacts with matter.
$V(\phi)$ is the scalar potential.
In this work, we will assume that the dimensionless BD coupling parameter $\omega$ to be a constant.
%coupling parameter $\omega$ measures the ratio of the scalar to tensor couplings to matter fields so that
%the larger the value of this parameter the lesser its contribution to the gravitational interaction.
The Lagrangian density associated to the ordinary matter
field is given by ${\cal L_{\rm matt}}={16\pi}\rho(\alpha)$,
 where $\rho$ is the fluid energy density \cite{PFLAG}.
%situations~\cite{BKM04,DDB07,BS07,B09}.
%It is easy to derive the Hamiltonian of the model as \rc{please write full derivation and give dimensional analysis}

By substituting the Ricci scalar associated to the
line element~(\ref{metric}) into the action~(\ref{lag1}),
neglecting the total time derivative
term and redefining an alternative dimensionless BD scalar field $\Phi$ and a new
dimensionless time coordinate $\eta$ as $\Phi:=L_\text{Pl}^2\phi$ and $\eta:=L_\text{Pl}^{-1}t$, respectively
(where  $L_\text{Pl}$  is the Planck's constant in natural units), the Lagrangian
of the model is obtained as
\begin{equation}\label{lag2}
{\cal L}=-N^{-1}e^{-3\alpha}\left[6\Phi\dot{\alpha}^2+6\dot{\alpha}\dot{\Phi}
-\omega\Phi^{-1}\dot{\Phi}^2+N^2L_\text{Pl}^4V(\Phi)\right]+16\pi NL_\text{Pl}^4 e^{3\alpha}\rho,
\end{equation}
where an overdot stands for differentiation with respect to dimensionless
time coordinate $\eta$ and we have assumed that the BD scalar field
to be homogeneous, i.e., to be a function of time only.
%$\Phi=\Phi(t)$.
Note that the scale factor, $\alpha$, is dimensionless.
Therefore, to have a well-defined Lagrangian, we defined an alternative dimensionless
BD scalar field\footnote{From now on, for simplicity, we will call it the BD scalar field.}
$\Phi$ and a new time coordinate $\eta$. The momenta associated to the scale factor and
the BD scalar field i.e., ${\rm P}_\alpha$ and ${\rm P}_\Phi$, can be obtained as
\begin{equation}\label{palpphi}
{{\rm P}}_{\alpha}=\f{\partial {\cal L}}{\partial \dot{\alpha}}=-\f{e^{3\alpha}}{N}\left[12\Phi\dot{\alpha}+6\dot{\Phi}\right],~~~~{{\rm P}}_{\Phi}=\f{\partial {\cal L}}{\partial \dot{\Phi}}=-\f{e^{3\alpha}}{N}\left[6\dot{\alpha}-2\f{\omega}{\Phi}\dot{\Phi}\right].
\end{equation}
Using Legender transformation, ${\cal H}={{\rm P}}_{\alpha}\dot{\alpha}+{{\rm P}}_{\Phi}\dot{\Phi}-{\cal L}$
together with substituting for $\dot{\alpha}$ and $\dot{\Phi}$ we finally obtain the classical Hamiltonian of the model as
\begin{eqnarray}\label{Ham-1}
{\cal H}=-\frac{Ne^{-3\alpha}}{2\chi\Phi}
\left(\frac{\omega}{6}{{\rm P}}_\alpha^2-\Phi^2{{\rm P}}_\Phi^2+\Phi {{\rm P}}_\alpha {{\rm P}}_\Phi\right)
+L_\text{Pl}^4Ne^{3\alpha}\left(V-16\pi\rho\right),
\end{eqnarray}
where $\chi\equiv 2\omega+3$. From now on we will consider the comoving gauge, namely, we will set $N=1$.
%Therefore, the equations of motion with respect to the Hamiltonian~(\ref{Ham-1})
In the commutative case, we consider the phase space coordinates
$\{\alpha,\Phi;{{\rm P}}_\alpha,{{\rm P}}_\Phi\}$, in which the Poisson algebra is
$\{\alpha,\Phi\}=0$, $\{{{\rm P}}_\alpha,{{\rm P}}_\Phi\}=0$,
$\{\alpha,{{\rm P}}_\alpha\}=1$ and $\{\Phi,{{\rm P}}_\Phi\}=1$.
Therefore, for this case, the equations of motion with respect to the
Hamiltonian~(\ref{Ham-1}) are easily derived as
\bea
\dot{\alpha}\!\!&=&\{\alpha,{\cal H}\}=-\frac{e^{-3\alpha}}{2\chi\Phi}\left(\frac{\omega}{3}{{\rm P}}_\alpha+\Phi {{\rm P}}_\Phi\right),\label{diff.eq1}\\
\dot{{{\rm P}}}_\alpha\!\!&=&\{{{\rm P}}_\alpha,{\cal H}\}=L_\text{Pl}^4e^{3\alpha}\left[-6V+16\pi\left(6\rho+\frac{d \rho}{d \alpha}\right)\right],\label{diff.eq2}\\
\dot{\Phi}\!\!&=&\{\Phi,{\cal H}\}=-\frac{e^{-3\alpha}}{2\chi}\left({{\rm P}}_\alpha-2\Phi {{\rm P}}_\Phi\right), \label{diff.eq3}\\
\dot{{{\rm P}}}_\Phi\!\!&=&\{{{\rm P}}_\Phi,{\cal H}\}=\frac{e^{-3\alpha}}{2\chi\Phi}\left({{\rm P}}_\alpha-2\Phi {{\rm P}}_\Phi\right){\rm P}_\Phi\nn
\!\!\!&-&\!\!\!\frac{e^{3\alpha}}{\Phi}\left(V+\Phi\frac{dV}{d\Phi}-16\pi\rho\right)L_\text{Pl}^4\label{diff.eq4},
\eea
where we have used the Hamiltonian constraint ${\cal H}=0$.

Now, we would consider a non-commutative scenario within the setting presented
above. In this regard, we can use two interesting different procedures for the deformation
of Poisson brackets, the Moyal product (or the star product) and the Generalized
Uncertainty Principle (GUP), see e.g. \cite{GUPrefs}. Our investigation
here is carried out by means of including the effects of non-commutativity on our classical
setting. We therefore introduce a deformed product
(star product) rule between two arbitrary observables of four-dimensional phase space as
\begin{eqnarray}\label{sh1}
(f*g)({{\rm X}},{{\rm P}})=\exp\left[\frac{1}{2}\Sigma^{ab}\partial^{(1)}_a
\partial^{(2)}_b\right]f({{\rm X}}_1,{{\rm P}}_1)g({{\rm X}}_2,{{\rm P}}_2)|_{{{\rm X}}_1={{\rm X}}_2={{\rm X}},\,{{\rm P}}_1={{\rm P}}_2={{\rm P}}},
\end{eqnarray}
where ${{\rm X}}=\{\alpha,\Phi\}$ and ${{\rm P}}=\{{{\rm P}}_\alpha,{{\rm P}}_\Phi\}$ are coordinates of
phase space and
\begin{equation}\label{R}
\Sigma_{ab}=\left(%
\begin{array}{cc}
\theta_{ij} & \delta_{ij}+\sigma_{ij} \\
-\delta_{ij}-\sigma_{ij} & \beta_{ij} \\
\end{array}%
\right),\end{equation}where $\theta$ and $\beta$ are the $2\times 2$
antisymmetric matrices which represent the non-commutativity in
coordinates and momenta, respectively and $\sigma_{ij}=-\frac{1}{8}(\theta_{ik}\beta_{kj}
+\beta_{ik}\theta_{kj})$ \cite{noncomshrefs}. The relation between the
above star-product of phase space functions and the usual Poisson brackets
becomes more clear if the formula (\ref{sh1}) is expressed as follows \cite{nonocomshrefs1}
\begin{equation}\label{R1}
f*g=fg+\frac{1}{2}\{f,g\}+\sum_{k=2}^{\infty}
\left(\frac{1}{2}\right)^k\frac{1}{k!}{\mathcal D}_k(f,g),
\end{equation}
where the bidifferential operator ${\mathcal D}_k$ is defined as
\begin{equation}\label{R2}
{\mathcal D}_k(f,g)({{\rm X}},{{\rm P}})=\left[\left(\frac{\partial}{\partial
{{\rm X}}_1}\frac{\partial}{\partial {{\rm P}}_2}-\frac{\partial}{\partial
{{\rm X}}_2}\frac{\partial}{\partial {{\rm P}}_1}\right)^k
f({{\rm X}}_1,{{\rm P}}_1)g({{\rm X}}_2,{{\rm P}}_2)\right]_{{{\rm X}_1
={\rm X}_2=q,\,{\rm P}_1={\rm P}_2={\rm P}}}.
\end{equation}
According to equations (\ref{sh1})-(\ref{R2}) we obtain the following
definition for the Moyal bracket as a deformed Poisson bracket
\begin{equation}\label{S}
\{f,g\}_{\rm M}=f*g-g*f=\{f,g\}+\sum_{k=2}^{\infty}
\left(\frac{1}{2}\right)^k\frac{1}{k!}\left[{\mathcal D}_k(f,g)
-{\mathcal D}_k(g,f)\right],\end{equation}which
looks like an $\Sigma$-commutation relation between two function $f$
and $g$. Hence,  the deformed
Poisson brackets between the phase space coordinates will be
found as
\begin{equation}\label{T}
\{{{\rm X}}_i,{{\rm X}}_j\}_\text{M}=\theta_{ij},\hspace{.5cm}\{{{\rm X}}_i,{{\rm P}}_j\}
_\text{ M}=\delta_{ij}+\sigma_{ij},\hspace{.5cm}\{{{\rm P}}_i,{{\rm P}}_j\}_\text{ M}=\beta_{ij}.
\end{equation}

As already mentioned in introduction, cosmology provides an attractive
setting for non-commutative models, both in the
realm of classical as well as quantum level.
As is shown in \cite{minwramsei,reviewnoncom1},
some non-trivial phenomena, such as UV/IR mixing, would appear in non-commutative quantum field theories.
This divergence mixing implies that physics at large distances is not disconnected from the physics at short scales and one can probe the physics of high energy regimes by low energy physics. It is then expected that even if the effects of non-commutativity are presented at a small scale, such effects might appear at an older time of the cosmos. Therefore, it justifies and hints the use of classical cosmology in the presence of non-commutativity.
In this paper, non-commutativity is achieved by an appropriate deformation of the usual (commutative) algebra of the classical phase space variables. Let us then introduce classical non-commutativity in the model by considering the Hamiltonian to have the same functional form as (\ref{Ham-1}), but is written on variables that satisfy the deformed Poisson brackets. For simplicity, we assume only
non-commutativity
 between configuration space variables, as
\begin{equation}
[\theta_{ij}] =\left(%
\begin{array}{cc}
0 & 1 \\
-1 & 0 \\
\end{array}%
\right)\theta,~~~~\beta_{ij}=0,
\end{equation}
where $\theta$ is a dimensionless positive constant. Therefore,  the deformed Poisson algebra
(\ref{T}) reduces to
\begin{eqnarray}\label{NC-Poisson}
\{\alpha,\Phi\}_\text{M}=\theta,\hspace{10mm} \{{{\rm P}}_\alpha,{{\rm P}}_\Phi\}_\text{M}=0,\\\nonumber
\{\alpha,{{\rm P}}_\alpha\}_\text{M}=1,\hspace{10mm} \{\Phi,{{\rm P}}_\Phi\}_\text{M}=1,
\end{eqnarray}
and the minisuperspace of the model is the deformed or Moyal plane. At the quantum
level, the above deformed Poisson algebra will be $[\alpha,\Phi] =i\theta$,
where one can immediately  obtain the uncertainty relation between coordinates
of Moyal plane
\begin{eqnarray}\label{sh2}
\Delta\alpha\Delta\Phi\geqslant\frac{\theta}{2},
\end{eqnarray}
 or equivalently in terms of the original BD scalar field as $\Delta\alpha\Delta\phi\geqslant\frac{
 L_\text{Pl}^{-2}}{2}\theta$. Also, one can show that
 the spectrum of the area of the triangles and distances in the $\{\alpha,\Phi\}$ Moyal plane are  respectively
given by \cite{Giovanni}
\begin{eqnarray}\label{sh2}
\begin{array}{cc}
A_n^\text{(triangle)}=\frac{\sqrt{3}}{2}\theta|n|,\,\,\,\,\,\, n\in \mathbb{Z}\\
d_j^2=4\theta\left(j+\frac{1}{2}\right),
\end{array}
\end{eqnarray}
where  $j$ is a  nonnegative integer.
The last equation shows that there is ``minimum-distance principle'' in the Moyal
minisuperspace. Therefore, as it has been shown in \cite{Hu}, there is
no classical limit at $\theta\neq0$. Explicitly, the classical limit
exists only if $\theta\rightarrow0$ at least as fast as $\hbar\rightarrow0$,
but this limit does not yield a classical commutative setting,
unless the limit of $\theta/\hbar$ vanishes as $\theta\rightarrow0$ \cite{Hu}.

{We should notice that in our herein non-commutative model by assuming that
the coordinates have a length dimension, the scale factor and the BD scalar
field are dimensionless quantities. It is then straightforward
to show that $\theta$ is also a dimensionless constant.
Therefore restricting the non-commutative parameter to be positive,
we take those values of this parameter for which $0\leqslant\theta<1$.}
Employing~(\ref{Ham-1}) and~(\ref{NC-Poisson}), we
obtain the equations of motion associated to the
non-commutative model as\footnote{In order to obtain the non-commutation
equations~(\ref{NC.H.eq1}) and (\ref{NC.H.eq2}), we have used the
formulas $\{\alpha,f(\alpha,\Phi)\}=\theta\frac{\partial f}
{\partial \Phi}$ and $\{\Phi,f(\alpha,\Phi)\}=-\theta\frac{\partial f}{\partial \alpha}$
which are calculated from the non-commutative relations~(\ref{NC-Poisson}), see, e.g., \cite{GSS11}.}
\begin{eqnarray}\label{NC.H.eq1}
\dot{\alpha}\!\!&=&\{\alpha,{\cal H}\}_\text{M}=\!\!-\frac{e^{-3\alpha}}{2\chi\Phi}
\left[\frac{\omega}{3}{{\rm P}}_\alpha+\Phi {{\rm P}}_\Phi
\!\!+\!\!\theta({{\rm P}}_\alpha-2\Phi {{\rm P}}_\Phi){{\rm P}}_\Phi\right]\\\nonumber
&+&\theta L_\text{Pl}^4\left(\frac{e^{3\alpha}}{\Phi}\right)
\left[V(\Phi)+\Phi\frac{dV(\Phi)}{d\Phi}-16\pi\rho\right]\nn
&=&\!\!-\frac{e^{-3\alpha}}{2\chi\Phi}
\left[\frac{\omega}{3}{{\rm P}}_\alpha+\Phi {{\rm P}}_\Phi\right]-\theta \dot{{{\rm P}}}_{\Phi},\nn
\dot{\Phi}\!\!&=&\{\Phi,{\cal H}\}_\text{M}=\!\!-\frac{e^{-3\alpha}}{2\chi}\left({{\rm P}}_\alpha-2\Phi {{\rm P}}_\Phi\right)\nn
&-&6\theta L_\text{Pl}^4e^{3\alpha}\left[V(\Phi)
-16\pi\left(\rho+\frac{1}{6}\frac{d\rho}{d\alpha}\right)\right]\nn
&=&\!\!-\frac{e^{-3\alpha}}{2\chi}\left({{\rm P}}_\alpha-2\Phi {{\rm P}}_\Phi\right)+\theta\dot{{{\rm P}}}_{\alpha},\label{NC.H.eq2}
\end{eqnarray}
where again, we have used the Hamiltonian constraint ${\cal H}=0$. Note that the equations of motion associated to the
momenta do not change under the phase space deformation~(\ref{NC-Poisson}), hence we have not rewritten them. We should also notice that
when $\theta$ tends to zero, all the obtained equations associated to
the non-commutative BD setting reduce to their corresponding ones in the usual BD theory.

%\rc{In the next section, we will analyze numerically the
%equations~(\ref{NC.H.eq1}) and (\ref{NC.H.eq2}),
%together with those for the momenta, namely Eqs.~(\ref{diff.eq2}) and (\ref{diff.eq4}).
%We will compare the results with their corresponding ones
%in the standard BD theory, as well as with general relativity.}

%Once the BD potential is
%fixed, these equations constitute a closed system of first order
%differential equations for the the phase space parameters $(\alpha,\Phi,P_{\alpha},P_{\Phi})$.

%\begin{figure}
%\caption{.}\label{KR}
%\end{figure}

\section{Noncommutative effects and the singularity removal}
\indent
\label{NC-SR}
%In order to study a dynamical collapse scenario within the noncommutative
%framework presented in section \ref{NC-BDT},
%%%%%%%%%%%%%%%%%%%%%%%%%%%%%%%%%%%%%%%%%%%%%5
Equations~(\ref{NC.H.eq1}) and (\ref{NC.H.eq2}) together with (\ref{diff.eq2}) and (\ref{diff.eq4}), govern the
dynamics of the collapse in the presence of non-commutativity.
Interestingly, comparing equations (\ref{diff.eq1}) and (\ref{NC.H.eq1}) suggests that
 applying the non-commutativity (\ref{NC-Poisson}) to the phase space
 corresponds to shifting the collapse rate in equation (\ref{diff.eq1})
 as $\dot{\alpha}\rightarrow\dot{\alpha}+\theta\dot{{{\rm P}}}_{\Phi}$.
 Similarly, by comparing equations (\ref{diff.eq3}) and (\ref{NC.H.eq2}),
 we observe that, in the presence of non-commutativity, the time
 derivative of the BD scalar field in equation (\ref{diff.eq3})
 undergoes a shift as $\dot{\Phi}\rightarrow\dot{\Phi}-\theta\dot{{{\rm P}}}_{\alpha}$.
 We will see that the presence of these additional terms alter the classical evolution of
 the collapse scenario and finally causes the singularity avoidance at the
semi-classical approximation. Let us assume a simple case where the scalar potential vanishes. Moreover, we take the matter content to be a perfect fluid with barotropic equation
of state (EoS) $p=w\rho$, in which $p$, $\rho$ and $w={\rm constant}$ are
the the pressure, energy density and the equation of state parameter, respectively.
%We should note that, for some of the following solutions, a few particular cases
%associated to special values of $\omega$ and $w$ will be discussed.
We assume that ordinary matter is conserved in the Jordan representation of the BD theory when non-commutative effects are present. Hence, we obtain
\be\label{COEQ}
\rho=\rho_ie^{3(1+w)(\alpha_i-\alpha)},
\ee
where $\rho_i$ and $\alpha_i$ are the initial values of energy density
and logarithm of the scale factor, respectively.

By substituting the above expression into the set of equations (\ref{diff.eq2}), (\ref{diff.eq4}), (\ref{NC.H.eq1}) and (\ref{NC.H.eq2}), we get
\bea
\dot{\alpha}\!\!&=&\!\!-\frac{e^{-3\alpha}}{2\chi\Phi}
\left[\frac{\omega}{3}{{\rm P}}_\alpha+\Phi {{\rm P}}_\Phi
\!\!+\!\!\theta({{\rm P}}_\alpha-2\Phi {{\rm P}}_\Phi){{\rm P}}_\Phi\right]-\f{16\pi
L_\text{Pl}^4 \rho_i\theta}{\Phi}e^{3\left[\alpha_i+w(\alpha_i-\alpha)\right]},\label{eqsrew9}\\
\dot{\Phi}\!\!&=&\!\!-\frac{e^{-3\alpha}}{2\chi}\left({{\rm P}}_\alpha-2\Phi {{\rm P}}_\Phi\right)+48\pi L_\text{Pl}^4\rho_i\theta (1-w)e^{3\left[\alpha_i+w(\alpha_i-\alpha)\right]},\label{eqsrew10}\\
\dot{{{\rm P}}}_{\alpha}&=&48\pi L_\text{Pl}^4\rho_i(1-w)e^{3\left[\alpha_i+w(\alpha_i-\alpha)\right]},\label{eqsrew5}\\
\dot{{{\rm P}}}_\Phi\!\!&=&\!\!\frac{e^{-3\alpha}}{2\chi\Phi}\left({{\rm P}}_\alpha-2\Phi {{\rm P}}_\Phi\right){{\rm P}}_\Phi
+\f{16\pi L_\text{Pl}^4 \rho_i}{\Phi}e^{3\left[\alpha_i+w(\alpha_i-\alpha)\right]}.\label{eqsrew7}
\eea
%Unfortunately, due to the complicated nature of the above set of field equations, it is very hard to find rigorous solutions analytically.

The above equations are a set of first order coupled non-linear
differential equations which govern the collapse dynamics and
evolution of BD scalar field. However, it is not
straightforward to find exact analytical solutions since these
equations are highly non-linear. Notwithstanding this difficulty,
we will discuss analytically these set of equations for two  special
cases, i.e. the dust $(w=0)$ and stiff $(w=1)$ fluids in appendix~\ref{APPBD}.
Let us then employ numerical methods in order to investigate solutions related
to these equations.  Numerical analysis for the above equations needs four initial
conditions, namely, $\alpha_i=\alpha(\eta_i)$, $\Phi(\eta_i)$, ${{\rm P}}_\Phi(\eta_i)$ and ${{\rm P}}_\alpha(\eta_i)$. We choose the initial values for $\{\alpha(\eta),\Phi(\eta),{{\rm P}}_\Phi(\eta)\}$ arbitrarily and the initial value for ${{\rm P}}_\alpha(\eta_i)$ is set subject to equation (\ref{eqsrew9}). In this sense, choosing the initial value for the collapse velocity as $\dot{a}_i=\dot{\alpha}(\eta_i)e^{\alpha(\eta_i)}$, we can solve equation (\ref{eqsrew9}) for ${{\rm P}}_\alpha(\eta_i)$ to get the fourth initial value hence finding a consistent solution for the system (\ref{eqsrew9})-(\ref{eqsrew7}). In the rest of this paper, we use the notation $\dot{a}=\dot{\alpha}e^{\alpha}$ for the collapse velocity and $\ddot{a}=e^{\alpha}(\ddot{\alpha}+\dot{\alpha}^2)$ for the collapse acceleration.

%With the proposed setup, due to the complicated structure of the equations,

\subsection{Scale factor behavior}
\begin{figure}
\includegraphics[scale=0.431]{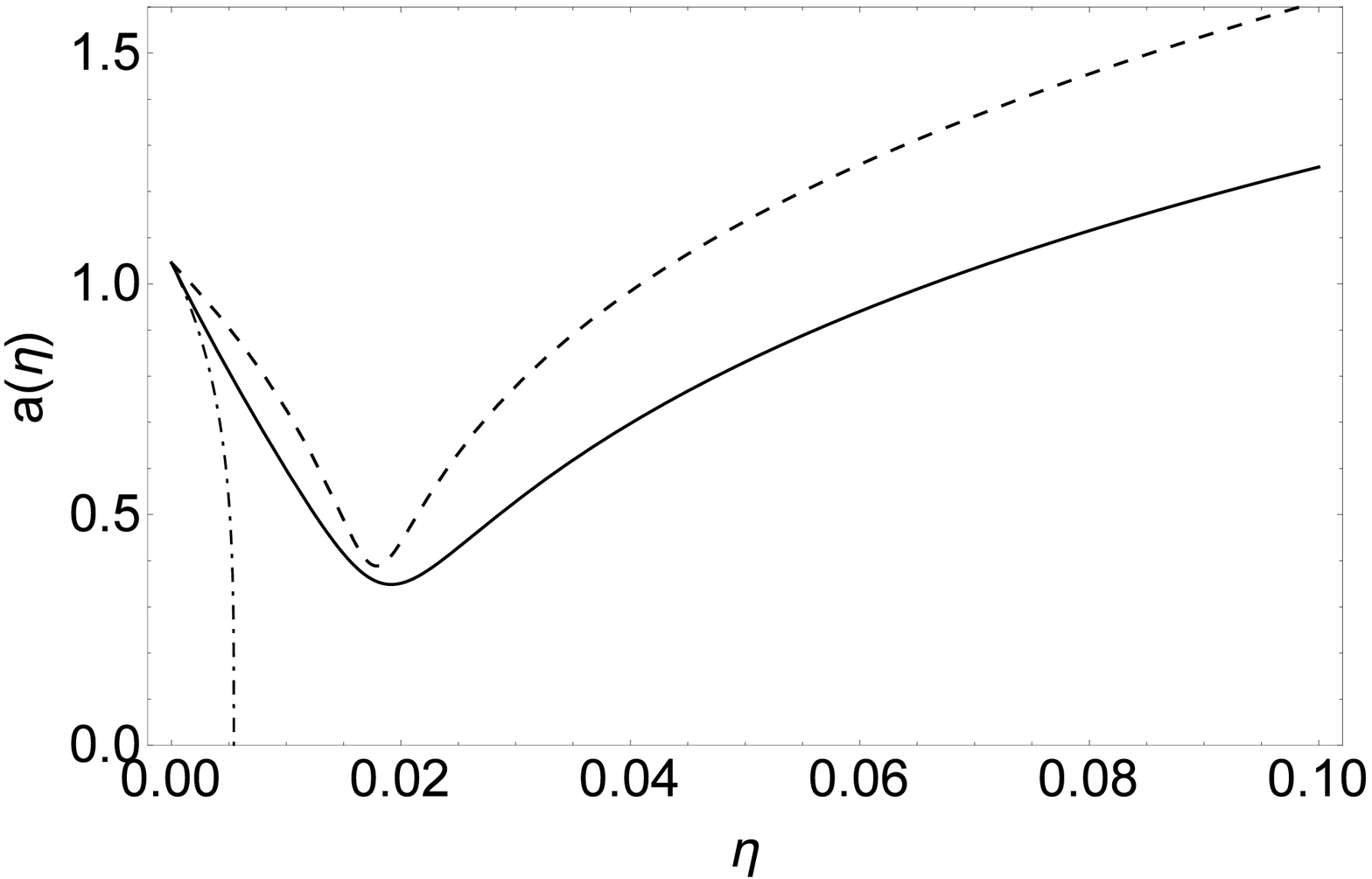}
\includegraphics[scale=0.37]{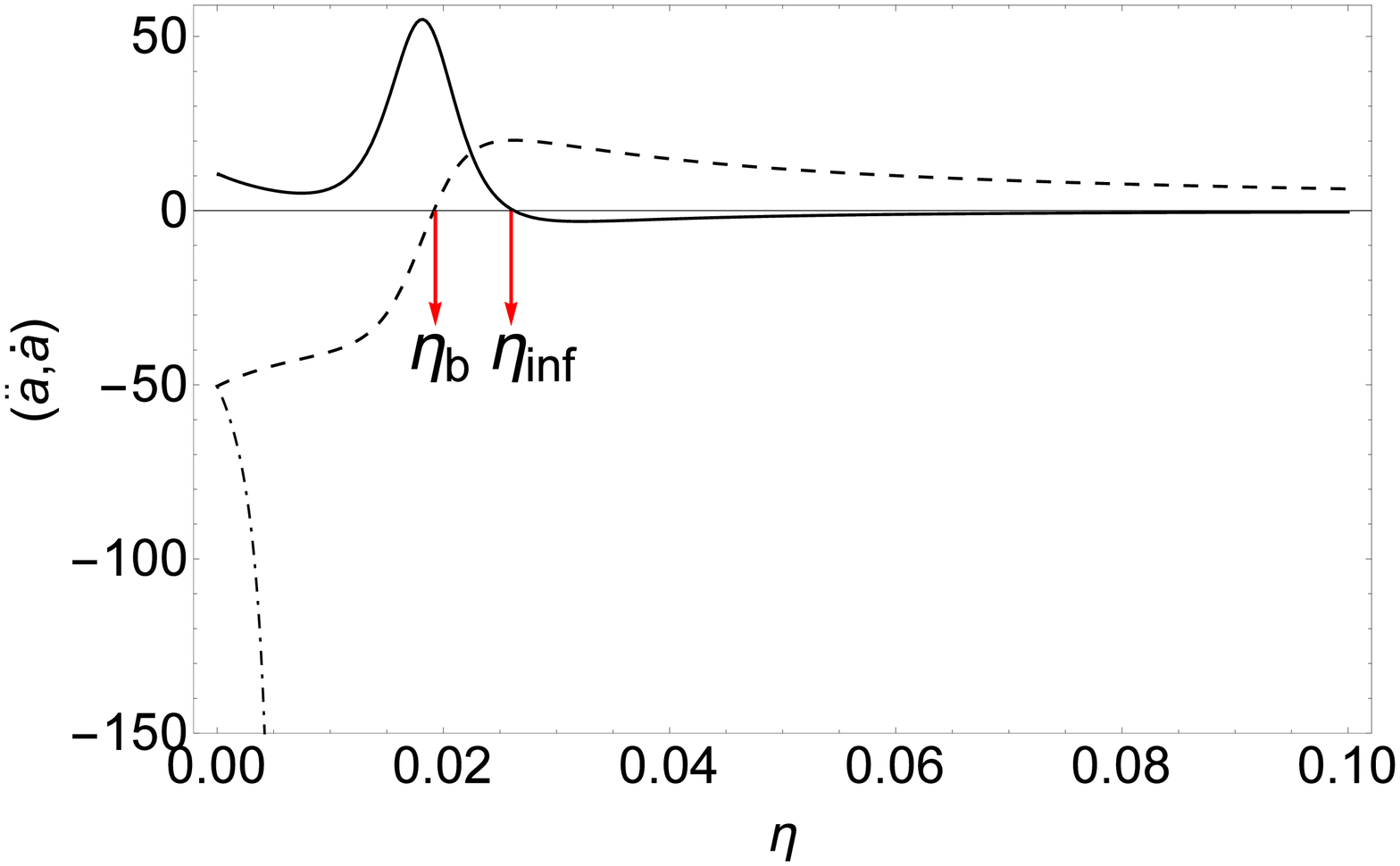}
\includegraphics[scale=0.4]{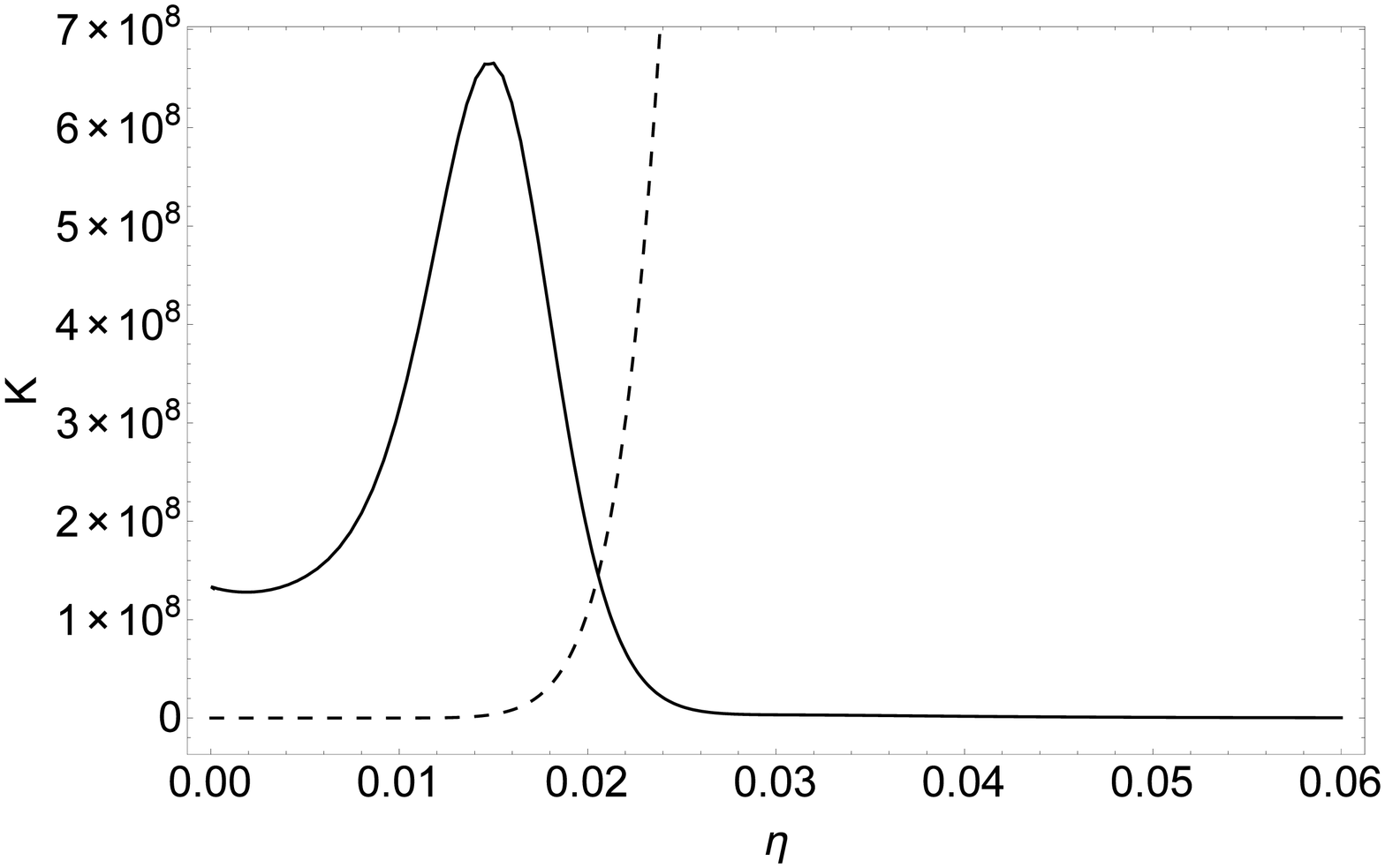}
\caption{Upper left panel: Time behavior of the scale factor for ${{\rm P}}_{\Phi}(\eta_i)=35.20$,
$\Phi(\eta_i)=16.68$, $L_\text{Pl}^4\rho_i=15.04$, $\alpha_i=0.0444$, $\omega=3.3$, $w=0$,
$\dot{a}_i=-50.40$, $\theta=0$ (dot-dashed curve), $\theta=0.316$ (full curve) and ${{\rm P}}_{\Phi}(\eta_i)=64.23$,
$\Phi(\eta_i)=18.53$, $L_\text{Pl}^4\rho_i=32.88$, $\alpha_i=0.0444$, $\omega=3.3$, $w=1$,
and $\theta=0.316$ (dashed curve).
Upper right panel: Time behavior of the collapse velocity $\theta=0$ (dot-dashed curve),
$\theta=0.316$ (dashed curve) and collapse acceleration (full curve) for the
same initial values of the parameters as above. Lower panel: Time behavior
of the Kretschmann scalar for ${{\rm P}}_{\Phi}(\eta_i)=35.20$, $\Phi(\eta_i)=16.68$,
$L_\text{Pl}^4\rho_i=15.04$, $\alpha_i=0.0444$, $w=0$,  $\dot{a}_i=-50.40$, $\omega=3.3$,
$\theta=0.316$ (full curve) and $\theta=0$ (dashed curve).}\label{sf}
\end{figure}

\begin{figure}
\includegraphics[scale=0.46]{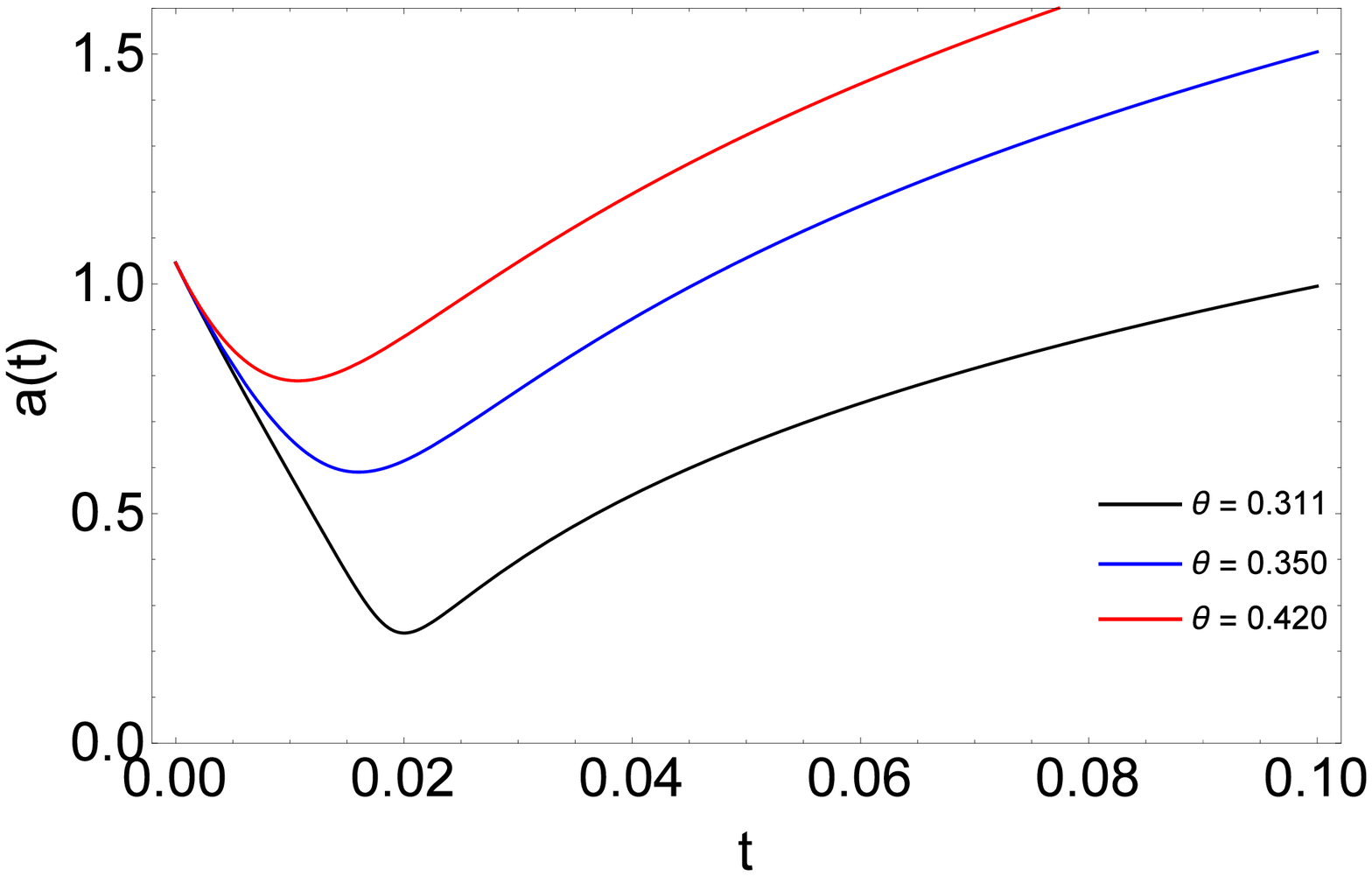}
\includegraphics[scale=0.46]{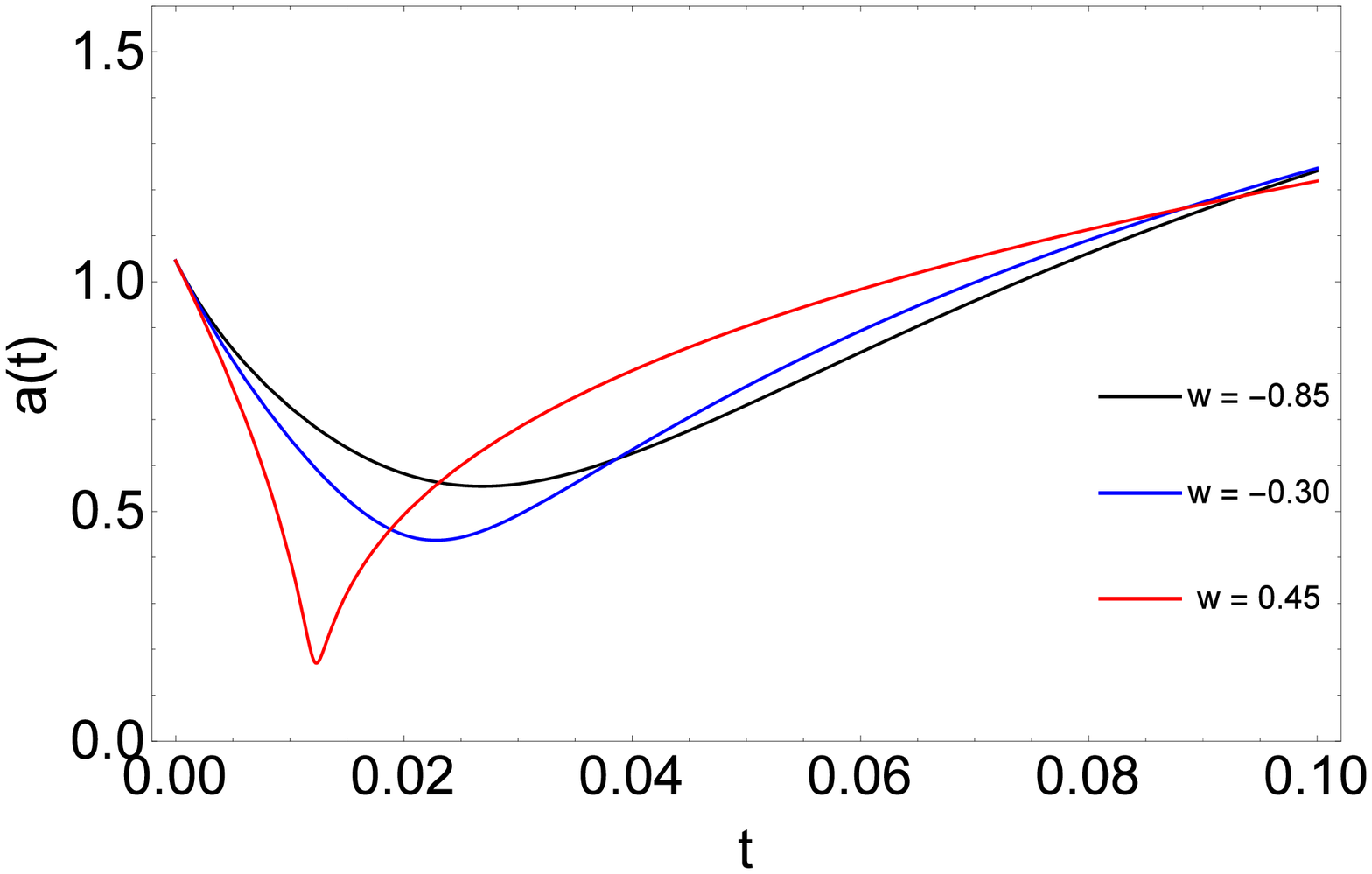}
\caption{Left panel: Time behavior of the scale factor for different
values of the non-commutative parameter and ${{\rm P}}_{\Phi}(\eta_i)=35.20$,
$\Phi(\eta_i)=16.68$, $L_\text{Pl}^4\rho_i=15.04$, $\alpha_i=0.0444$, $\omega=3.3$,
$w=0$,  $\dot{a}_i=-50.40$. Right panel: Time behavior of the scale
factor for different values of EoS parameter for $\theta=0.316$.
}\label{f19}
\end{figure}

\begin{figure}
\includegraphics[scale=0.36]{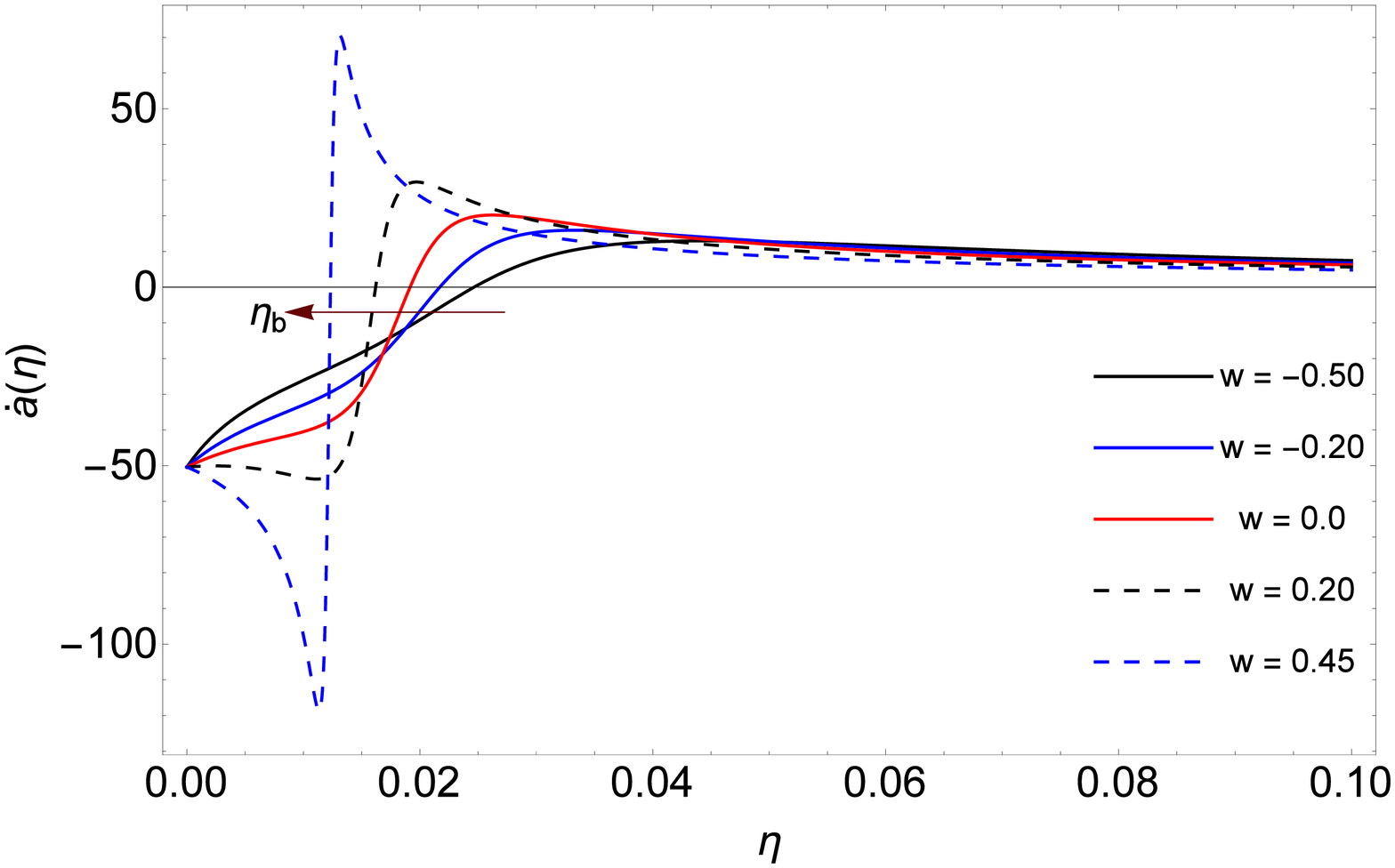}
\includegraphics[scale=0.38]{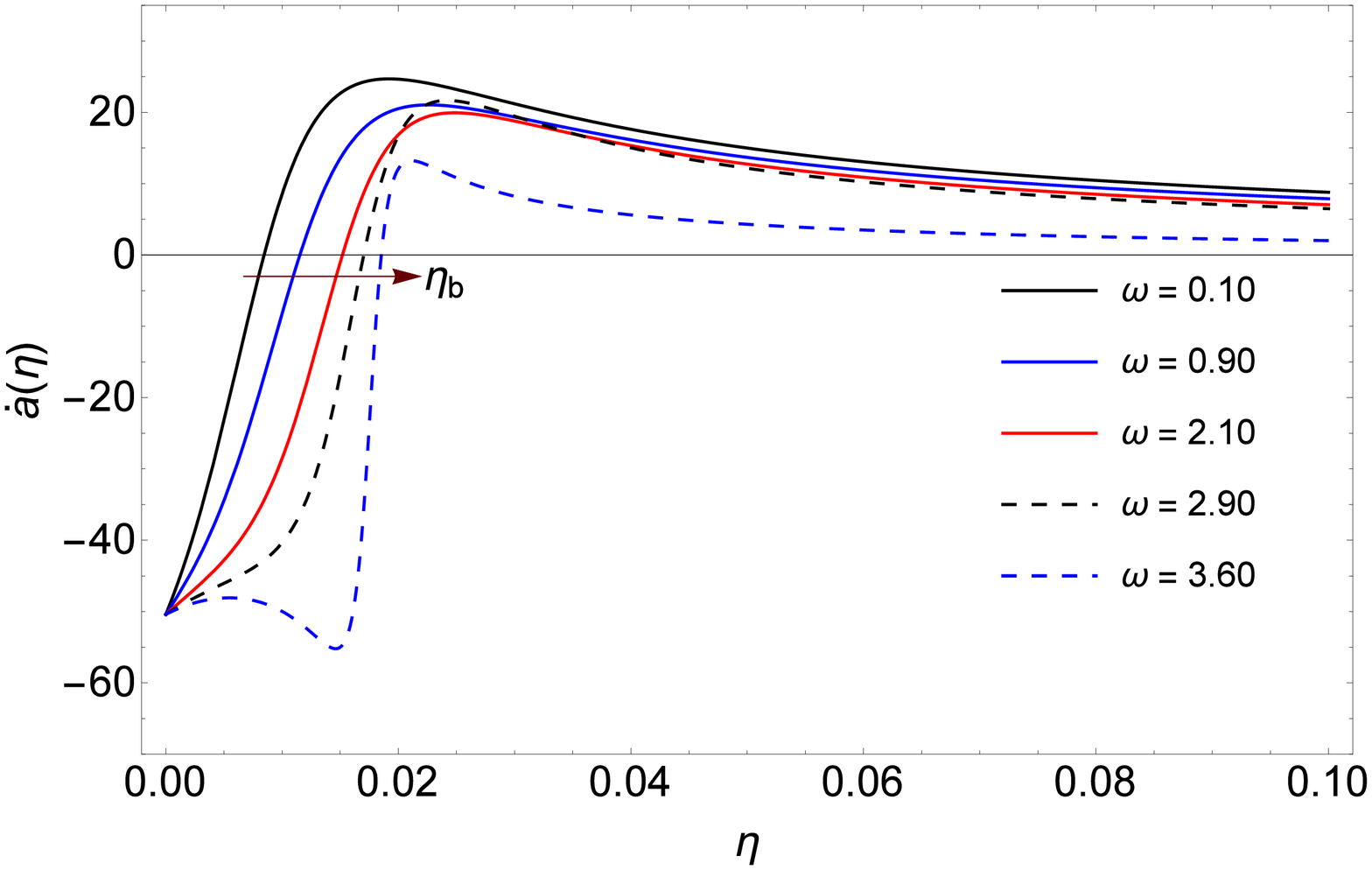}
\includegraphics[scale=0.4]{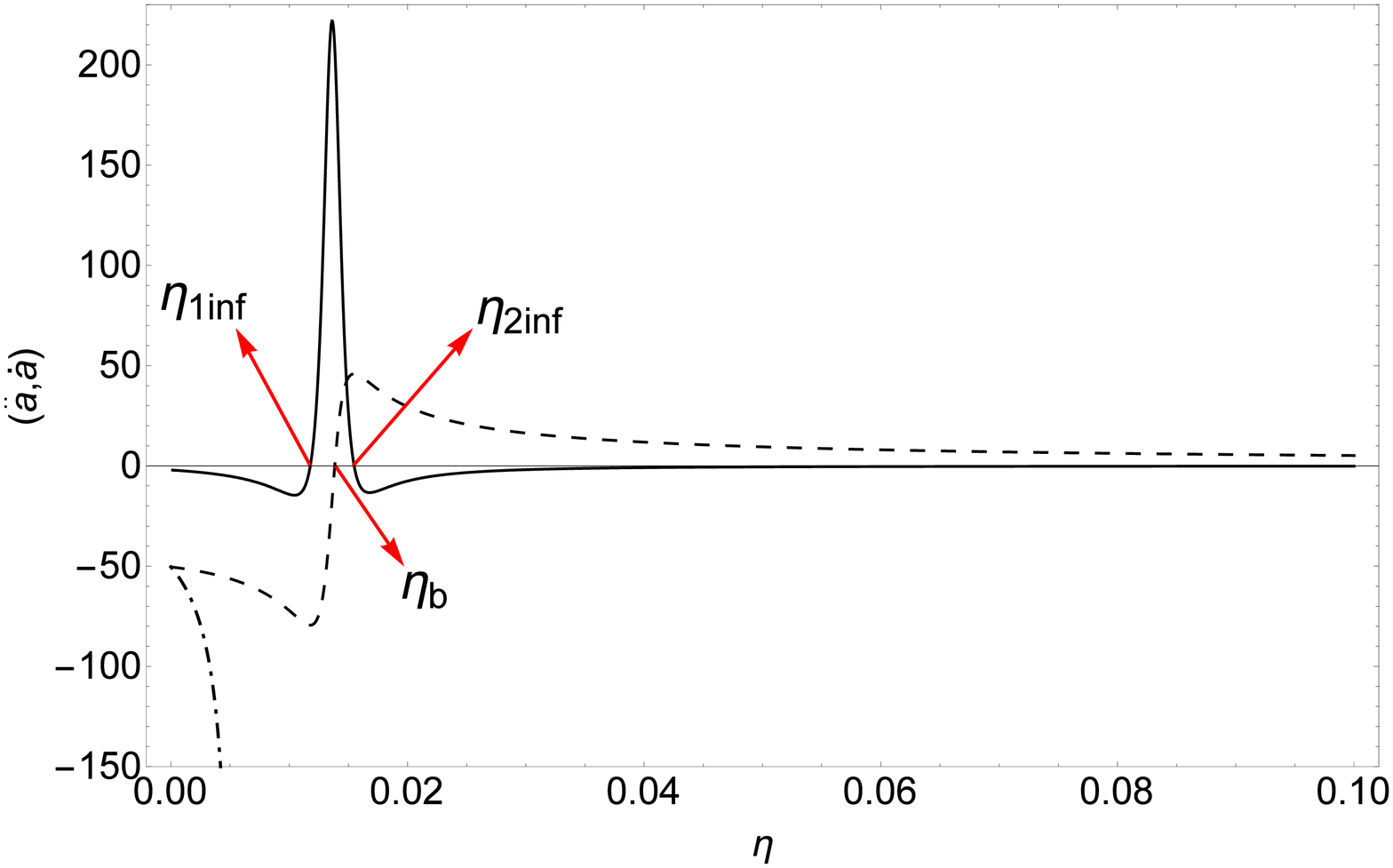}
\caption{Left panel: Time behavior of the collapse velocity for different values of the EoS parameter and ${\rm P}_{\Phi}(\eta_i)=35.20$, $\Phi(\eta_i)=16.68$, $L_{{\rm Pl}}^4\rho_i=15.04$, $\alpha_i=0.0444$, $\dot{a}_i=-50.40$, $\omega=3.3$ and $\theta=0.316$. The arrow shows the direction of decrease in the bouncing time as the EoS parameter increases.
Right panel: Time behavior of the collapse velocity for different values of the BD coupling parameter and $w=0.10$. The other initial data has been chosen with the same values as above. The arrow shows the direction of increase in the bouncing time as $\omega$ increases. Lower panel: Time behavior of the collapse velocity (dashed curve) and its acceleration (full curve). The initial data has been taken same as above with $w=0.35$,  $\theta=0.316$ (full and dashed curves) and $\theta=0$ (dot-dashed curve).}\label{womega}
\end{figure}

The upper left panel in Fig.~\ref{sf} shows the time behavior of the scale
factor for dust collapse\rlap.\footnote{The choice of dust is for simplicity, though we will also probe the
effects of the non-zero pressure, see, e.g., Figs. \ref{f19} and \ref{womega}.}
%(we have considered a vanishing potential so the BD scalar field is massless).
In the absence of non-commutative effects, the collapse scenario
terminates at a singularity as the dotted curve shows. When non-commutative effects of the
type considered in this paper are present, the evolution of the scale factor deviates
from the commutative case as the collapse proceeds. The collapse process
halts at a minimum value of the scale factor where a non-singular bounce occurs.
%which is greater than zero (see the solid curve).
%Such a scenario prohibits the singularity formation and leaves
%the scale factor regular and continuous throughout the dynamical
%evolution of the collapsing body (see the full curve).

In more detail, as the upper right panel in Fig.~\ref{sf} shows,
in the presence of non-commutative
effects, the collapse begins its evolution with a
decelerating contracting phase ($\dot{a}<0$ and $\ddot{a}>0$),
%proceeding within this regime (see the full curve)
and halts at the bounce time ($\eta_{{\rm b}}$), where the speed of collapse
(see the dotted curve) vanishes and the scale factor reaches
its minimum value. Then, $\dot{a}$ turns to positive values where an
expanding regime begins. After a time at which a soft bounce occurs, the collapse
experiences an accelerated expanding phase ($\ddot{a}>0$ and $\dot{a}>0$) till the time at which the
acceleration vanishes at the inflection point ($\eta_{{\rm inf}}$) and
then turns its sign to negative values. After this,
the collapse goes under a decelerating expanding phase ($\ddot{a}<0$ and $\dot{a}>0$).
The speed of collapse asymptotically vanishes at late times.

The lower panel of Fig.~\ref{sf} shows the behavior of Kretschmann scalar
\begin{equation}\label{KRSC}
K=R^{a b c d}R_{a b c d}=12\left[\ddot{\alpha}^2+2\dot{\alpha}^2+2\dot{\alpha}\ddot{\alpha}\right],
\end{equation}
where we see that this quantity behaves regularly when non-commutative effects are present and
diverges for $\theta=0$ signaling the formation of a  singularity. The lower right panel of Fig.~\ref{sf} shows the behavior of BD scalar field for the case of stiff fluid. The near bounce behavior of the scalar field has been discussed in appendix~\ref{APPBD}.

The left panel of Fig.~\ref{f19} shows the role of non-commutative parameter on
the minimum value of the scale factor (for the case of dust collapse) at which the bounce occurs.
It is seen that as $\theta$ increases, the minimum value of
the scale factor at the bounce time decreases. On the other hand,
%as the noncommutative
%parameter takes smaller values, not only the minimum value
%of the scale factor decreases (more noncommutativity is needed for the volume
%to be contracted), but also
it takes more time for the bounce to happen. It is worth mentioning that, a suitable choice of initial configuration of the collapse would lead to such a behavior for the scale factor for almost all values of the non-commutative parameter that satisfy $0\leq\theta<1.$

In the right panel of Fig.~\ref{f19},
we present the behavior of the scale factor for different values of the
EoS parameter in the presence of non-commutative effects.
It is seen that the EoS parameter controls the time
interval at which the bounce occurs (hereafter we call it $\Delta \eta_{{\rm b}}$) or more
precisely the softness of the bounce\footnote{The time
interval $\Delta \eta_{{\rm b}}$ for a soft bounce is much larger than a fast bounce.}. As the EoS parameter changes
its sign from negative to positive values, $\Delta \eta_{{\rm b}}$
decreases leading to a fast bounce. Furthermore, the more negative
the value of EoS parameter the less the minimum value for the scale
factor at the bounce. It is worth mentioning that though the overall
behavior of the curves in the right panel of Fig.~\ref{f19}
is similar (all the curves exhibit a bouncing scenario), the collapse dynamics and the location of  horizons
varies in response to the change in EoS and BD coupling parameters. Let us first deal with the collapse velocity. In the left and right panels of Fig.~(\ref{womega}), we have plotted the collapse
velocity for several values of EoS and BD coupling parameters. We see that, as the
EoS increases from negative to positive values (fixing the $\omega$ parameter, see
the upper panel), the collapse stops and turns faster to the expansion, with the bounce time
getting advanced. Conversely, as the BD coupling parameter increases (fixing the EoS,
see the right panel), the contracting phase turns faster to expansion with the bouncing time getting retarded.

The lower panel of Fig.~\ref{womega} shows how the dynamical evolution of the
collapsing body experiences one more phase compared to the dust case. Let us be more precise.
For $w>0$, the collapse scenario begins with an accelerated contracting phase ($\ddot{a}(\eta)<0$ and $\dot{a}(\eta)<0$ in the time interval $0<\eta<\eta_{{\rm 1inf}}$) till the first inflection point is reached at which $\ddot{a}(\eta_{{\rm 1inf}})=0$ and the collapse velocity has gained its maximum value in negative direction. The collapse then proceeds with a decelerating contracting phase before the bounce occurs, i.e., $\ddot{a}(\eta)>0$ and $\dot{a}(\eta)<0$, where $\eta_{{\rm 1inf}}<\eta<\eta_{{\rm b}}$. Just after the bounce occurs, the collapse turns to an accelerating
 expanding phase for which $\ddot{a}(\eta)>0$ and $\dot{a}(\eta)>0$ that happens in the time interval $\eta_{{\rm b}}<\eta<\eta_{{\rm 2inf}}$. Finally, the dynamical evolution of the object ends through a decelerating expanding phase where $\ddot{a}>0$ and $\dot{a}<0$ and $\eta>\eta_{{\rm 2inf}}$.

%We shall discuss this issue in the next subsections.

%The different behaviors of the scale factor for both vanishing
%and non-vanishing noncommutative parameters is also reflected in the
%behavior of the apparent horizon.
\subsection{Dynamics of the apparent horizon}
In order to study the evolution of
the apparent horizon, we begin by re-writing the spacetime metric (\ref{metric}) into the double null form as
\be\label{DNF}
ds^2=-2d\xi^+d\xi^{-}+R^2(\eta,r)d\Omega^2,
\ee
with the null one-forms defined as
\bea\label{NOFS}
d\xi^+&=&-\f{1}{\sqrt{2}}\left[N(\eta)d\eta-e^{\alpha(\eta)} dr\right],\nn
d\xi^-&=&-\f{1}{\sqrt{2}}\left[N(\eta)d\eta+e^{\alpha(\eta)}dr\right],
\eea
and we have transformed the spatial sector of the metric (\ref{metric})
from Cartesian coordinates to spherical coordinates with the area radius
$R(\eta,r)=re^{\alpha(\eta)}$. The null vector fields dual to the null one-forms can be obtained easily as
\bea\label{NVFS}
\partial_+=\f{\partial}{\partial\xi^+}&=&\f{1}{\sqrt{2}}\left[\f{\partial_\eta}{N(\eta)}-\f{\partial_r}{e^{\alpha(\eta)}}\right],\nn
\partial_-=\f{\partial}{\partial\xi^-}&=&\f{1}{\sqrt{2}}\left[\f{\partial_\eta}{N(\eta)}+\f{\partial_r}{e^{\alpha(\eta)}}\right].
\eea
The condition for radial null geodesics obeying equation $ds^2 = 0$ implies the existence of two kinds
of future pointing radial null geodesics, which correspond
to $\xi^+={\rm constant}$ and $\xi^-={\rm constant}$.
The expansion factors along these geodesics are given by
\be\label{expf}
\Theta_{\pm}=\f{2}{R}\partial_{\pm}R.
\ee
The spacetime is said to be trapped, un-trapped or marginally trapped if, respectively \cite{SHAY}
\bea\label{THE}
\Theta_+\Theta_->0,~~~\Theta_+\Theta_-<0,~~~\Theta_+\Theta_-=0,
\eea
where the equality characterizes the outermost boundary of the trapped region, the apparent horizon. Therefore, a shell labeled by the comoving radial coordinate $r$ will get trapped if $\dot{R}_{{\rm ah}}^2(\eta,r)=1$, or equivalently
\be\label{AHC}
R_{{\rm ah}}(\eta,r)=\f{1}{|\dot{\alpha}(\eta)|}=\f{a(\eta)}{|\dot{a}(\eta)|}.
\ee

\begin{figure}
\includegraphics[scale=0.431]{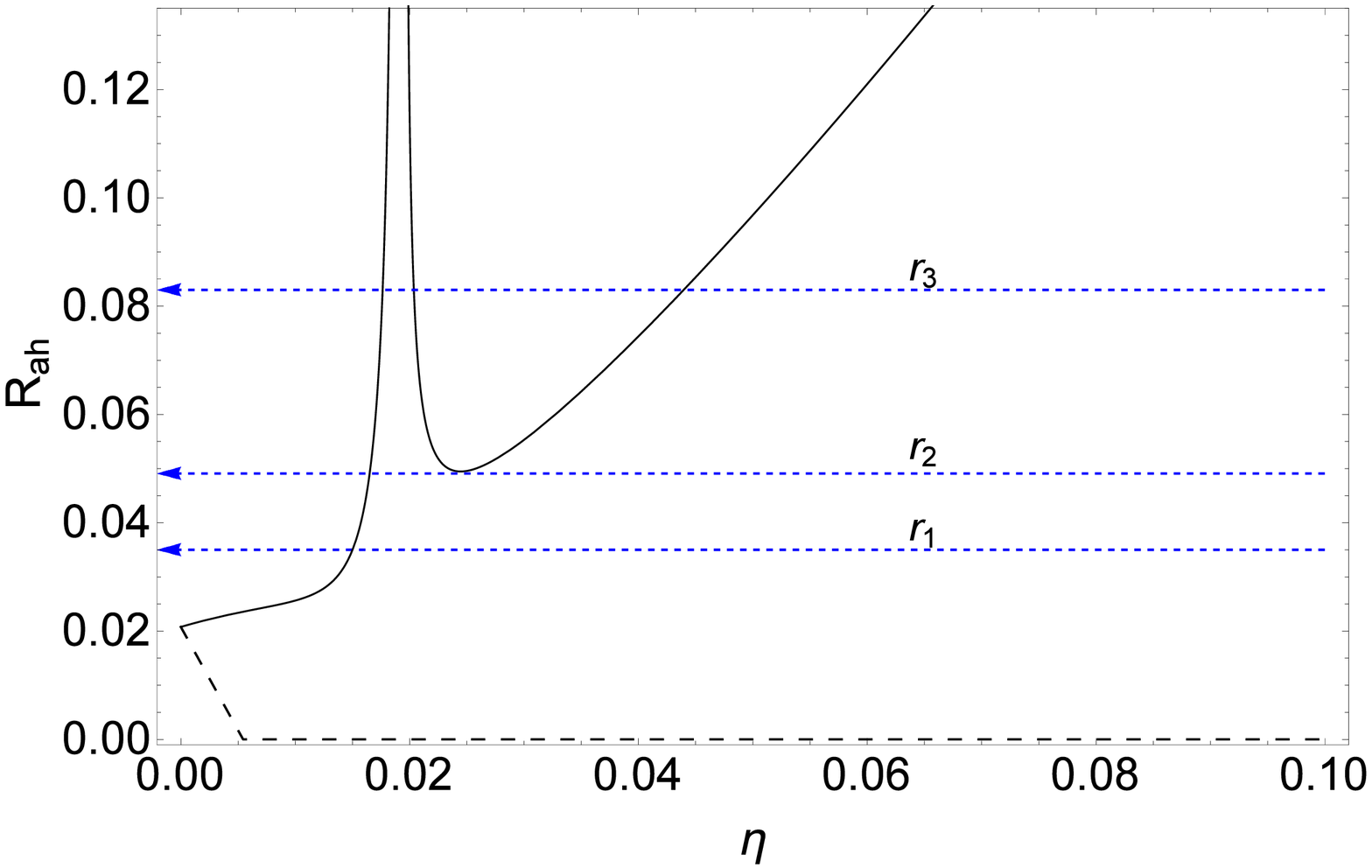}
\includegraphics[scale=0.334]{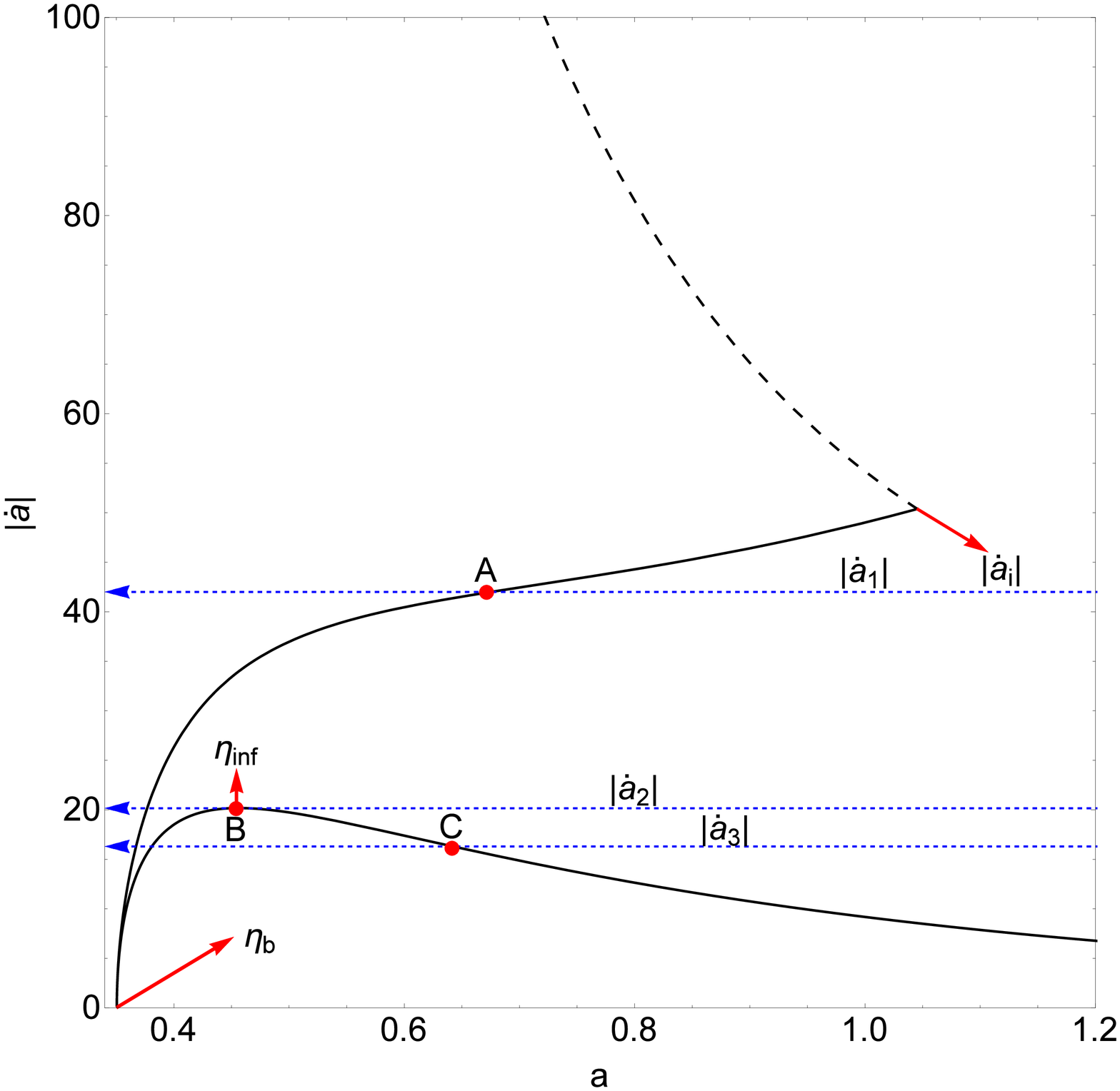}
\caption{Left panel: Time behavior of the apparent horizon curve for ${{\rm P}}_{\Phi}(\eta_i)=35.20$,
$\Phi(\eta_i)=16.68$, $L_\text{Pl}^4\rho_i=15.04$, $\alpha_i=0.0444$, $\omega=3.3$,
$w=0$, $\dot{a}_i=-50.40$, $\theta=0.316$ (full curve) and $\theta=0$ (dashed curve).
Right panel: The absolute value of the collapse velocity versus the scale factor for the same initial values of the left panel and $\theta=0.316$ (full curve) and $\theta=0$ (dashed curve).
The dashed arrows show the values of collapse velocity for which the
horizon equation is satisfied. The initial data has been set in the same way as the left figure.}\label{rahh}
\end{figure}

The left panel in Fig.~\ref{rahh} shows the behavior of the physical areal radius, $R_{ah}$, of the apparent
horizon for the case of dust fluid. It is seen that when the effects of non-commutativity are present (full curve),
the radius of the apparent horizon increases for a while and diverges just before the
bounce where a contracting phase governs the scenario.
It then decreases in the expanding phase to a minimum and
monotonically increases at late times. From the above equation we can extract
that, depending on the behavior of the collapse velocity, the horizons
could either form or are avoided. This depends on how the initial radius of
the collapsing object, i.e., $R(\eta_i,r_{\Sigma})=r_{\Sigma}$, is chosen:
\begin{itemize}
  \item Having taken the boundary of the collapsing cloud in such a way that
$r_{\Sigma}=r_{\rm {1}}$, then Eq.~(\ref{AHC}) implies that once the
absolute value of collapse speed and the scale factor reach the values $\left\{|\dot{a}_1(\eta)|,a_1(\eta)\right\}$ (see also point A in the right panel of Fig.~\ref{rahh}), the horizon equation is satisfied just once and a dynamical horizon would form to meet the boundary in the contracting phase. This means that the uttermost boundary of the trapped region will intersect the boundary of the collapsing cloud, i.e., $R_{{\rm ah}}(\eta,r_{\rm 1})=r_\Sigma$.%R_{{\rm ah}}(\eta,r_1)=Dashing[{.01, .015}]
\item For $r_{\Sigma}=r_{\rm {2}}$
or correspondingly when the absolute value of collapse speed and the scale factor reach the values $\left\{|\dot{a}_2(\eta)|,a_2(\eta)\right\}$ (point B), two horizons could form; the first one in the collapsing phase and the second one in the expanding phase.
  \item However, for $r_{{\rm 1}}<r_{\Sigma}<r_{{\rm 2}}$, since the horizon equation is satisfied once, only one horizon
could form in the contracting regime and the post-bounce regime is
free of horizon formation.
\item Finally, for $r_{\Sigma}>r_{{\rm 2}}$, the horizon equation is fulfilled three times where the absolute value of collapse speed and the scale factor assume the values $\left\{|\dot{a}_3(\eta)|,a_3(\eta)\right\}$ (point C). The first one appears in the contracting phase, the second one in
the accelerated expanding regime and the third one forms in the
decelerated expanding regime to cover the bounce. We then conclude
that the boundary surface of the collapsing body can be taken
sufficiently small so that equation (\ref{AHC}) is never
satisfied, i.e., no horizon would develop to meet the boundary as the speed of collapse is always bounded (see the full curve in the right panel of Fig.~\ref{rahh}). As a result, the formation of the horizon can be prevented in the presence of non-commutative effects.
However, for $\theta=0$ (see dashed curve in the left panel of Fig. \ref{rahh}), the radius of the dynamical horizon decreases
monotonically and vanishes at a finite amount of time. Thus, there can
not be found a minimum radius for the boundary of the collapsing
volume (since the collapse velocity is unbounded; see the dashed curve in the right panel of Fig. \ref{rahh})
in order to avoid the horizon formation.
\end{itemize}

%A point that deserves more elaboration here (see Fig.~\ref{rahh}) is that one
%may hypothetically consider the surface boundary of the collapsing cloud, say $r_{{\rm 1b}}$, to be greater than the location of the points living on the dashed curve and then claim $r_{{\rm 1b}}>r_{{\rm ah}}^{\theta=0}$, i.e., the horizon would not meet the boundary. This makes physically no sense, since in the collapse setting one tries to take the initial size (or correspondingly the initial mass) of the object as small as possible, so that in the later stages of the collapse the gravitational attraction will prevent propagating of light, i.e., the horizon formation is failed.
%This is astrophysically reasonable as we know from the famous Chandrasekhar limit for white dwarfs and Tolman-Oppenheimer-Volkoff limit for neutron stars.

%At first sight, we note that the initial hyper surface from which the collapse commences must not be a portion of trapped region, that is, there should not be any trapping of light at initial epoch. This is what the regularity condition demands. Therefore, we require $r_{\Sigma}<r_{{\rm ah}}(t_i)$ in order that the regularity condition be satisfied.
%\subsection{Collapse velocity}

\subsection{Horizon location for non-dust case}
\begin{figure}
\includegraphics[scale=0.39]{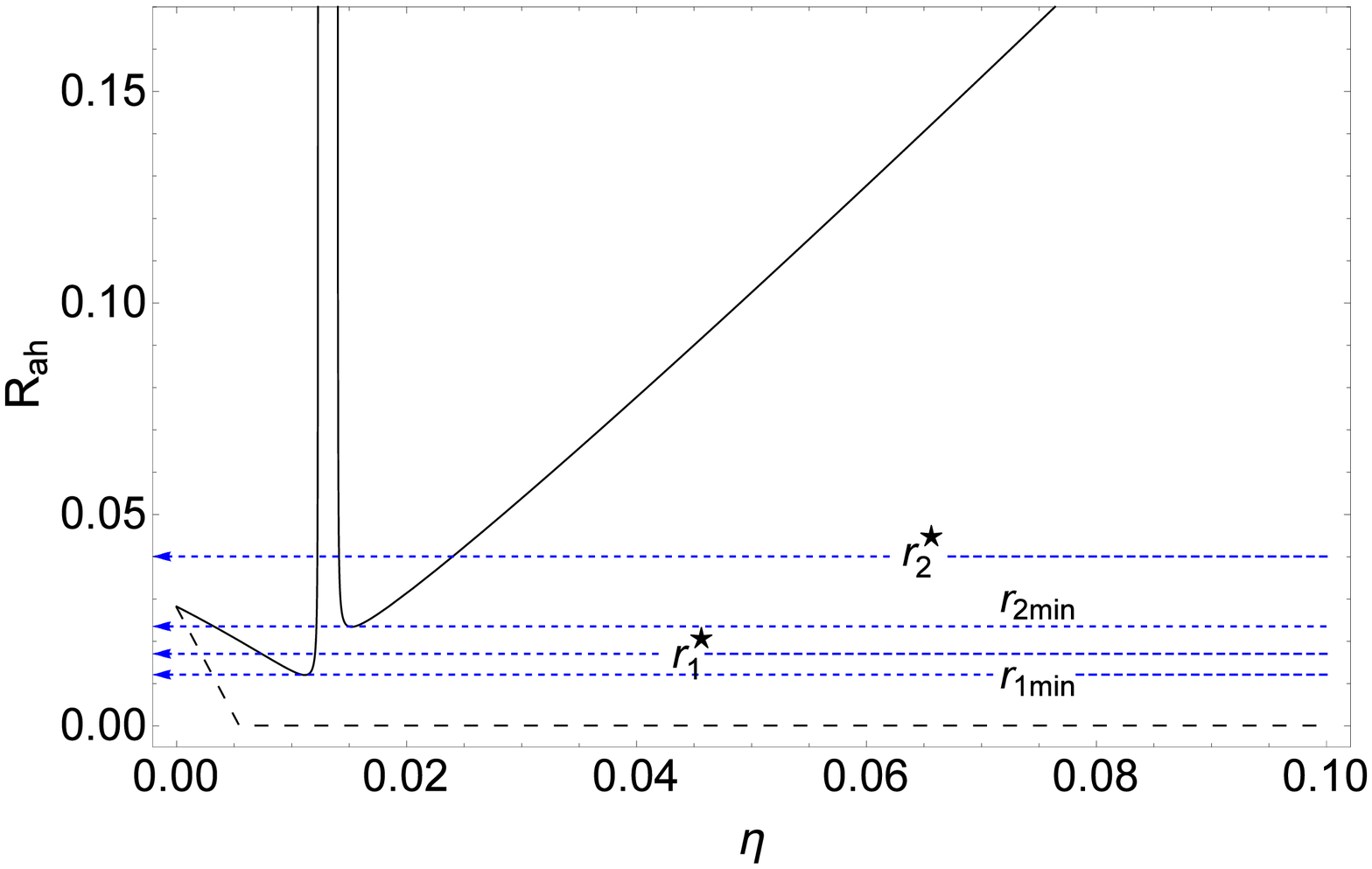}
\includegraphics[scale=0.35]{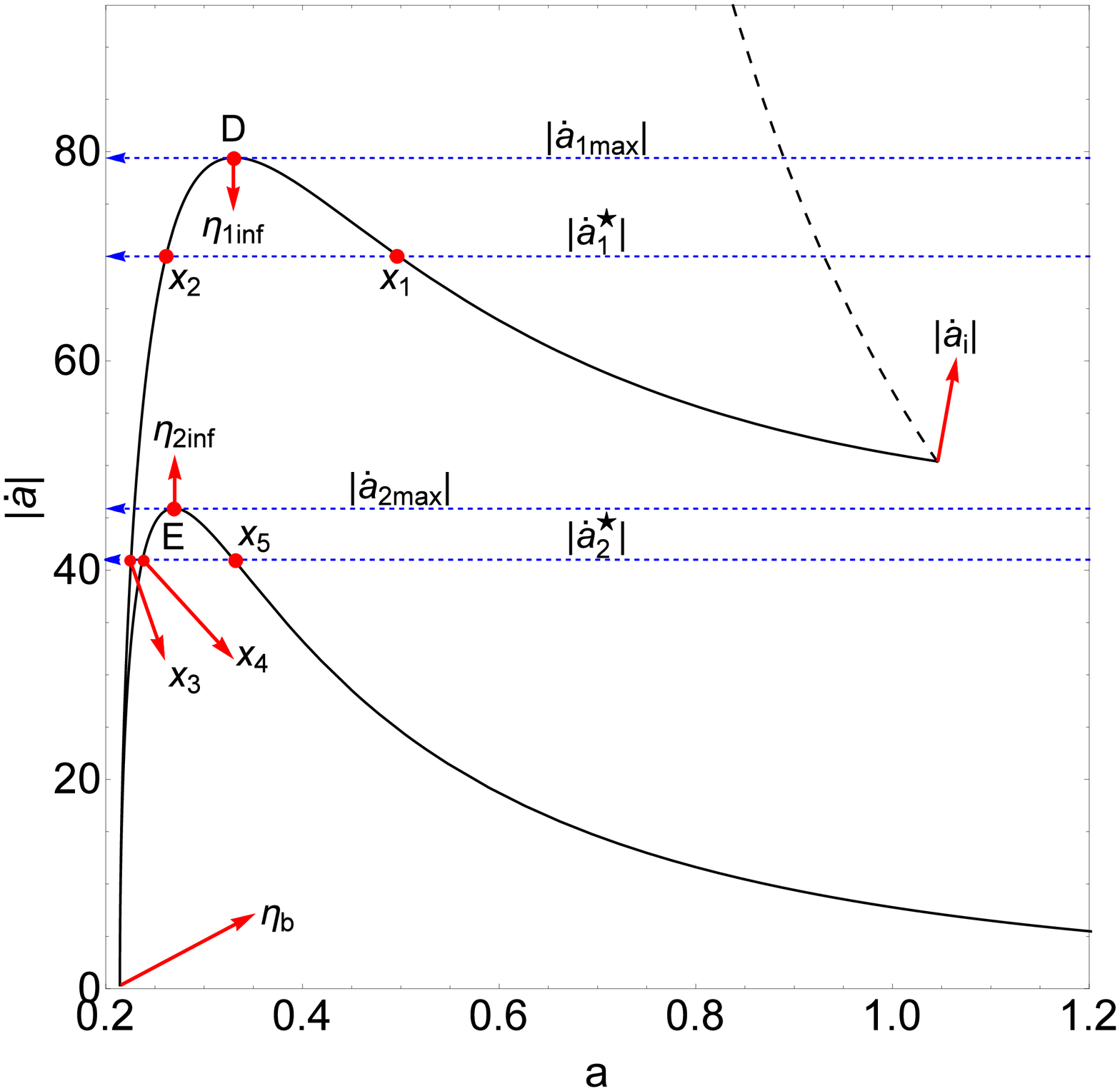}
\caption{Left panel: The behavior of the apparent horizon curve for ${\rm P}_{\Phi}(\eta_i)=35.20$, $\Phi(\eta_i)=16.68$, $L_\text{Pl}^4\rho_i=15.04$, $\alpha_i=0.0444$, $\dot{a}_i=-50.40$, $\omega=3.3$, $w=0.35$ and $\theta=0.316$ (full curve) and $w=0.35$ and $\theta=0$ (dashed curve). Right panel: The absolute value of the collapse velocity versus the scale factor for $\omega=3.3$, $w=0.35$ and $\theta=0.316$ (full curve) and $\theta=0$ (dashed curve). The initial data have been chosen same as above.}\label{rahaaw}
\end{figure}

Let us now check the location of dynamical horizons for $w>0$.
The left panel in Fig.~\ref{rahaaw} shows that the apparent horizon
decreases for a while till the first inflection point, at which the
absolute value of speed of collapse reaches its maximum value, at the corresponding value of the scale factor, in the
contracting regime, $\left\{|\dot{a}_{{\rm 1max}}|,a_1\right\}=\left\{|\dot{a}(\eta_{{\rm 1inf}})|,a(\eta_{{\rm 1inf}})\right\}$
(see also point D in the right panel). It then increases and diverges to infinity at
the bounce time. After the bounce, the apparent horizon decreases again to
its local minimum value at the second inflection point where the absolute
value of the speed of the collapse reaches its maximum (Point E) in the post bounce
regime, i.e., $\left\{|\dot{a}_{{\rm 2max}}|,a_2\right\}=\left\{|\dot{a}(\eta_{{\rm 2inf}})|,a(\eta_{{\rm 2inf}})\right\}$. The boundedness
of the speed of collapse signals the fact that, depending on the suitable
choice of the boundary surface of the collapsing body, formation of dynamical horizons can be avoided. Therefore, in view of the Fig.~\ref{rahaaw}, we may deduce the following considerations:
 \begin{itemize}
   \item If we take the boundary to be $r_{\Sigma}\leq r_{\rm{1min}}$, then no
   horizon would form throughout the dynamical evolution of the object, while the
   equality leads to the formation of only one horizon in the contracting phase.
   \item If we take $r_{\Sigma}=r_{1}^{\star}$ then, two dynamical horizons would appear once the absolute value of the
   speed of collapse get the value $|\dot{a}_{1}^{\star}|$ at the corresponding values for the scale factor; the first horizon forms in the accelerated contracting regime once the point ${\sf x}_1$ is reached and the second one forms at ${\sf x}_2$, after the first inflection point, where the collapse undergoes a decelerated contracting regime.

   \item For $r_{\Sigma}< r_{\rm{2min}}$ the expanding regime is free of horizon formation, since for this choice of the boundary surface, the horizon equation is never satisfied in the expanding regime.
   %\item  while for $r_{\Sigma}=r_{\rm{1min}}$ the bounce is necessarily covered by a dynamical horizon.
   \item Finally, for $r_{\Sigma}> r_{\rm{2min}}$, say $r_{\Sigma}=r_{2}^{\star}$, three horizons may form; the first one in the decelerated contracting regime (once the curve in $(|\dot{a}|,a)$ plane reaches the point ${\sf x}_3$). The second one occurs in the inflationary expanding regime, once the speed of collapse and the scale factor reach the point ${\sf x}_4$ and finally, the third one would form at ${\sf x}_5$, after the second inflection point, i.e., in the decelerating expanding regime.
    \end{itemize}

   However, when the non-commutative effects are absent, the horizon formation
   cannot be avoided as te speed of the collapse is unbounded and a dynamical horizon would always form to cover the resulted singularity (see the dashed curve).

%For $w<0$ where the bounce point is passed smoothly, there exists one inflection point (same as the middle panel of figure (\ref{sf}) for dust case) with the same dynamical evolution as the dust case.

%From equation (\ref{NC.H.eq2})
%it is seen that for large values of the BD parameter, the first term can
%be neglected but this does not give rise to the GR limit of the model due
%to the presence of the noncommutative parameter. However, as figure (\ref{g18})
%%\rc{depending on the initial data set from which the collapse commences,}
\subsection{Oscillatory bounce, special cases and complementary perspective}
\begin{figure}
\includegraphics[scale=0.49]{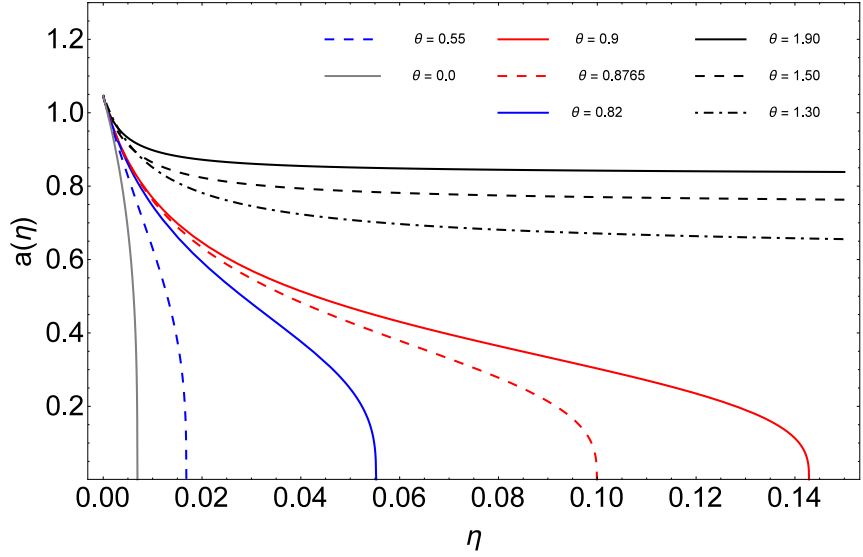}
\caption{Time behavior of the scale factor in large $\omega$ limit for
different values of the non-commutative parameter and ${\rm P}_{\Phi}(\eta_i)=35.20$,
$\Phi(\eta_i)=16.68$, $L_\text{Pl}^4\rho_i=15.04$, $\alpha_i=0.0444$, $w=0$,  $\dot{a}_i=-50.40$, $\omega=60000$.}\label{g18}
\end{figure}

In Fig.~\ref{g18}, we have plotted the time behavior of the scale
factor associated to dust cloud collapse for large values of the BD coupling parameter. We find
that there is a critical value $\theta_{{\rm c}}$ for the non-commutative parameter such
 that when $\theta>\theta_{{\rm c}}$, the scale
factor reaches a minimum value and stays at this value till late times (family of black curves). However, when this
parameter is less than $\theta_{{\rm c}}$, the collapse culminates in a
singularity formation with quite different dynamics during its
evolution (see the family of red and blue curves) in comparison to the case
where $\theta=0$ (gray curve).  Although different initial collapse settings imply different numerical
values for $\theta_{{\rm c}}$ (for instance, the initial conditions chosen for plotting
Fig.~\ref{g18} give $\theta_{{\rm c}}\approx0.87$), it should be noted that the mentioned behavior is
completely general and does not depend on the initial conditions at all.

\begin{figure}
\includegraphics[scale=0.45]{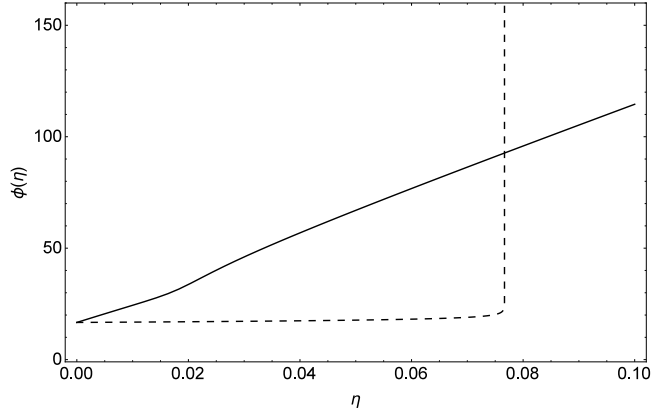}
\caption{Time behavior of the BD scalar field
for ${{\rm P}}_{\Phi}(\eta_i)=35.20$, $\Phi(\eta_i)=16.68$, $L_\text{Pl}^4\rho_i=15.04$, $\alpha_i=0.0444$, $w=0$,
$\dot{a}_i=-50.40$, $\omega=3.3$, $\theta=0.316$ full curve and $\theta=0$ (dashed curve).
%Right panel: Time behavior of the conformally
%transformed scale factor in Einstein frame for the same parameters as above.
}
\label{EF}
\end{figure}

Fig.~\ref{EF} presents the time behavior of BD scalar field for
both non-commutative and commutative settings for a dust cloud. While for the former the BD
field increases monotonically, for the latter, it undergoes a sudden
divergence at the singularity time.

%\rc{Are the following terms necessary? (This part is not necessary, Shahram)}In BD cosmology, for probing whether
%or not an inflation occurs in the
%conformal Einstein frame, we must consider not only the
%behavior of $a(t)$, but also we should take into account the behavior of the
%quantity $a(t)L_\text{Pl}/L_{\rm Pl}(t)$, where $L_\text{Pl}$ (Planck's length
%at present epoch) and $L_{\rm Pl}(t)=\sqrt{\frac{1}{\Phi(t)}}$  are the
%constant comoving and the Planck lengths, respectively, see, e.g.~\cite{GSS111}. \rc{In the right
%panel of Fig.~\ref{EF}, we have also plotted the time evolution of the quantity $\Phi(t)^{\f{1}{2}}a(t)$,
%Therefore, at distances comparable to the noncommutative parameter the
%quantum effects of gravity could arise to finally remove the singularity that appears in classical framework} \bl{What does the sentence tell? It is ambiguous. Please be careful Mehrdad!}
%where we see that the collapse process in the Einstein frame also turns to a bounce at
%different minimum value for the conformally transformed scale factor.
%However, in the absence of non-commutative effects the
%collapse scenario terminates in the singularity formation for the mentioned quantity. \rc{Please correct or in case not necessary, delete the terms as they could be challenging for referee.}

\begin{figure}
\includegraphics[scale=0.45]{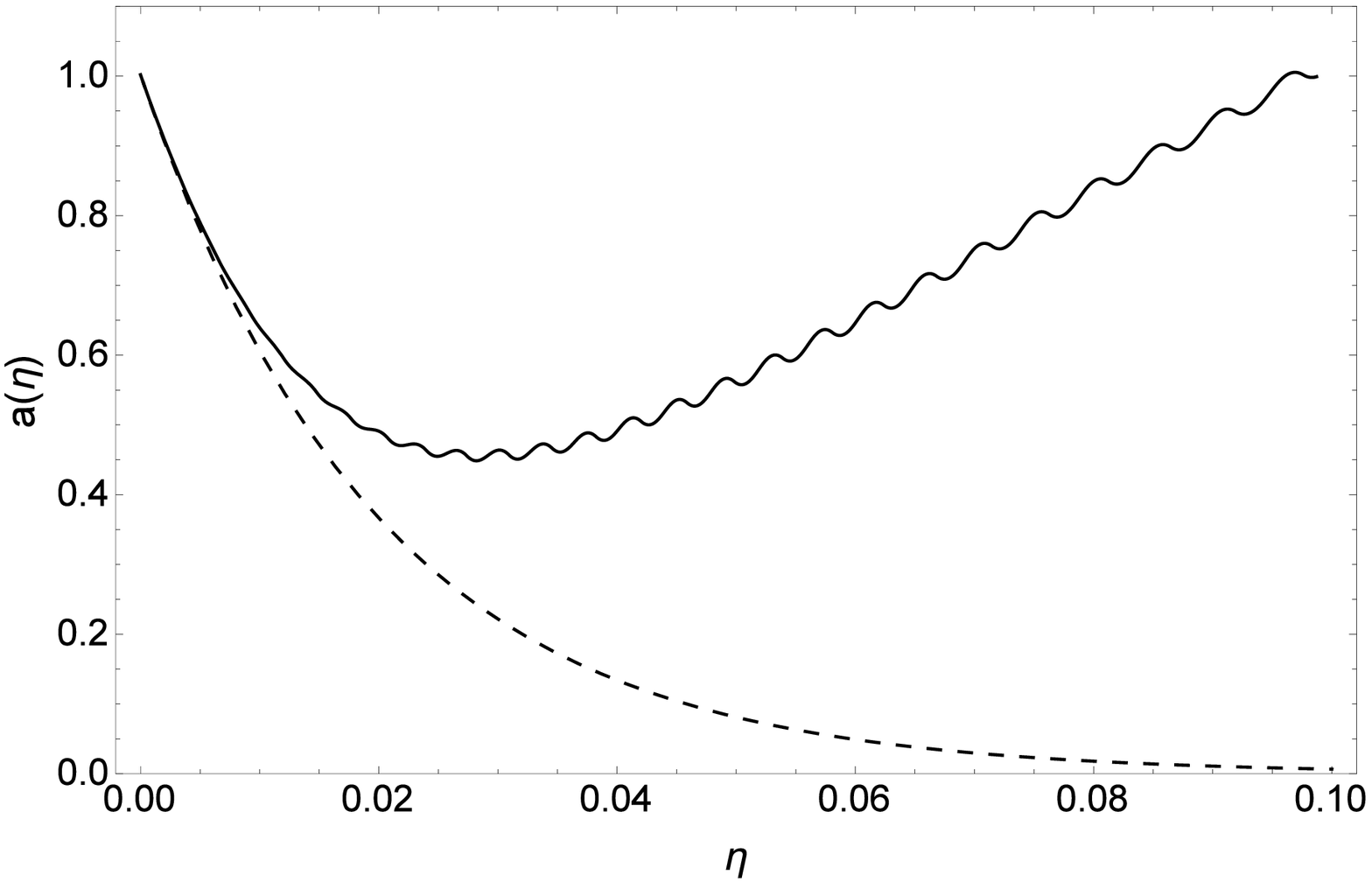}
\includegraphics[scale=0.428]{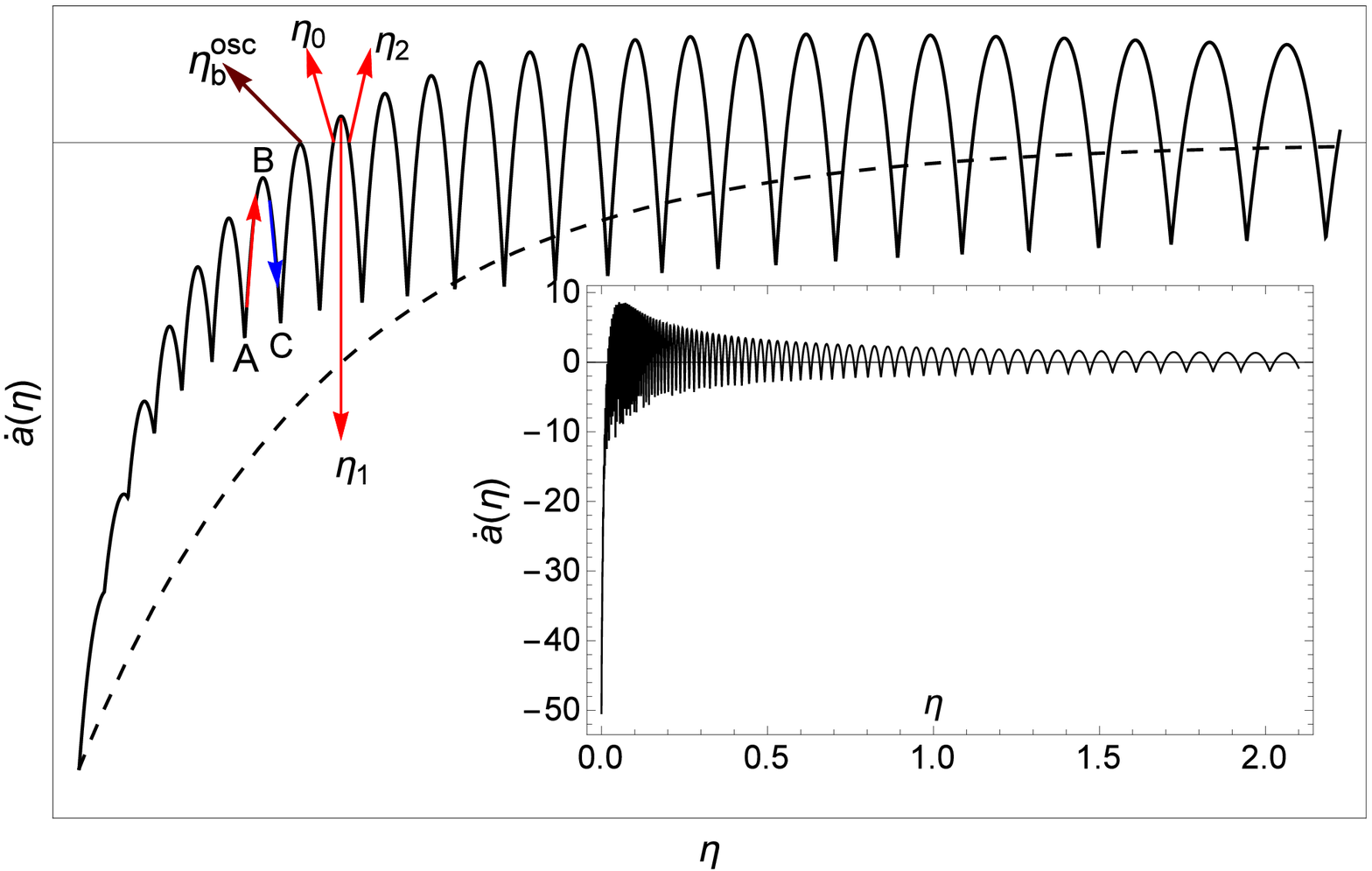}
\includegraphics[scale=0.41]{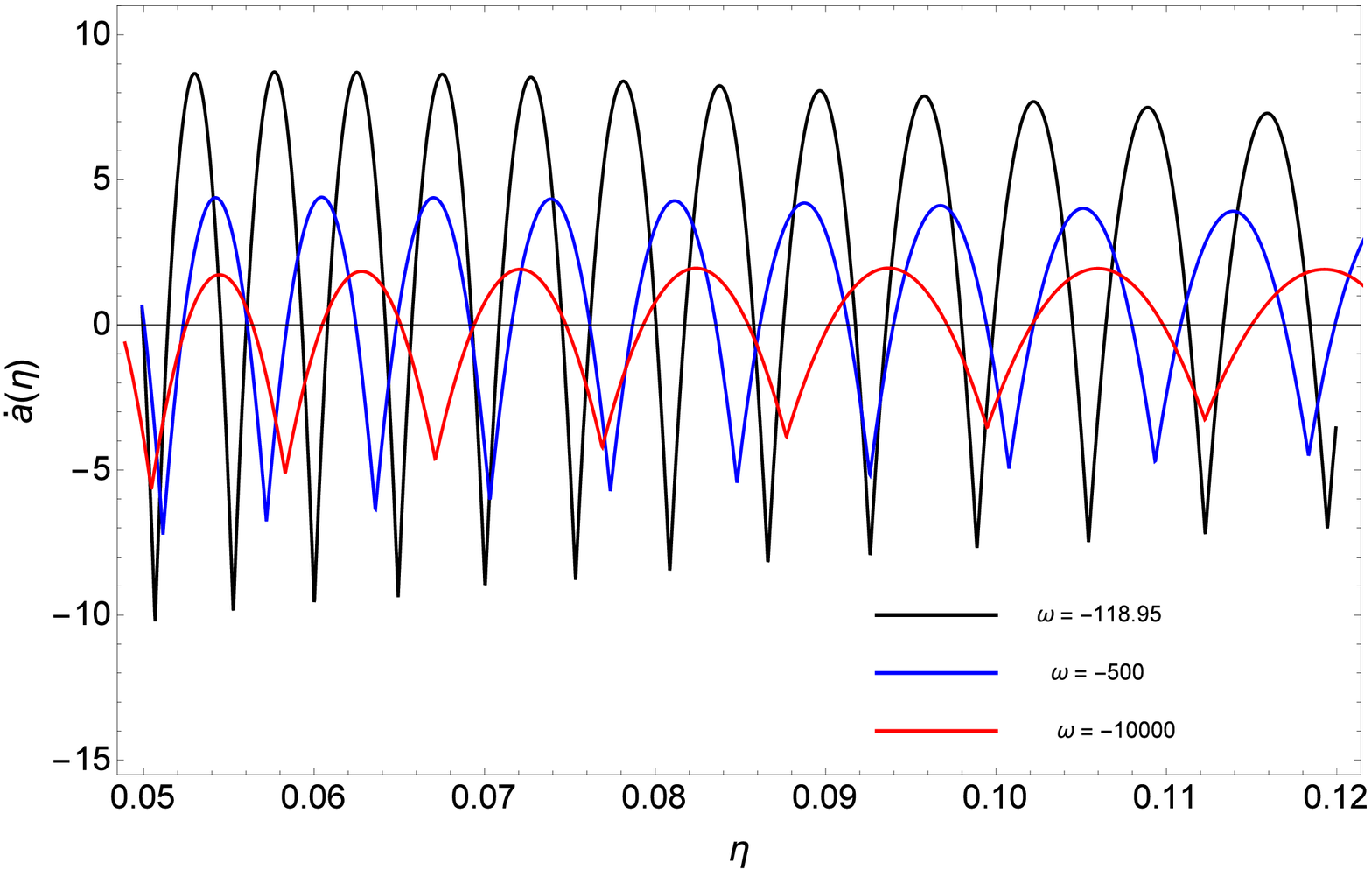}
\caption{Left panel: Time behavior of the scale factor for ${{\rm P}}_{\Phi}(\eta_i)=45.95$,
$\Phi(\eta_i)=9.15$, $L_\text{Pl}^4\rho_i=29.60$, $\alpha_i=0.002$, $\omega=-118.95$, $w=0$,
$\dot{a}_i=-50.40$, $\theta=0$ (dashed curve) and $\theta=0.316$ (full curve).
Right panel: The speed of collapse versus time for the same initial values as
above when $\theta=0$ (dashed curve) and $\theta=0.316$ (full curve). Lower panel: The frequency of the oscillation of collapse velocity for different values of BD coupling parameter.}\label{omega-}
\end{figure}

Another class of solutions concerns the collapse of dust fluid for negative
values of the BD coupling parameter.
%\rc{Such values have been found in the effective low energy models of the BD theory that arise from Kaluza-Klein and superstring theories \cite{negomkks}. %Furthermore, these negative values for the BD coupling parameter are required to explain the structure formation and accelerated expansion of the Universe %\cite{negomaeu}. We are therefore interested to investigate the collapse dynamics and its final outcome for negative values of $\omega$ parameter}.
The left plot in Fig. \ref{omega-} presents the behavior of scale factor for both non-commutative and commutative settings.
As the full curve shows, the scale factor decreases for a while, reaching a
time at which an oscillatory bounce begins. It then increases by keeping this
oscillatory behavior at the later times. The dashed curve shows the singular
behavior for a vanishing value of the non-commutative parameter. The right
panel shows the behavior of the collapse velocity, where we see that it begins
oscillating in the negative direction before the bounce, at which the speed of
collapse vanishes and proceeds this oscillatory behavior around its vanishing
value. The inset shows the late time behavior of the collapse velocity, where
it is seen that at early stages of the collapse the frequency of oscillation
of $\dot{a}$ is high and then it decreases at late times. Furthermore, the
envelope of the oscillatory phase shows a damping as the time advances, which
corresponds to the case where the collapsing body settles at rest at late
times. Such a behavior in the dynamics of the collapse is due to the highly
non-linear characteristic of the non-commutative equations.

\begin{figure}
\includegraphics[scale=0.65]{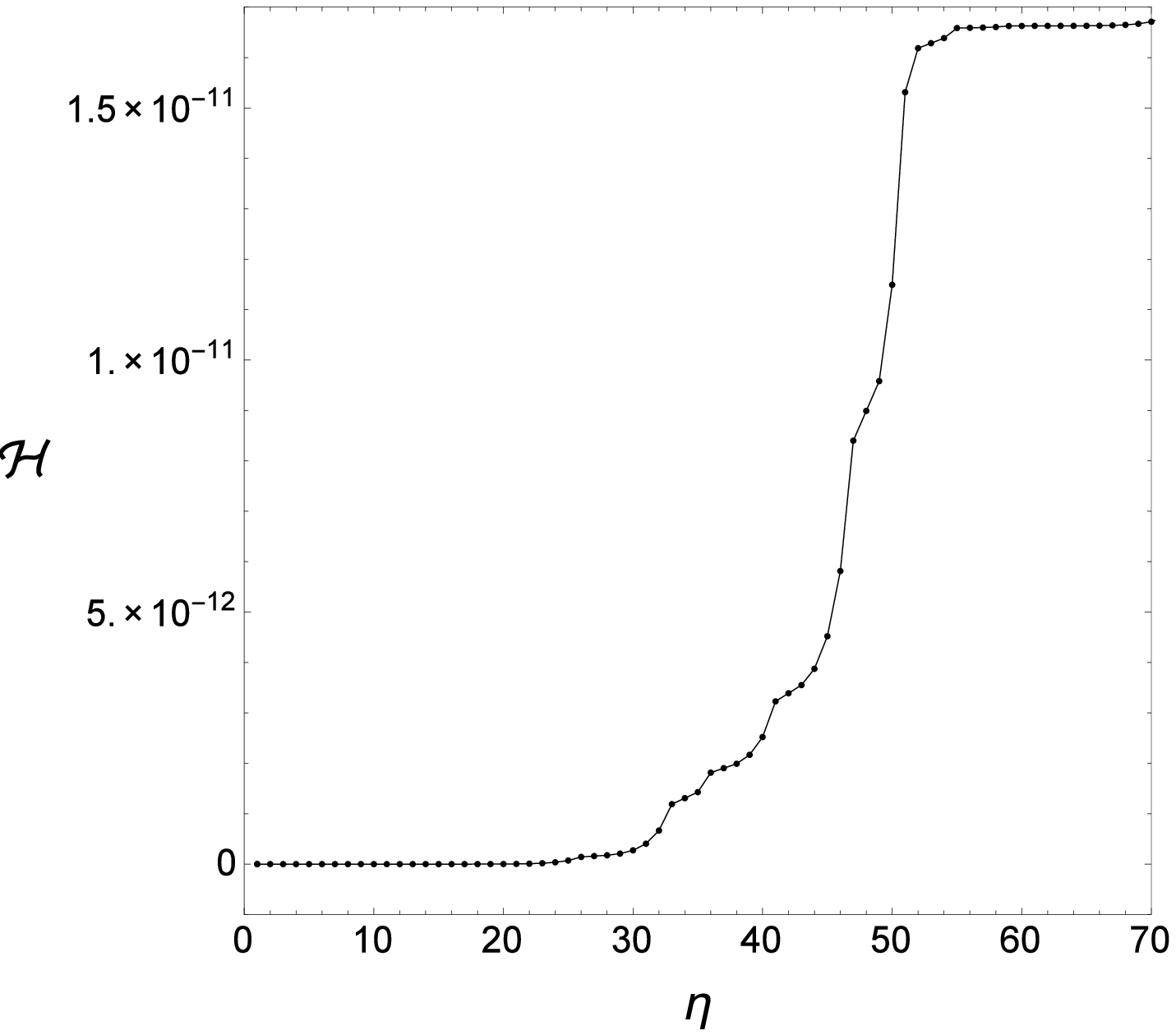}
\caption{The numerical error
of the  time evolution of Hamiltonian constraint (\ref{Ham-1}). }\label{ha}
\end{figure}

Finally, in Fig.~\ref{ha}, we have shown the time evolution of
Hamiltonian constraint, where we see that this constraint is fulfilled with the accuracy of $10^{-12}$ or less.

\subsection{Exterior solution}
The collapse setting studied so far deals with the interior of the collapsing cloud.
In order to complete the model we need to match the interior spacetime to a suitable exterior one. Let us take $({\rm V}^{\pm},g^{\pm})$ as four dimensional spacetimes and $\Sigma$ as a timelike three dimensional hypersurface that divides spacetime into two regions ${\rm V}^+$ (exterior) and ${\rm V}^-$ (interior). The line element for the exterior spacetime is taken as that of generalized Vaidya metric \cite{vaidyaext} which in retarded (exploding) null coordinates is given by
\begin{equation}
ds^2_{+}=-h(\bar{r},v)dv^2-2dvd\bar{r}+\bar{r}^2(d\vartheta^2+\sin^2\vartheta d\phi^2),
\end{equation}
where $h(\bar{r},v)$ being the exterior metric function with $\bar{r}$ and $v$ being the Vaidya radius and retarded null coordinate, respectively. We label the exterior coordinates as $\{Z_+^{\mu}\}\equiv\left\{v,\bar{r},\vartheta,\phi\right\}$. The interior line element (\ref{metric}) in spherical coordinates reads
\begin{equation}\label{metric1}
ds_-^{2}=-d\eta^2+e^{2\alpha(\eta)}\left(dr^2+r^2d\vartheta^2+r^2\sin^2\vartheta d\phi^2\right),
\end{equation}
where we have labeled the interior coordinates as $\{Z_-^{\mu}\}\equiv\{{\eta,r,\vartheta,\phi}\}$. The hypersurface $\Sigma$ with intrinsic coordinates $\{x^a\}\equiv\{\eta,\vartheta,\phi\},~ (a=0,2,3)$ results from the isometric gluing of two hypersurfaces $\Sigma^+$ and $\Sigma^-$, which, respectively, bound the exterior and interior regions. Utilizing generalized Israel-Darmois junction conditions \cite{gidjcbd} we proceed to match the above line element to the exterior one through the boundary surface $r=r_{\Sigma}$. The induced metrics on $\Sigma^+$ and $\Sigma^-$ will take the following form, respectively
\be\label{ININn}
ds_{\Sigma^{-}}^2=-d\eta^2+e^{2\alpha(\eta)}r_{\Sigma}^2(d\vartheta^2+\sin^2\vartheta d\phi^2),
\ee
and
\be \label{INEXn}
ds_{\Sigma^{+}}^2=-\left[h\big(\bar{r}(\eta),v(\eta)\big)\dot{v}^2+2\dot{\bar{r}}
\dot{v}\right]d\tau^2+\bar{r}^2(\eta)(d\vartheta^2+\sin^2\vartheta d\phi^2).
\ee
The junction conditions for the first fundamental forms (induced metrics) give
\be\label{FFF1n}
h\big(\bar{r}(\eta),v(\eta)\big)\dot{v}^2+2\dot{\bar{r}}\dot{v}=1,~~~~~~\bar{r}(\eta)=r_{\Sigma}e^{\alpha(\eta)},
\ee
where an overdot denotes $d/d\eta$. In order to calculate the extrinsic curvature tensors of $\Sigma^+$ and $\Sigma^-$, we need to find the unit vector fields normal to these hypersurfaces. We then get
\be\label{NVF+-n}
n^{-}_{\mu}=\left[0,e^{\alpha(\eta)},0,0\right],~~~~~~n^{+}_{\mu}=\frac{1}{\left[h(\bar{r},v)\dot{v}^2+2\dot{\bar{r}}\dot{v}\right]^{\frac{1}{2}}}\left[-\dot{\bar{r}},\dot{v},0,0\right].
\ee
The extrinsic curvature tensors associated with $\Sigma^\pm$ are given by
\be\label{EXCn}
K^{\pm}_{ab}=-n_{\mu}^{\pm}\left[\frac{\partial^2Z_{\pm}^{\mu}}{\partial x^a\partial x^b}+\Gamma^{\mu\pm}_{\nu\sigma}\frac{\partial Z_{\pm}^{\nu}}{\partial x^a}\f{\partial Z_{\pm}^{\sigma}}{\partial x^b}\right],
\ee
where $Z_+^\mu(x^a)$ and $Z_-^\mu(x^a)$ are parametric relations for the hypersurfaces $\Sigma^+$ and $\Sigma^-$ and $\Gamma^{\mu\pm}_{\nu\sigma}$ are the components of connections associated to interior and exterior line elements. The junction conditions in BD theory have the form \cite{gidjcbd}
\bea
-\left[K^a\,\!\!_b\right]+\left[K\right]\delta^a\,\!\!_b&=&\f{8\pi}{\Phi}\left(S^a\,\!\!_b-\f{S}{3+2\omega}\delta^a\,\!\!_b\right),\label{ksjump}\\
\left[\Phi_{,n}\right]&=&\f{8\pi S}{\chi},\label{phisjump} ~~~\left[\Phi\right]=0,
\eea
where the notation $\left[\Psi\right]=\Psi^+|_\Sigma-\Psi^-|_\Sigma$ stands for the jump of given field across the hypersurface $\Sigma$, $n$ labels the coordinate normal to this surface and $S_{ab}={\rm diag}(\sigma,p,p)$ is the energy-momentum tensor of matter fields (except the BD scalar field) on the shell located at $\Sigma$ where $\sigma$ and $p$ are the surface density of mass-energy and surface pressure, respectively. The quantities $K$ and $S=2p-\sigma$ are traces of $K^a\,\!\!_b$ and $S^a\,\!\!_b$, respectively. We also note that equation (\ref{ksjump}) can be rewritten in an equivalent form as
\be\label{eq39equiv}
S^a\,\!\!_b=\f{\Phi}{8\pi}\left(\f{\omega+1}{\omega}\left[K\right]\delta^a\,\!\!_b-\left[K^a\,\!\!_b\right]\right).
\ee
A straightforward but lengthy calculation leads to the following expressions for the components of extrinsic curvature tensors, as
\bea
K^{-\eta}\,\!\!_\eta&=&0,~~~K^{-\vartheta}\,\!\!_\vartheta=K^{-\phi}\,\!\!_\phi=\f{1}{r_{\Sigma}e^{\alpha}},\\
K^{+\eta}\,\!\!_\eta&=&\frac{\dot{v}^2\left[hh_{,\bar{r}}\dot{v}+h_{,v}\dot{v}+3h_{,\bar{r}}\dot{\bar{r}}\right]+2\left(\dot{v}\ddot{\bar{r}}-\dot{\bar{r}}\ddot{v}\right)}{2\left(h\dot{v}^2+2\dot{\bar{r}}\dot{v}\right)^{\frac{5}{2}}},\nonumber\\
K^{+\vartheta}\,\!\!_\vartheta&=&K^{+\phi}\,\!\!_\phi=\frac{h\dot{v}+\dot{\bar{r}}}{\bar{r}\left(h\dot{v}^2+2\dot{\bar{r}}\dot{v}\right)^{\f{1}{2}}}.
\eea
Next, we proceed to compute the components of equation (\ref{eq39equiv}). We then get
\bea
S^\eta\,\!\!_\eta=-\sigma=\f{\Phi_{\Sigma}}{8\pi}\left(\f{\omega+1}{\omega}\left[K\right]-\left[K^\eta\,\!\!_\eta\right]\right),~~~~S^\vartheta\,\!\!_\vartheta=S^\phi\,\!\!_\phi=p=\f{\Phi_\Sigma}{8\pi}\left(\f{\omega+1}{\omega}\left[K\right]-\left[K^\vartheta\,\!\!_\vartheta\right]\right).
\eea
The first part leaves us with the following expression for surface density in terms of extrinsic curvature components, as
\be\label{surfdens}
\sigma=-\f{\Phi_{\Sigma}}{8\pi}\left\{\f{K^{+\eta}\,\!\!_\eta}{\omega}+\f{2(\omega+1)}{\omega}\left(K^{+\vartheta}\,\!\!_\vartheta-K^{-\vartheta}\,\!\!_\vartheta\right)\right\},
\ee
while for the second part we get
\bea\label{surpre}
p=\f{\Phi_{\Sigma}}{8\pi}\left\{\f{\omega+1}{\omega}K^{+\eta}\,\!\!_\eta+\f{\omega+2}{\omega}\left(K^{+\vartheta}\,\!\!_\vartheta-K^{-\vartheta}\,\!\!_\vartheta\right)\right\},
%2p-\sigma&=&\f{\chi\Phi_\Sigma}{8\pi\omega}\left\{K^{+t}\,\!\!_t+2\left(K^{+\vartheta}\,\!\!_\vartheta-K^{-\vartheta}\,\!\!_\vartheta\right)\right\},
\eea
where only non-vanishing components of the extrinsic curvature tensors have been considered. These equations can also be re-written in the following form as
\bea\label{rewriktttheta}
K^{+\eta}\,\!\!_\eta=\f{8\pi}{\chi\Phi_\Sigma}\left(2(p+\sigma)+(\sigma+2p)\omega\right),~~~~K^{+\vartheta}\,\!\!_\vartheta-K^{-\vartheta}\,\!\!_\vartheta=-\f{8\pi}{\chi\Phi_\Sigma}(p+\sigma(\omega+1)).
\eea
The jump for the normal derivative of the BD scalar field across $\Sigma$ is found as
\bea\label{jumpbdsf}
\left[\Phi_{,n}\right]&=&n^{\!\!+a}\Phi^+_{,a}-n^{\!\!-a}\Phi^-_{,a}=n^{\!\!+v}\Phi^+_{,v}+n^{\!\!+\bar{r}}\Phi^+_{,\bar{r}}-n^{\!\!-r}\Phi^-_{,r}\nn
&=&\f{1}{\left(h\dot{v}^2+2\dot{\bar{r}}\dot{v}\right)^{\f{1}{2}}}\left\{(\dot{\bar{r}}+h\dot{v})\Phi^+_{,\bar{r}}-\dot{v}\Phi^+_{,v}\right\},
\eea
where use has been made of the contravariant components of the normal vector fields
\be\label{contnorms}
n^{\!\!-\mu}=[0,e^{-\alpha},0,0],~~~~~~~n^{\!\!+\mu}=\f{1}{\left(h\dot{v}^2+2\dot{\bar{r}}\dot{v}\right)^{\f{1}{2}}}[-\dot{v},\dot{\bar{r}}+h\dot{v},0,0].
\ee
Thus, from equation (\ref{phisjump}), the jump in the normal derivative of BD scalar field and its continuity across $\Sigma$ we obtain
\be\label{jumpbdsff}
\f{1}{\left(h\dot{v}^2+2\dot{\bar{r}}\dot{v}\right)^{\f{1}{2}}}\left\{(\dot{\bar{r}}+h\dot{v})\Phi^+_{,\bar{r}}-\dot{v}\Phi^+_{,v}\right\}=\f{\Phi_{\Sigma}}{\omega}\left\{K^{+\eta}\,\!\!_\eta+2\left(K^{+\vartheta}\,\!\!_\vartheta-K^{-\vartheta}\,\!\!_\vartheta\right)\right\},~~~~\Phi^+|_{\Sigma}=\Phi^-|_{\Sigma}=\Phi_\Sigma.
\ee
Equations (\ref{FFF1n}), (\ref{rewriktttheta}) and (\ref{jumpbdsff}) fully determine the dynamics of the boundary, the exterior metric function and BD scalar field in the exterior region, once we know the matter distribution on the boundary. Assuming the boundary surface is devoid of mass-energy density and pressure, i.e., $\sigma=0$ and $p=0$, we get
%\be\label{devoidofsp}
%-\omega(2\omega+3)\left(K^{+\vartheta}\,\!\!_\vartheta-K^{-\vartheta}\,\!\!_\vartheta\right)=0,
%\ee
$K^{+\vartheta}\,\!\!_\vartheta=K^{-\vartheta}\,\!\!_\vartheta$ and $K^{+\eta}\,\!\!_\eta=0$ if $\omega\neq-3/2$. Therefore, the continuity of $(\vartheta,\vartheta)$ and $(\eta,\eta)$ components of extrinsic curvature across $\Sigma$ leaves us with following relations for exterior quantities, as
\bea\label{extquant}
h\dot{v}+\dot{\bar{r}}&=&1,\label{aa2z}\\
\dot{v}^2\left[(hh_{,\bar{r}}+h_{,v})\dot{v}+3h_{,\bar{r}}\dot{\bar{r}}\right]&+&2\left(\dot{v}\ddot{\bar{r}}-\dot{\bar{r}}\ddot{v}\right)=0,\label{aa1z}
\eea
where use has been made of relations given in (\ref{FFF1n}). Taking derivatives of (\ref{aa2z}) and the first part of (\ref{FFF1n}) we get
\be\label{A17n}
2\dot{\bar{r}}\ddot{v}-\dot{h}\dot{v}^2=0,~~~~~~2\dot{\bar{r}}\ddot{\bar{r}}+\dot{h}\dot{v}\left(2\dot{\bar{r}}+h\dot{v}\right)=0,
\ee
whence we have
\be\label{RECn}
\ddot{\bar{r}}\dot{v}=-\ddot{v}\left(2\dot{\bar{r}}+h\dot{v}\right).
\ee
Substituting for $\ddot{\bar{r}}\dot{v}$ and $\dot{\bar{r}}\ddot{v}$ from the above relations into (\ref{aa1z}) we finally get
\be\label{KTTPFn}
K^{+\eta}\,\!\!_\eta=-\frac{h_{,v}\dot{v}^2}{2\dot{\bar{r}}}=0,
\ee
which implies that $h(\bar{r},v)=h(\bar{r})$. Solving equations (\ref{aa2z}) and the first part of (\ref{FFF1n}) we obtain the four-velocity of the boundary as
\be\label{4Vn}
{\rm U}^{\alpha}=\left(\dot{v},\dot{\bar{r}},0,0\right)=\left[\f{1+\sqrt{1-h}}{h},-\sqrt{1-h},0,0\right],
\ee
where we have chosen the minus sign for $\dot{\bar{r}}$ as we deal with a collapse setting. The continuity of the BD scalar field implies that this field must be homogeneous in the exterior region, i.e., $\Phi^{+}_{,\bar{r}}=0$. Taking this into account together with the first part of equation (\ref{jumpbdsff}) gives $\Phi^{+}_{,v}=0$ which implies the BD scalar field outside the collapsing cloud is constant. We therefore conclude that the exterior region is a static spacetime with dynamical boundary. The process of finding the exterior metric function on the boundary passes through utilizing the Hamiltonian constraint, the equation of motion of the scale factor (\ref{eqsrew9}), the junction condition for induced line elements (the second part of (\ref{FFF1n})) and the trajectory equation for the boundary. Such a procedure may not be a simple task. However, we may intuitively deduce that the effects of phase space deformation could appear in the exterior spacetime as we solve the radial component of the four-vector velocity to find $h(\dot{\alpha})$. From the second part of (\ref{FFF1n}) and radial component of the four-vector velocity we get $r_{\Sigma}^2\dot{a}^2=1-h$. The location of horizon from the exterior view is given by the condition $h=0$. Therefore, the horizon would interest the boundary surface if the collapse velocity satisfies
\be\label{horrr21}
r_{\Sigma}=\f{1}{|\dot{a}|},
\ee
whereby it is seen that if the collapse velocity is bounded, the boundary
of the collapsing cloud can be chosen so that the formation of horizon is
avoided. Such a scenario could occur within the non-commutative setting we
presented here, whereas for commutative case, the collapse velocity
diverges and the horizon would always form to cover the resulted singularity.
%%%%%%%%%%%%%%%%%%%%%%%%%%%%%%%%%%%%%%%%%%%%%%%%%%%%%%%%%%%%%%%%%%%
\section{concluding remarks}
\indent
\label{Concl.}
In this paper, we have investigated the collapse process of a homogeneous perfect fluid
in the context of the BD theory in deformed phase space.
Let us clarify (further to our comments in in section~\ref{Intro.}) why we
have investigated in this paper a collapse scenario in a
BD noncommutative setting instead of a corresponding one in GR.
On the one hand, a collapse scenario in
noncommutative GR has already been discussed in previous
publications~\cite{Nicollinipapers}.
%In summary, the results in those papers point
%to \rc{[ PLEASE PUT A very brief SUMMARY HERE]}
On the other hand, from another complementary perspective, let us add the following.
It is well known that scalar tensor theories such as the BD theory can agree with GR in
the post-Newtonian limit~\cite{SST95}. However, it should be emphasized that in
a strong field setting\footnote{As two appropriate examples of such strong field
settings, we can mention the generation of the gravitational waves and the formation
of the black holes and singularities during gravitational
collapse, see, e.g.,~\cite{SST95} and references therein.}, those theories may yield very different
predictions. That may mean a few
experimental and observational features, but also important
different structural implications of these theories.
In particular, we can allude to the formation of a singularity and
black hole during gravitational collapse.
This has been specifically investigated in
the herein work. Considerable differences and relevant
features were extracted, demonstrating how a noncommutative
BD setting is much different from the corresponding standard BD theory as well as GR.
%\rc{[ PLEASE PUT A very very  brief SUMMARY of those differences here]}.
Intrinsic to such difference is the fact that a scalar tensor gravitation
involves more degrees of freedom, therefore, it yields a larger number
of solutions than GR~\cite{SST95}. Moreover, the employed noncommutative
parameter in the BD theory also couples to variables which
are absent in the GR, and therefore, the range of
solutions and possible scenarios is much wider.
It was this broad scope of possibilities we investigated herein, regarding a collapse scenario in a
BD noncommutative setting.

%We should argue that why the collapse scenario has been motivated in a BD noncommutative setting instead of a corresponding
%one in GR.
%We already have discussed regarding this important comment in section~\ref{Intro.}.
%It should be a good idea to look at this point in another perspective:
%it has been shown that scalar tensor theories can agree with GR in the post-Newtonian limit~\cite{SST95}.
%What should be emphasized is to study a few strong field settings in which these theories
%may yield different predictions.
%Such strong field settings may not only
%suggest a few new experimental and observational viewpoints which might
%distinguish the difference between the predictions of the mentioned theories
%but they may also illustrate the structure of these theories.
%Among mentioned strong field settings, we can allude the formation of the
%singularity and black hole during gravitational collapse which has been
%investigated in the herein work. On the other hand, as the scalar tensor
%gravitation involves more degrees of freedom, therefore, it yields a
%larger number of solutions than GR~\cite{SST95}. Moreover, the employed noncommutative
%parameter in the BD theory may also couple to the variables which are absent in the GR and
%therefore may exceed the number of the solutions.
%We have found that introducing a special type of non-commutativity between phase space coordinates
%could alter the final fate of the collapse process of a perfect fluid in BD theory.
%}

Assuming the interior geometry of the collapsing cloud to be that of a
spatially flat FLRW spacetime, we employed a particular type of
non-commutativity between the phase space coordinates and examined
its effects on the collapse dynamics. More precisely, we have introduced
in~(\ref{NC-Poisson}) a modified Poisson algebra %Then, in order to discuss the dynamics, we employed and obtained the  modified field equations that govern
in the Hamiltonian formalism.
%\rc{To derive the
%equations of motion,
%for the noncommutative framework,
%we have not carried out an expansion on the non-commutative parameters (which has been usually
%applied for Moyal product of functions) but instead,
%we have further used the symplectic formalism
%(see, e.g.,~\cite{GSS11}) associated to the effective classical mechanics.} \bl{Mehrdad, Please give more precise explanation on this sentence and improve the English.}
%\bl{Exterior space? [I should check the papers by Malafarina]}
%As a construction of the collapsing star,
%We have taken a perfect fluid with a barotropic EoS as a seed for the
%collapse process.
%Moreover, we have assumed that the scalar potential vanishes.

%Finding analytical solutions of the noncommutative equations of motion
%were very complicated, so we analyzed them instead by means of numerical methods.
Our numerical analysis shows that there are two different type of  solutions which depend effectively on the
sign of the BD coupling parameter. In the case where the BD coupling parameter is negative,
 oscillatory behaviors appear. However, by
 %will summarize and
%discuss them at the end of this section, while, in the next paragraphs,
%we are going to sum up the solutions obtained by
assuming positive values for $\omega$,
%For a dust cloud collapse, our numeric simulation (see Fig.\ref{sf}) shows
when the constant non-commutative parameter is absent,
the collapse scenario is terminated at a spacetime singularity,
%a point at which the Kreschmann scalar diverges.
whilst, for small values of the non-commutative parameter,
%we have seen
%different dynamical behaviors for the quantities involved in the herein model; namely,
there is a nonzero minimum value for the scale factor where the collapse halts and then an expanding phase begins.
%Precisely at $t_{\rm b}$, the speed of collapse
%vanishes and collapse turns to an expansion. For all the times
%$t$ where $t<t_{\rm b}$, the collapse evolves through a
%decelerating contracting phase, while
%for $t>t_{\rm b}$ it undergoes an inflationary expanding
%regime till the time at which the acceleration vanishes, $t=t_{\rm inf}$. After $t_{\rm inf}$, it experiences a
%decelerating expansion for a while and in the late times the
%speed of collapse tends to zero where the collapsing cloud is asymptotically at rest.
%In addition, for small values of the noncommutative
%parameter, the dynamical behavior of the scale factor and
%the Kretschmann scalar are continuous and regular
%throughout the dynamical evolution of the object.
%This signals the avoidance of spacetime singularity
%which does not appear when the noncommutative effects are absent.

For small values of the non-commutative parameter\rlap,\footnote{In order
to study appropriately the behavior of the quantities involved within the collapse setting
 and have a correct comparison among them as well as with the standard
 commutative models, we have taken the non-commutative parameter to be
 the same for all numerical plots. However, it is important to note that, for the
 variety of values, which can be taken by $\theta$, they have also been examined both for
 the general case (according to left panel of Fig. \ref{f19}) and also for
 particular case (large values of the BD coupling parameter), as seen in Fig. \ref{g18}.}
 both the EoS and BD coupling parameters can effectively control
the dynamics of the collapse setting (see Fig.~\ref{f19} and Fig. \ref{womega}).
More precisely, for small values of $\theta$, the softness of
the bounce depends effectively on the values taken by $w$, i.e., matter pressure, and $\omega$.
%Further to our consideration on singularity avoidance in dust case, we note that
%The role of pressure of the fluid cannot be underestimated.
%within a typical collapse setting.
%In this sense, the inclusion of pressure of the matter undergoing gravitational
%collapse could drastically change the fate of such a scenario \cite{PRCOLL}.
For a non-dust case, as the pressure of the fluid tends to positive
values, the collapse dynamics is altered such that the location and the
number of the dynamical horizons are changed. Such a situation similarly
happens as the BD coupling parameter increases (for a fixed value of EoS parameter).

 %\rc{This implies that the geometry
% outside of the collapsing object could be that of a generalized Vaidya spacetime \cite{VSP},
% where the outgoing flux emanates and reaches a far away observer.}

%The upper panel of Fig. \ref{womega} shows that as the matter pressure undergoing the
%collapse process becomes more positive, the collapse evolves to a bounce
%faster ($\Delta t_{\rm b}$ decreases) and the bounce occurs
%sooner ($t_{\rm b}$ decreases). The more the pressure becomes negative,
%the stronger the resistance of the matter content against the pull of gravity,
%which leads to a soft bounce ($\Delta t_{\rm b}$ increases) at a delayed
%time ($t_{\rm b}$ increases). Moreover, when we fix the EoS
%parameter, an increase in the BD coupling parameter leads to
%the occurrence of a faster bounce, with the bounce time getting retarded
%($t_{\rm b}$ increases). Conversely, the lesser the values that the BD coupling
%parameter takes, the milder the bounce occurs (a smooth transition from
%contracting phase to expanding phase, i.e., $\Delta t_{\rm b}$ increases), with the bounce time getting
%advanced, i.e., $t_{\rm b}$ decreases (see the middle panel of Fig. \ref{womega}).
\par
The strength of the scalar to tensor coupling to the matter is encoded in
the BD coupling parameter so that the smaller the value of $\omega$ parameter, the
larger the contribution of the scalar field to gravitational interaction. As we found in the herein model, the coupling of the scalar to tensor field to matter content could affect
the softness of the bounce and the time interval during which the bounce occurs.
In this manner we could say that the stronger the contribution of the BD scalar
field to the gravitational interaction (i.e., the $\omega$ parameter decreases), the softer the bounce occurs.
\par
The behavior of the apparent horizon curve for a dust
collapse has been also analyzed according to Figs.~\ref{rahh} and \ref{rahaaw}.
It is seen that the dynamics of the apparent horizon in the case in
which non-commutative effects are present is quite different to the case
where these effects are absent. More concretely, when the non-commutative parameter takes small values, the
formation or otherwise of the dynamical horizons depend crucially on the behavior of the collapse velocity.

\par
%The existence of a minimal radius below with which the whole scenario
%if free of horizon formation may be significance. In this sense, we
%would like to argue this issue from two perspectives. Firstly, we note that the BD theory
%can be regarded as a special case of the scalar-tensor theories.
%In the latter, $\omega$ can be assumed to be a function of the scalar field, which takes small values
%(hence, contribution of the scalar field to the gravitational interaction)
%at the early universe while it is propagating, through an attractor mechanism, to the large values today
%\cite{CMWF}. In our model (which, due to the presence of the noncommutative parameter,
%can be considered as a generalized version of the standard BD theory), we may consider
%a sufficiently dense matter distribution undergoing gravitational
%collapse in BD theory during the early evolution of the universe
%and in the presence of noncommutative effects, such that these
%effects could provide a non-singular bounce at a finite time.
%In such a scenario if the surface boundary of the collapsing
%body is small enough, the horizons are failed to cover the bounce.
%From this viewpoint, the remnants of a collapsing object that has
%turned to a bounce may be observed. Hence, a laboratory may be
%provided which might astrophysically detect the possible effects of noncommutativity.
%\pr
 %From another viewpoint,
\par
 Recent observations of radiation damping in mixed binary systems have put limitations on the BD coupling parameter as $\omega>40000$ from the Cassini measurements \cite{ABWZ}. %This makes the necessity for analyzing the present model in the large $\omega$ limit\rlap.\footnote{When $\omega$ goes to infinity, the BD theory may reduce to the GR~\cite{BDtoGR}.} Let us have a brief review on the dynamics of a dust cloud collapse in large $\omega$ limit in the presence of noncommutative effects.
In our scenario, we have shown that when $\omega$ takes very large
values\footnote{When $\omega$ goes to infinity, the BD theory may
reduce to the GR~\cite{BDtoGR}.}, we can find a critical value for
the non-commutative parameter, i.e., $\theta_{\rm c}$, which completely
disassociates two different dynamical behaviors for the collapsing
object, see Fig.~\ref{g18}. More precisely, for $\theta>\theta_{\rm c}$, the scale factor
decreases till reaching a minimum value and stays at this value as the time
evolves. Whilst, for $\theta<\theta_{\rm c}$, the collapse culminates in a
spacetime singularity with completely different dynamics compared to the case
where the non-commutative effects are absent. It should be noted that
the value of $\theta_{\rm c}$ depends on the initial values which are taken by
the present parameters of the model.
%Therefore, based upon the large values for
%the BD coupling parameter obtained from the observations at present epoch, and our numerical analysis has been done for large
%values of this parameter (see Fig. \ref{g18}), we may argue the following: if the initial radius
%of the collapsing body is chosen suitably, in the sense that the formation of the
%horizon is avoided, then, what will remain as the end product of the collapse
%scenario will be a dense object with constant radius (see the family of black
%curves in Fig.~\ref{g18}), which can be observationally seen through astrophysical
%apparatus. Hence, again, it might be provided a possible chance to detect the effects of noncommutativity.
%In order to analyze a proposed model in the
%context of the BD cosmology and gravity, it is
%worthwhile to study the behavior of the quantities in the large omega limit.
%In this paper, we have not skipped this necessary process.
%
\par
We have also studied the collapse associated to a
pressureless fluid for negative values of the BD coupling parameter.
As the left panel in Fig.~\ref{omega-} indicates, the
%the behavior of
%the quantities depend not only on the absolute value of
%the BD coupling parameter, but also
%are affected by the sign of this parameter which, for negative
%values of $\omega$, we see oscillatory behaviors.
% Unlike the commutative case in which the scale factor always monotonically decreases till
 %it reaches the singularity, for the noncommutative case, the scale
 scale factor starts its decreasing behavior till reaching the
 bounce, beyond which an increasing oscillatory behavior commences.
 %For the mentioned oscillatory behavior of the scale
 %factor, we found expectedly that
 The collapse velocity experiences
 an oscillatory phase such that at the earlier times, it oscillates
 toward its first vanishing point at the time $\eta=\eta_{{\rm b}}^{{\rm osc}}$
 (see the right panel of Fig. \ref{omega-}). Note that the speed of
 collapse remains negative for $\eta<\eta_{{\rm b}}^{{\rm osc}}$.
 It is therefore the acceleration of the collapse that changes
 its sign rapidly, signaling that the collapsing object experiences
 a series of decelerating (see the red arrow heading upward from points A to B)
 and accelerating (see the blue arrow heading downward from the points B to C)
 contracting regimes with a soft jump from the former to the latter (at point B)
 but a quick jump from the latter to the former (at point A or C). For $\eta>\eta_{{\rm b}}^{{\rm osc}}$
 the scenario enters an accelerated expanding phase for the first
 time (at $\eta=\eta_{{\rm 0}}$) and remains in this regime in the time
 interval $\eta_{{\rm 0}}<\eta<\eta_{{\rm 1}}$. It then goes again under a
 decelerated expanding phase between $\eta_{{\rm 1}}<\eta<\eta_{{\rm 2}}$
 till it enters an accelerated contracting phase which occurs for $\eta>\eta_{{\rm 2}}$.
 This oscillatory behavior continues around the zero point velocity.
 We should note that the frequency of the oscillation of collapse velocity decreases with time.
 \par
 Moreover, the corresponding envelop of the oscillatory phase is also damping with time. It is also worth mentioning that there exists a difference between the behavior of collapse velocity for this case and the case where $\omega>0$. In the former, though the scale factor vanishes at a finite amount of time (singular behavior), the speed of collapse is limited (see the dashed curve in the right panel of Fig.~\ref{omega-}) so that by a suitable choice of the boundary of the collapsing body, the formation of the apparent horizon can be avoided. Thus, the collapse may culminate in a naked singularity. For the latter, the collapse velocity diverges in the limit of approach to the singularity and the apparent horizon would always form to cover the resulting singularity. The lower panel in Fig.~(\ref{omega-}) shows the collapse velocity for different values of BD coupling parameter. It is seen that the larger the negative value of $\omega$ parameter, the smaller the frequency of oscillation.

To complete the collapse model, we performed matching the interior spacetime with that of generalized Vaydia spacetime using the generalized Israel-Darmois junction conditions. We observed that as long as there is no surface stress energy on the boundary of the collapsing cloud, the extrinsic curvature tensor is continuous across the boundary. This makes the exterior spacetime to be static with a dynamical boundary so that the horizon could intersect the boundary depending on the initial size of the collapsing body. Thus, if the initial size of the collapsing object is taken as small as enough, horizon formation could be avoided. We then conclude that the process of gravitational collapse in the presence of non-commutativity would lead to a non-singular bounce, that is uncovered by the horizon and so it can
 be causally connected to an external observer.
 \par
% Besides, the collapse setting presented here, a bouncing scenario for BD theory with negative coupling parameter has been also reported in cosmological settings, where it is found that the universe experiences periodic oscillations about a finite non-singular minimum~\cite{BKM04}. Furthermore, the oscillatory behavior of the collapsing object in our model can be astrophysically interesting, if we deal with radial oscillations of spherical stars \cite{OSCRA}. This could potentially provide us a useful laboratory for possible observational features of non-commutativity.

%the \rc{quasars? pulsars?}. \rc{It should be noted that since the collapsing body is spherically symmetric, homogeneous and isotropic, such an oscillatory behavior can be only realized through the $p-modes$ (pressure modes) of a pulsating neutron star \cite{}}.

We would like to compare our results to other collapse scenarios as reported in the literature.
%Within the present model, we have found that the collapse dynamics
%in the presence of noncommutative effects exhibits a non-singular
%bouncing process for the positive values of the BD coupling parameter
%and an oscillatory bouncing scenario for negative ones.
%Whereas, in the commutative case, the gravitational collapse
%of a dust fluid within the framework of BD theory
%(for both positive and negative coupling parameters) would lead
%to the singularity formation covered by a horizon \cite{SST95}.
Work along this line has been done within the $f({\mathcal R})$ gravity models
which correspond to the BD theory with a vanishing coupling parameter, using
the metric formalism. In this context, it has been shown that the collapse
process of a perfect fluid leads to a spacetime singularity which can be
either hidden behind a horizon or visible to external observers \cite{BDSINCOL}.
%Moreover, according to our previous work~\cite{RZMM14} in which we have
%investigated a collapse scenario whose matter content is only a minimally
%coupled scalar field in the presence of a special type of non-commutativity,
%we would like to argue here that when these effects are in game, the spacetime
%singularity as the end-state of gravitational collapse of a minimally coupled
%scalar field evolving through a background BD field  \cite{TwoKorea}, could also
%be possibly removed.
Besides the model presented here, non-singular bouncing
scenarios have also been reported in the literature such as $f({\mathcal R})$ theories in
Palatini formalism \cite{frb}, generalized teleparallel gravity theories \cite{gtpb},
bouncing models in the presence of interacting spinning particles in the
framework of Einstein-Cartan theory in both cosmological \cite{bretchet}
and astrophysical scenarios \cite{khodam}. Non-singular bouncing scenarios
have also been reported in loop quantum cosmology for a massless scalar
field \cite{loopc} and in the presence of
anisotropy \cite{loopi} (see also \cite{loop3} and references therein).

Finally, we should be aware of the following notes about the herein model:
(i) All the results of this paper have been obtained for a simple case in
which we have introduced a constant non-commutative parameter within the standard BD theory, a
homogeneous matter distribution and a spatially flat FLRW line-element for the interior
region suitably matched to a generized Vaidya spacetime as the exterior solution.
%If we would like to modify the achievements, we can extend
It would be interesting to extend the herein model by introducing other
Poisson brackets instead of~(\ref{NC-Poisson}), other kinds of
matter or geometry and/or other extended scalar-tensor theories.
 (ii) As it has been shown in section (\ref{NC-BDT}), the new quantities $\Phi$ and $\eta$ have been redefined
  such that we obtained dimensionless quantities. In this
 rescaled setting, it is important to note that, in order to get an
 appropriate noncommutative scenario, we should not think that $\theta$ must
 take very small values (of orders the Planck length), but instead, it is
 enough to consider $\theta$ such that it is restricted to the interval $0\leq\theta<1$.
  (iii) As the field equations in the Einstein representation do not
contain second derivatives of $\Phi$, thus, this representation can be appropriate to
extend predictions from GR to BD theory, especially in the vacuum case~\cite{SST95}.
However, deriving the equations associated to the Einstein frame
for our model (the same as obtaining the perturbed equations) is not easy to perform and we have not
investigated them.

%%Finally, we should mention that all the results of this paper have been obtained
%% by introducing a constant noncommutative parameter, a
%%homogeneous matter fluid, and the standard BD theory.
%\bl {It is important to note the the scenario
%presented in this paper may be figured out either as a setting beyond the
%standard BD theory in the classical regime or as a semiclassical setting.} If we would
%like to modify the achievements, we can
%%To extend the herein model we can also investigate other
%%Poisson brackets, other kinds of matter and/or (generalized) scalar-tensor theories.

%\rc{Such behaviors resemble
% those obtained in our }

%It is worth mentioning at this stage that the collapse dynamics for both positive and negative values of the $\omega$
\section{ACKNOWLEDGMENTS}
S. M. M. Rasouli is grateful for the support of
grant SFRH/BPD/82479/2011 from the Portuguese
Agency Funda\c c\~{a}o para a Ci\^encia e Tecnologia.
This research work was supported by the grant PEst-OE/MAT/UI0212/2014.

\appendix

\section{On the Field equations in Non-Commutative Brans-Dicke Theory}\label{APPBD}
\begin{figure}
\includegraphics[scale=0.36]{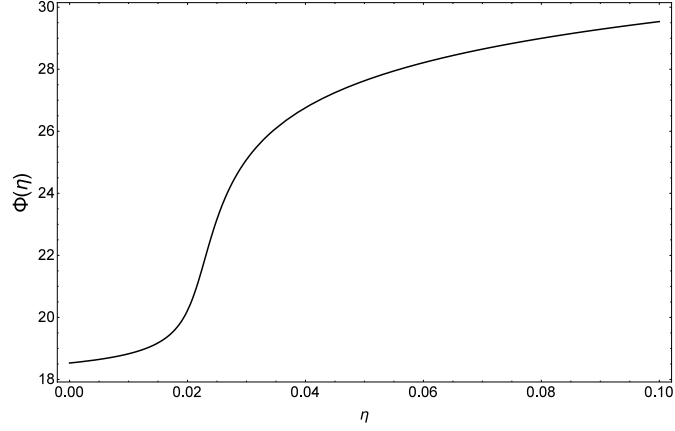}
\caption{Time behavior of the BD scalar field for ${{\rm P}}_{\Phi}(\eta_i)=64.23$,
$\Phi(\eta_i)=18.53$, $L_\text{Pl}^4\rho_i=32.88$, $\alpha_i=0.0444$, $\omega=3.3$, $w=1$,
and $\theta=0.316$.}\label{phiw11}
\end{figure}

For stiff fluid $(w=1)$, equation (\ref{eqsrew5}) leaves us with a constant of motion as
${{\rm P}}_{\alpha}={\rm constant}$. Multiplying equation (\ref{eqsrew10})
by ${{\rm P}}_\Phi$ and equation (\ref{eqsrew7}) by $\Phi$, after adding up the results, we get
\be\label{add1315}
\f{d}{d\eta}\left(\Phi {{\rm P}}_\Phi\right)=16\pi L_\text{Pl}^4\rho_ie^{3(2\alpha_i-\alpha)}.
\ee
The right hand side of the above equation can be expanded near the bounce, where $\dot{\alpha}(\eta_{{\rm b}})\approx0$ and $\alpha(\eta_{\rm b})=\alpha_{{\rm b}}<\alpha_i$. We then get, up to the first order
\be\label{expnbeq}
\f{d}{d\eta}\left(\Phi {{\rm P}}_\Phi\right)\!\bigg|_{\alpha=\alpha_{{\rm b}}}\approx16\pi
L_\text{Pl}^4\rho_ie^{6\alpha_i}\left(1-3\alpha_{{\rm b}}\right)+{\mathcal O}(\alpha_{{\rm b}})^2,
\ee
whence integration in the neighborhood of the bounce point gives
\be\label{expannbint}
\Phi {{\rm P}}_\Phi\approx16\pi L_\text{Pl}^4\rho_ie^{6\alpha_i}\left(1-3\alpha_{{\rm b}}\right)\eta+C_0,
\ee
%\bigg|_{t=t_{{\rm b}}}
where $C_0$ is an integration constant.
Substituting the above approximation into equation (\ref{eqsrew10}) and after a few algebra we find near the bounce
\be\label{phinb}
\Phi\approx-\f{e^{-3\alpha_{{\rm b}}}}{2\chi}\left[({{\rm P}}_{\alpha}-2C_0)\eta-16\pi
L_\text{Pl}^4\rho_ie^{6\alpha_i}\left(1-3\alpha_{{\rm b}}\right)\eta^2\right].
\ee
It is seen that the BD scalar field has a parabolic behavior near the bounce
point. This behavior has been sketched in Fig.~\ref{phiw11} where we see that the scalar field behaves parabolically near the bounce time $\eta_{{\rm b}}=0.01915$.

 Next, we can evaluate equation~(\ref{eqsrew9}) at the bounce point which yields
\bea\label{e12nbp}
\f{\omega}{3}\Phi {{\rm P}}_\alpha+\Phi^2 {{\rm P}}_\Phi+\theta {{\rm P}}_\alpha\left(\Phi {{\rm P}}_\Phi\right)
-2\theta\left(\Phi {{\rm P}}_\Phi\right)^2+32\pi L_\text{Pl}^4\chi\theta\rho_i e^{6\alpha_i}\Phi=0.
\eea

Now, with the help of equations (\ref{expannbint}) and (\ref{phinb}), we arrive at the following cubic equation for the bounce time as
\bea\label{cubiceqbounce}
\left(\epsilon\eta_b^3-\delta\eta_b^2\right)L_{{\rm Pl}}^8+\left(\gamma\eta_b^2+\lambda\eta_b\right)L_{{\rm Pl}}^4+\zeta\eta_b+\sigma=0,
\eea
%\bea\label{cubiceqbou}
%\epsilon\delta \eta_{{\rm b}}^3+\left[\delta\left(C_0+\f{\omega {{\rm P}}_\alpha}{3}\right)+\epsilon\gamma-2\theta\epsilon^2+\lambda\delta\right]\eta_{{\rm b}}^2+\left[\gamma\left(C_0+\f{\omega {{\rm P}}_\alpha}{3}\right)+\theta {{\rm P}}_\alpha\epsilon-4\theta \epsilon C_0+\lambda\gamma\right]\eta_{{\rm b}}+\theta {{\rm P}}_\alpha C_0-2\theta C_0^2=0,\nn
%\eea
where
\bea\label{bgdeplam}
\epsilon&=&\f{128}{\chi}\rho_i^2\pi^2(3\alpha_b-1)^2{\rm e}^{3(4\alpha_i-\alpha_b)},~~~~~~~~~~~~~~~~~~~~~\delta=256\rho_i^2\pi^2\theta(3\alpha_b-1){\rm e}^{12\alpha_i}\bigg[{\rm e}^{-3\alpha_b}+2(3\alpha_b-1)\bigg],~~\nn
\gamma&=&\f{8\pi\rho_i}{3\chi}(1-3\alpha_b)\left[{\rm P}_\alpha(\omega-3)+9C_0\right]{\rm e}^{3(2\alpha_i-\alpha_b)},~~\lambda=32\pi\rho_i\theta{\rm e}^{6\alpha_i}\bigg[\left(C_0-\f{P_\alpha}{2}\right){\rm e}^{-3\alpha_b}+2(3\alpha_b-1)\left(C_0-\f{P_\alpha}{4}\right)\bigg],\nn
\zeta&=&\f{{\rm e}^{-3\alpha_b}}{6\chi}(2C_0-P_\alpha)(\omega{\rm P}_\alpha+3C_0),~~~~~~~~~~~~~~~~~~~\sigma=C_0(C_0-2{\rm P}_\alpha)\theta.
\eea
%\beta&=&\f{{\rm P}_1}{6}(2\omega+3),~~~~
Setting $C_0={{\rm P}}_\alpha/2$ and taking the terms in (\ref{cubiceqbounce}) up to $L_{{\rm Pl}}^4$, we get the solution for the bounce time as (we assume $\omega\neq-3/2$)
\bea\label{quadeqtb}
\eta_b=0,~~~~~\eta_b=12\theta{\rm e}^{3\alpha_b}.
\eea
The first solution is not acceptable as for a physically reasonable collapse setting we require that $\eta_{{\rm b}}>\eta_i$. We therefore take the second one as the bounce time. Furthermore, the case of vanishing $\theta$ does not display a bouncing scenario.
%The bouncing time can be estimated for
%For $\theta=0$ the bounce time
%we can find from equation (\ref{eqsrew9}), the time at which the bounce occurs. We note that if we take the initial value of the scale factor as $\alpha_i=\sqrt{3\alpha_{{\rm b}}}/2$, then, $\Phi P_{\Phi}=C_0$. Therefore, near the bounce equation (\ref{eqsrew9}) reads
%\be\label{eqtbapp}
%\f{\omega}{3}P_1+C_0+\f{2\chi\theta e^{3\alpha_{{\rm b}}}}{(P_1-2C_0)t_{{\rm b}}}\left(2C_0^2-P_1\right)+32\pi\rho_i\theta\chi e^{6\alpha_i}=0,
%\ee
%whereby the bounce time is obtained as
%\be\label{boutim}
%t_{{\rm b}}=\frac{6 C_0 \theta  (2 \omega +3) e^{3 \alpha _{\rm b}}}{3 C_0+96 \pi  \theta  (2 \omega +3) \rho _i e^{6 \alpha _i}+P_1 \omega }.
%\ee
%The bounce time for large values of the BD coupling parameter reads
%\be\label{btomegalargeapp}
%t_{{\rm b}}=\frac{12 \theta  e^{3 \alpha _b} \left(2 C_0^2-P_1\right)}{\left(2 C_0-P_1\right) \left(192 \pi  \theta  e^{6 \alpha _i} \rho _i+P_1\right)}
%\ee
%Th%erefore, in the GR limit, there still exists a non-singular scenario for the gravitational collapse of a stiff matter in the presence of noncommutativity.
%It is seen that for $\theta=0$ the bouncing time vanishes which is not physically reasonable. Since we require that $t_{{\rm b}}>t_i$ for a physically reasonable collapse setting and thus the case of vanishing $\theta$ does not display a bouncing scenario.

For a dust fluid, Eq.~(\ref{eqsrew5}) can be immediately integrated to give
\be\label{sol14}
{{\rm P}}_{\alpha}=48\pi L_\text{Pl}^4\rho_ie^{3\alpha_i}\eta+{{\rm P}}_{0\alpha}.
\ee
Multiplying equation (\ref{eqsrew10}) by ${{\rm P}}_\Phi$ and equation (\ref{eqsrew7}) by $\Phi$ and after adding up the results we have
\be\label{ddtphiphidust}
\f{d}{d\eta}\left(\Phi {{\rm P}}_\Phi\right)=16\pi L_\text{Pl}^4 \rho_ie^{3\alpha_i}\left(1+3\theta {{\rm P}}_\Phi\right).
\ee
By differentiating equation (\ref{eqsrew10}) together with employing
Eqs. (\ref{sol14}) and (\ref{ddtphiphidust}), after a straightforward but
lengthy calculation we arrive at the evolution equation for BD scalar field as
\be\label{BDEVOLSC}
\ddot{\Phi}+3\dot{\alpha}\left(\dot{\Phi}-48\pi\rho_i L_\text{Pl}^4\theta e^{3\alpha_i}\right)+\f{8\pi\rho_i L_\text{Pl}^4e^{3(\alpha_i-\alpha)}}{2\omega+3}\left(1-6\theta {{\rm P}}_\Phi\right)=0.
\ee
Similarly, we have found the acceleration equation for the collapse scenario as
\bea\label{accnoncom}
\ddot{\alpha}&+&\left(\dot{\alpha}+\f{16\pi\rho_i L_\text{Pl}^4\theta}{\Phi}e^{3\alpha_i}\right)\left(3\dot{\alpha}+\f{\dot{\Phi}}{\Phi}\right)+\f{e^{-3\alpha}}{2(2\omega+3)\Phi}\left[16\pi\rho_i  L_\text{Pl}^4e^{3\alpha_i}(1+\omega)+64\pi L_\text{Pl}^4\rho_i\theta {{\rm P}}_\Phi e^{3\alpha_i}+\theta({{\rm P}}_\alpha-2\Phi {{\rm P}}_\Phi)\dot{{{\rm P}}}_\Phi\right]\nn
&-&16\pi L_\text{Pl}^4\rho_i\theta e^{3\alpha_i}\f{\dot{\Phi}}{\Phi^2}=0,
\eea
where we have neglected the terms containing $\theta^2$. In order to complete the above set of equations, we need to substitute for ${ {\rm P}}_\Phi$ and its derivative from equation (\ref{eqsrew7}). However, such a complicated process can not be done analytically and needs numerical considerations.


\begin{thebibliography}{99}
\bibitem{vfsct} V. Faraoni, {\it Cosmology in Scalar Tensor Gravity,} KLUWER ACADEMIC PUBLISHERS (2004).
\bibitem{BD61}C. Brans and R. H. Dicke, {\it Phys. Rev.} \textbf{124}, 925 (1961).
\bibitem{godel} K. G${\rm\ddot{o}}$del, Rev. Mod. Phys. {\bf 21} 447 (1949).
\bibitem{ppw} I. Ozvath, E. Sch${\rm \ddot{u}}$cking, {\it in Recent Developments in General Relativity}, (Pergamon, New York, 1962).
\bibitem{Jordan-rep} P. Jordan, \emph{Projective Relativity} (Friedrich Vieweg und Sohn, Braunschweig, 1955).\\
P. Jordan,  Z. Phys.  {\bf 157} 112 (1959);\\
D.  R. Brill, {\it Evidence for Gravitational Theories},  Enrico  Fermi Course  XX, Ed.  C. Moiler (Academic Press,  New York 1962) p. 50-68;\\
Y. Fujii and K.-Ichi Maeda, {\it The Scalar-Tensor Theory of Gravitation}, Cambridge University Press 2003;\\
S. Capozziello and V. Faraoni, {\it Beyond Einstein Gravity, A Survey of Gravitational Theories for Cosmology and Astrophysics}, Springer Science + Business Media 2011;\\ V. Faraoni, E. Gunzig and P. Nardone, {\it Fundam. Cosmic Phys.} \textbf{20}, 121 (1999).
\bibitem{SSEX} C. M. Will, {\it Theory and Experiment in Gravitational Physics} (Cambridge University Press, Cambridge, England, 1993);\\ Living Rev. Relativity {\bf 9} 3 (2006).
\bibitem{nuhyidgas} T. Matsuda and H. Nariai, Prog. Theor. Phys. {\bf 49} 1195 (1973).
\bibitem{SHNANA} M. Shibata, K. Nakao and T. Nakamura, Phys. Rev. D {\bf 50} 7304 (1994).
\bibitem{H72} S. W. Hawking, Commun. Math. Phys. \textbf{25}, 167 (1972).
\bibitem{sofa-1109.6324} T. P. Sotiriou and V. Faraoni, Phys. Rev. Lett. {\bf 108} 081103 (2012).
%\bibitem{OS39} J. R. Oppenheimer and H. Snyder, Phys. Rev. \textbf{56}, 455 (1939).
\bibitem{SST95}M. A. Scheel, S. L. Shapiro and S. A. Teukolsky,
Phys. Rev. D, \textbf{51} 4208 (1995); Phys. Rev. D \textbf{51}, 4236 (1995).
\bibitem{K96} G. Kang, Phys. Rev. D \textbf{54}, 7483 (1996).
\bibitem{thorndyk1971} K. S. Thorne and J. J. Dykla, Astrophys. J. {\bf 166} L35 (1971).
\bibitem{J99} T. Jacobson, Phys. Rev. Lett. \textbf{83}, 2699 (1999).
\bibitem{KKMC86} O. J. Kwon, Y. D. Kim, Y. S. Myung, B. H. Cho and Y. J. Park, Phys. Rev. D \textbf{34}, 333 (1986).

\bibitem{CL93} M. Campanelli and C. O. Lousto, Int. J. Mod. Phys. D \textbf{2}, 451 (1993).

\bibitem{AC94} A. G. Agnese and M. La Camera, Phys. Rev. D \textbf{51}, 2011 (1994).
\bibitem{crit-bdcollapse} T. Chiba and J. Soda, Prog. Theor. Phys. {\bf 96} 567 (1996).
\bibitem{4jap-bdcol-97} T. Harada, T. Chiba, K.-I. Nakao and T. Nakamura, Phys. Rev. D {\bf 55} 2024 (1997).
\bibitem{NBA98} K. K. Nandi, Bhattacharjee, S. M. K. Alam and J. Evans, Phys. Rev. D \textbf{57}, 823 (1998).
\bibitem{K99} H. Kim, Phys. Rev. D, \textbf{60} 024001 (1999).
\bibitem{bhbdcosmol-2000} H. Saida and J. Soda, Class. Quant. Grav. {\bf 17} 4967 (2000).
\bibitem{SB01} N. Sakai and J. D. Barrow, Classical Quantum Gravity \textbf{18}, 4717 (2001).
\bibitem{HGC02} T. Harada, C. Goymer and B. J. Carr, Phys. Rev. D \textbf{66}, 104023 (2002).
\bibitem{sarkarstbh2005} A. Bhadra, K. Sarkar, Gen. Rel. Grav. {\bf 37} 2189 (2005).
\bibitem{GW07} Y. Gong and A. Wang, Phys. Rev. Lett. \textbf{99}, 211301 (2007).
\bibitem{NS09} B. Nayak and L. P. Singh, Phys. Rev. D \textbf{80} 023529 (2009).
\bibitem{HY10} D. Hwang and D. Yeom, Classical Quantum Gravity \textbf{27}, 205002 (2010).
\bibitem{F10} V. Faraoni, Entropy 2010 \textbf{12}, 1246 (2010).
\bibitem{nsbdexistence} N. Bedjaoui, P. G. LeFloch, J. M. Martin-Garcia and J. Novak, Class. Quant. Grav. {\bf 27} 245010 (2010).
\bibitem{dilnearns} J. R. Morris, Gen. Relativ. Grav. {\bf 44} 437 (2011)
\bibitem{RBD14} P. Rudra, R. Biswas and U. Debnath, Astrophys. and Space Sci., \textbf{354} 2101 (2014); arXiv:1307.6603 [physics.gen-ph]. Astrophys. and Space Sci., {\bf 339} 135 (2012); arXiv:1203.1454 [gr-qc].
\bibitem{bhstkorea} Y. S. Myung and T. Moon, Phys. Rev. D {\bf 89} 104009 (2014).
\bibitem{shman-ltbbd} M. Sharif and R. Manzoor, Mod. Phys. Lett. A {\bf 29} 1450192 (2014).
\bibitem{jnovak98} J. Novak, Phys. Rev. D {\bf 57} 4789 (1998).
\bibitem{Snyder1947} H. S. Snyder, Phys. Rev {\bf 71} 38 (1947); Phys. Rev {\bf 72} 68 (1947).
\bibitem{reviewnoncom} R. Banerjee, B. Chakraborty, S. Ghosh, P. Mukherjee and S. Samanta, Found. Phys. {\bf 39} 1297 (2009);\\ I. Hinchliffe, N. Kersting and Y. L. Ma, Int. J. Mod. Phys. A {\bf 19} 179 (2004).
\bibitem{reviewnoncom1} M. R. Douglas and N. A. Nekrasov, Rev. Mod. Phys. {\bf 73} 977 (2001);\\ R. J. Szabo, Phys. Rep. \textbf{378}, 207 (2003).
\bibitem{Pbook98} J. Polchinski, \textit{String Theory}, (Cambridge University Press, Cambridge, 1998).
\bibitem{MRS00} M. Heller, W. Sasin, arXiv:gr-qc/9906072; M. Patel, arXiv:math/0008023 [math.GM]; J. W. Moffat, Phys. Lett. B {\bf 493} 142 (2000); C.-S. Chu, K. Furuta and T. Inami, Int. J. Mod. Phys. A {\bf 21} 67 (2006); I. Bars, hep-th/0109132.
\bibitem{LV} S. M. Carroll, J. A. Harvey, V. A. Kostelecky, C. D. Lane,
and T. Okamoto, Phys. Rev. Lett. \textbf{87}, 141601 (2001);\\ C. E.
Carlson, C. D. Carone, and R. F. Lebed, Phys. Lett. \textbf{B518},
201 (2001); \textbf{549}, 337 (2002).
\bibitem{RV03} J. M. Romero and J. D. Vergara, Mod. Phys. Lett. A \textbf{18}, 1673 (2003).
\bibitem{noncomhosc} V. P. Nair and A. P. Polychronakos, Phys. Lett. B
{\bf 505} 267 (2001), arXiv:hep-th/0011172; J. Gamboa, M.
Loewe and J. C. Rojas, Phys. Rev. D {\bf 64} 067901 (2001),
arXiv:hep-th/0010220; S. Bellucci, A. Nersessian and C. Sochichiu, Phys. Lett. B {\bf 522} 345 (2001), arXiv:hep-th/0106138; R. Banerjee, Mod. Phys. Lett A {\bf 17} 631
(2002), arXiv:hep-th/0106280; B. Muthukumar and P.
Mitra, Phys. Rev. D {\bf 66} 027701 (2002), arXiv:hep-th/0204149 ; S. Samanta, Mod. Phys. Lett. A {\bf 21} 675
(2006), arXiv:hep-th/0510138; S. Gangopadhyay and F.
G. Scholtz, arXiv:0812.3474 [math-ph].
\bibitem{HYatomnc} M. Chaichian and M. M. Sheikh-Jabbari, Phys. Rev. Lett
{\bf 86} 2716 (2001), arXiv:hep-th/0010175; X. Calmet, Eur.
Phys. J. C {\bf 41} 269 (2005), arXiv:hep-th/0401097; Z. Gur-alnik R. Jackiw, Pi, S.-Y. and A. P. Polychronakos, Phys.
Lett. B {\bf 517} 450 (2001), arXiv:hep-th/0106044.
\bibitem{qHall} J . Belissard, Lect. Notes in Physics {\bf 257} 99 (1986); J. Belissard, A. Van Elst and H. SchulzBaldes, cond-mat/9301005.
\bibitem{Landaupr} J. Gamboa, M. Loewe, F. Mendez and J. C. Rojas, Mod. Phys. Lett. A {\bf 16} 2075 (2001);\\ J. Gamboa, M. Loewe, J. C. Rojas, Phys. Rev. D {\bf 64} 067901 (2001).
\bibitem{SW99} N. Seiberg and E. Witten, JHEP {\bf 9909} 032 (1999);\\ A. Connes, M. R. Douglas and A. Schwarz, JHEP {\bf 9802} 003 (1998).
\bibitem{noncomgrav} H. Garcia-Compean, O. Obregon, C. Ramirez and M. Sabido, Phys. Rev. D {\bf 68} 044015 (2003); Phys. Rev. D {\bf 68} 045010 (2003); A. H. Chamseddine, J. Math. Phys. {\bf 44} 2534 (2003); V. O. Rivelles, Phys. Lett. B {\bf 558} 191 (2003); M. Maceda, J. Madore, P. Manousselis, and G. Zoupanos, Eur. Phys. J. C \textbf{36} 529 (2004); P. Aschieri, M. Dimitrijevic, F. Meyer and J. Wess, Class. Quant. Grav. {\bf 23} 1883 (2006); X. Calmet and A. Kobakhidze, Phys. Rev. D {\bf 72} 045010 (2005); L. Alvarez-Gaume, F. Meyer and M. A. Vazquez- Mozo, Nucl. Phys. B {\bf753} 92 (2006); S. Estrada-Jimenez, H. Garcia-Compean, O. Obregon and C. Ramirez, Phys. Rev. D {\bf 78} 124008 (2008); P. Aschieri and L. Castellani, J. Geom. Phys. {\bf 60} 375 (2010).

\bibitem{NCNC111} J. M. Romero and J. A. Santiago, Mod. Phys. Lett.
A \textbf{20} 781 (2005), arXiv:hep-th/0310266.
\bibitem{NONCOMINF1400} R. Brandenberger and P.-M. Ho, Phys. Rev. D \textbf{66}
023517 (2002), arXiv:hep-th/0203119; Q.-G. Huang and M. Li, J. Cosmol.
Astropart. Phys. \textbf{11} 001 (2003), arXiv:0308458 {[}astro-ph{]};
JHEP \textbf{06} 014 (2003), arXiv:hep-th/0304203; H. Kim, G. S. Lee,
and Y. S. Myung, Mod. Phys. Lett. A \textbf{20} 271 (2005), arXiv:hep-th/0402018;
H. Kim, G. S. Lee, H. W. Lee, and Y. S. Myung, Phys. Rev. D \textbf{70}
043521 (2004), arXiv:hep-th/0402198; A. Kempf and L. Lorenz, Phys. Rev. D \textbf{74}
103517 (2006); A. Ashoorioon, A. Kempf,
and R. B. Mann, Phys. Rev. D \textbf{71} 023503 (2005); Y. S. Myung, Phys. Lett. B \textbf{601}
1 (2004), arXiv:hep-th/0407066; Dao-jun Liu and Xin-zhou Li, Phys.
Rev. D \textbf{70} 123504 (2004), arXiv:0402063 {[}astro-ph{]}; G.
Calcagni, Phys. Rev. D \textbf{70} 103525 (2004), arXiv:hep-th/0406006;
Phys. Lett. B \textbf{606} 177 (2005), arXiv:hep-th/0406057; Rong-Gen
Cai, Phys. Lett. B \textbf{593} 1 (2004), arXiv:hep-th/0403134; C.-S.
Chu, B. R. Greene, and G. Shiu, Mod. Phys. Lett. A \textbf{16} 2231
(2001), arXiv:hep-th/0010207; S. Tsujikawa, R. Maartens, and R. Brandenberger,
Phys. Lett. B \textbf{574} 141 (2003), arXiv:0308169 {[}astro-ph{]};
G. Calcagni and S. Tsujikawa, Phys. Rev. D \textbf{70} 103514 (2004),
arXiv:0407543 {[}astro-ph{]}; S. M. M. Rasouli and P. V. Moniz, Phys. Rev. D \textbf{90}, 083533 (2014); S. M. M. Rasouli, M. Farhoudi and N. Khosravi, Gen. Rel. Grav. \textbf{43}, 2895 (2011).
\bibitem{NONCOMQC12} H. Garcia-Compean, O. Obregon, and C. Ramirez,
Phys. Rev. Lett. 88, 161301 (2002), arXiv:hep-th/0107250.
\bibitem{BP0412} G. D. Barbosa and N. Pinto--Neto, Phys. Rev. D \textbf{70}; N. Khosravi, S. Jalalzadeh, H. R. Sepangi, Gen. Rel. Grav. {\bf 39} 899 (2007).
\bibitem{GUP} A. Bina, S. Jalalzadeh, A. Moslehi, Phys. Rev. D \textbf{81}
023528 (2010), arXiv:1001.0861 {[}gr-qc{]}; A. Bina, K. Atazadeh,
S. Jalalzadeh, Int. J. Theor. Phys. \textbf{47} 1354 (2008), arXiv:0709.3623
{[}gr-qc{]}; S. Pramanik, S. Ghosh, arXiv:1301.4042 {[}hep-th{]}.
\bibitem{CCPNONCOM} W. Kim and E. J. Son, Phys. Rev. D {\bf 75} 024025 (2007); L. N. Chang, D. Minic, N. Okamura, and T. Takeuchi, Phys. Rev. D \textbf{65} 125028 (2002).
\bibitem{CCPNONCOM1} N. Khosravi, S. Jalalzadeh and H. R. Sepangi, JHEP {\bf 01} 134 (2006).
\bibitem{bbbcsinnc} M. Maceda, J. Madore, P. Manousselis, G. Zoupanos, Eur. Phys. J. C {\bf 36} 529 (2004); F. Finelli, JCAP 0310 (2003) 011; A. Chaney, L. Lu and A. Stern, Phys. Rev. D {\bf 92} 064021 (2015).
\bibitem{BP04} G. D. Barbosa and N. Pinto--Neto,  Phys. Rev. D \textbf{70} 103512 (2004).
\bibitem{SFCOSDEFPH} L. O. Pimentel and C. Mora, Gen. Rel. Grav., \textbf{37} 817 (2005).
\bibitem{GUZSASO} W. Guzman, M. Sabido, J. Socorro, Phys. Rev. D, 76 (2007), 087302; S. M. M. Rasouli, N. Saba, M. Farhoudi, J. Marto and P. V. Moniz, ``In
ationary Universe in a Deformed Phase Space Setting'', in progress.
\bibitem{compacnjs} N. Khosravi, S. Jalalzadeh and H.R. Sepangi, Int. J. Mod. Phys. D, \textbf{16} 1187 (2007); JHEP, \textbf{0601} 134 (2006).

\bibitem{MN97} T. Matsufuji and S. Naka, Progress of Theoretical Physics, \textbf{98}, 3 (1997).

\bibitem{S16} S. Kobayashi, Int. J. Mod. Phys. A, \textbf{31}, 1650080 (2016).






\bibitem{BBDP09} C. Bastos, O. Bertolami, N.C. Dias and J.N. Prata, Phys. Rev. D \textbf{80} (2009) 124038.
\bibitem{Nicollinipapers} P. Nicolini, arXiv:hep-th/0510203; P. Nicolini, A. Smailagic, and E. Spallucci, Phys. Lett. B {\bf 632} 547 (2006);
T. G. Rizzo, JHEP 09 (2006) 021; S. Ansoldi, P. Nicolini, A. Smailagic and E. Spallucci, Phys. Lett. B {\bf 645} 261 (2007); P. Mukherjee and A. Saha, Phys. Rev. D {\bf 77} 064014 (2008);
R. Casadio  and P. Nicolini, JHEP 11 (2008) 072; E. Spallucci, A. Smailagic and P. Nicolini, Phys. Lett. B {\bf 670} 449 (2009); S. Ansoldi, {\it Spherical black holes with regular center: a review of existing models including a recent realization with Gaussian sources,} arXiv:0802.0330 [gr-qc]; P. Nicolini, Int. J. Mod. Phys. A {\bf 24} 1229 (2009); I. A. Guerrero, D. Batic and M. Nowakowski, Class. Quant. Grav. {\bf 26} 245006 (2009); A. Smailagic and E. Spallucci, Phys. Lett. B {\bf 688} 82 (2010); C. Bastos, O. Bertolami, N. C. Dias and J. N. Prata, J. Phys. Conf. Ser. {\bf 314} 012042 (2011); J. R. Mureika and P. Nicolini, Phys. Rev. D {\bf 84} 044020 (2011); R. Garattini and B. Majumder, Nucl. Phys. B {\bf 884}  125 (2014).
\bibitem{RZMM14} S. M. M. Rasouli, A. H. Ziaie, J. Marto, and P. V. Moniz, Phys. Rev. D \textbf{89}, 044028 (2014).
\bibitem{BDSINCOS} L. E. Gurevich, A. M. Finkelstein and V. A. Ruban, Astrophys. Space Sci. {\bf 22} 231 (1973); O. Hrycyna and M. Szydlowski, JCAP {\bf 12} 016 (2013).
\bibitem{BDSINCOL} A. H. Ziaie, K. Atazadeh and Y. Tavakoli, Class. Quant. Grav. {\bf 27} 075016 (2010); A. H. Ziaie, K. Atazadeh and S. M. M. Rasouli, Gen. Relativ. Gravit {\bf 43} 2943 (2011); A. H. Ziaie, A. Ranjbar and H. R. Sepangi, Class. Quantum Grav. {\bf 32} 025010 (2015).
\bibitem{McVittieMTW} C. W. Misner, K. S. Thorne and J. A. Wheeler, {\it Gravitation}, Freeman (1973); G. C. McVittie, Astrophys. J. {\bf 140} 401 (1964).

%%%%%%%%%%%%%%%%%%%%%%%%%%%%%%%%%%%%%%%%%%%%%%%%%%%55



%\bibitem{QG} M. B. Green, J.H. Schwarz and E. Witten, \textit{Superstring Theory}, (Cambridge University Press, Cambridge, 1988); G. �t Hooft, Classical Quantum Gravity
%\textbf{16}, 3263 (1999); C. Rovelli, \textit{Quantum Gravity}, (Cambridge University Press, Cambridge, 2004);
%T. Thiemann, \textit{Modern Canonical Quantum General Relativity}, (Cambridge University Press, Cambridge, 2007).
\bibitem{SS01} S. Sen and A. A. Sen, Phys. Rev. D \textbf{63} 124006 (2001).
\bibitem{RFS11-R14} S. M. M. Rasouli, M. Farhoudi and H. R. Sepangi,
Classical Quantum Gravity \textbf{28}, 155004 (2011);
S.M. M. Rasouli, M. Farhoudi and P. V. Moniz, Classical Quantum Gravity \textbf{31}, 115002 (2014).
S. M. M. Rasouli, Prog. Math. Rel., Gravit. Cosmol. \textbf{60}, 371 (2014).
\bibitem{stm99-DRJ09-RJ10}P. S. Wesson, \textit{Space--Time--Matter: Modern Kaluza--Klein Theory} (World Scientific, Singapore, 1999);
N. doroud, S. M. M. Rasouli and S. Jalalzadeh, Gen. Rel. Grav. \textbf{41}, 2637 (2009);
S. M. M. Rasouli and S. Jalalzadeh, Ann. Phys. (Berlin) \textbf{19}, 276 (2010).
%\bibitem{FGN99}
\bibitem{PFLAG} J. D. Brown, Classical Quantum Gravity {\bf 10} 1579 (1993);\\ O. Bertolami, F. S. N. Lobo, J. Paramos, Phys.  Rev. D {\bf 78} 064036 (2008).
\bibitem{GUPrefs} M. R. Setare, Phys. Rev. D {\bf 70} 087501 (2004); M. R. Setare, Int. J. Mod. Phys. A {\bf 21} 1325 (2006); M. V. Battisti and G. Montani, Phys. Lett. B {\bf 656} 96 (2007); M. V.  Battisti and G. Montani Phys. Rev. D {\bf 77} 023518 (2008).
\bibitem{noncomshrefs}
A. E. F. Djema\"i, H. Smail, Commun. Theor. Phys. {\bf 41} (2004) 837;
 A. E. F. Djema\"i, Int. J. Theor. Phys. {\bf 35} (1996) 519.
\bibitem{nonocomshrefs1} G. Esposito and C. Stornaiolo, {\it Int. J. Geom. Meth. Mod. Phys.} {\bf 4} (2007)
349 (arXiv: hep-th/0607114)\\ J. M. Gracia-Bondia, F. Lizzi, G. Marmo
and P. Vitale, {\it J. High Energy Phys.} {\bf 0204} (2002) 026
(arXiv: hep-th/0112092).
\bibitem{minwramsei} S. Minwalla, M. Van Raamsdonk, and N. Seiberg, JHEP 02 (2000) 020; M. Van Raamsdonk and Seiberg N., JHEP, {\bf 03} 035 (2000); S. Minwalla, M. Van Raamsdonk and N. Seiberg, JHEP
0002: 020 (2000); A. Micu and M. M. Sheikh-Jabbari, JHEP {\bf 0101} 025 (2001).
\bibitem{Giovanni} G. Amelino-Camelia, G. Gubitosi and F. Mercati, Phys.
Lett. B  676 (2009) 180.
\bibitem{Hu} A. Eftekharzadeh and B. L. Hu, Braz. J. Phys. 35 (2005) 333.
\bibitem{GSS11}W. Guzm\'{a}n, M. Sabido and J. Socorro, {\it Phys. Lett. B} \textbf{697}, 271 (2011).
\bibitem{SHAY} S. Hayward, Phys. Rev. D \textbf{49} 6467 (1994),
arXiv:9303006[gr-qc]; ibid, Phys. Rev. D \textbf{53} 1938 (1996), arXiv:9408002 [gr-qc]; C. W. Misner and D. H. Sharp, Phys. Rev. \textbf{136} B571
(1964); S. A. Hayward, Phys. Rev. D \textbf{49} 831 (1994), arXiv:9303030 [gr-qc].
%\bibitem{GSS111} V. Faraoni, \lq\lq{}Cosmology in Scalar-Tensor Gravity,\rq\rq{} Kluwer Academic Publishers (2004).
%\bibitem{negomkks} S. J. Kolitch and D. M. Eardley, Ann. Phys. {\bf 241} 128 (1995).
%\bibitem{negomaeu} J. P. Baptista, J. C. Fabris and S. V. B. Gonclaves, Astrophys. Space Sc. {\bf 296} 315 (1996);\\
O. Bertolami and P. J. Martins, Phys. Rev. D {\bf 61} 064007 (2000).
\bibitem{vaidyaext} N. O. Santos, Mon. Not. Roy. Astron. Soc. \textbf{216} 403 (1985);
A. Wang and Y. Wu, Gen. Rel. Grav., \textbf{31} (1999) 107; S. D. Maharaj, G. Govender and M. Govender, Gen. Relativ. Gravit. \textbf{44} (2012) 1089.
\bibitem{gidjcbd} K. G. Suffern, J. Phys. A: Math. Gen. {\bf 15} (1982) 1599; C. Barrabes and G. F. Bressange, Class. Quantum Grav. {\bf 14} 805 (1997); F. Dahia and C. Romero, Phys. Rev. D {\bf 60} 104019 (1999).
%\bibitem{thetapl} Everton M. C. Abreu, M. J. Neves, Nucl. Phys. B {\bf 884} 741 (2014); C. Bastos  and O. Bertolami, Phys. Rev. D {\bf 78} 023516 (2008); P. Wang, H. Yang and X. Zhang, Phys. Lett. B {\bf 718} 265 (2012).
%\bibitem{PRCOLL} F. I. Cooperstock, S. Jhingan, P. S. Josh and T. P. Singh, Class. Quantum Grav. \textbf{14} 2195 (1997); Subenoy Chakraborty, Sanjukta Chakraborty, Ujjal Debnath, Int. J. Mod. Phys. D \textbf{14} 1707 (2005); P. S. Joshi and R. Goswami, arXiv:gr-qc/0504019.
J. D. Barrow and P. Parsons, Phys. Rev. D \textbf{55} (1997) 1906.
\bibitem{ABWZ} J. Alsing, E. Berti, C. M. Will and H. Zaglauer, Phys. Rev. D \textbf{85} (2012) 064041.
%\bibitem{BKM04}J.D. Barrow, D. Kimberly and J. Magueijo, {\it Class. Quant. Grav.} \textbf{21}, 4289 (2004).
\bibitem{BDtoGR}A. Barros and C. Romero, Phys. Lett. A \textbf{173}, 243 (1993);
N. Banerjee and S. Sen, Phys. Rev. D \textbf{56}, 1334 (1997);
V. Faraoni, Phys. Lett. A \textbf{245}, 26 (1998);
V. Faraoni, Phys. Rev. D \textbf{59}, 084021 (1999).

%\bibitem{OSCRA} E. N. Glass and L. Lindblom, Astrophysical Journal
Supplement Series \textbf{53} (1983) 93; K. D. Kokkotas and J. Ruoff, Astron. Astrophys. \textbf{366} 565 (2001).
%\bibitem{FRG} A. H. Ziaie, K. Atazadeh, S. M. M. Rasouli, Gen Relativ Gravit \textbf{43} 2943 (2011).
%\bibitem{TwoKorea} D.-il Hwang, D.-han Yeom, Class. Quant. Grav. \textbf{27} 205002 (2010).
\bibitem{frb} C. Barragan, G. J. Olmo and H. Sanchis-Alepuz, Phys. Rev. D \textbf{80} (2009) 024016.
\bibitem{gtpb} Yi-Fu Cai, Shih-Hung Chen, J. B. Dent, S. Dutta and E.
N. Saridakis, Class. Quantum Grav. \textbf{28} (2011) 215011.
\bibitem{bretchet} S. D. Brechet, M. P. Hobson, A. N. Lasenby, Class. Quantum Grav. \textbf{25} (2008) 245016.
\bibitem{khodam} A. H. Ziaie, P. V. Moniz, A. Ranjbar, H. R. Sepangi, arXiv:1305.3085 [gr-qc].
\bibitem{loopc} A. Ashtekar, T. Pawlowski and P. Singh, Phys. Rev. Lett.
\textbf{96} (2006) 141301; ibid, Phys. Rev. D \textbf{73} (2006) 124038; ibid , Phys.
Rev. D \textbf{74} (2006) 084003.
\bibitem{loopi} A. Ashtekar and E. Wilson-Ewing, Phys. Rev. D \textbf{79} (2009) 083535; ibid Phys.
Rev. D \textbf{80} (2009) 123532; E. Wilson-Ewing, Phys. Rev. D \textbf{82} (2010) 043508; M. Martin-Benito, G. A. Mena Marugan and T. Pawlowski, Phys. Rev. D \textbf{78} (2008) 064008; ibid Phys. Rev. D \textbf{80} (2009) 084038.
\bibitem{loop3} A. Ashtekar and P. Singh, Class. Quantum. Grav. \textbf{28} (2011) 213001. \textbf{366} (2001) 565;V. K. Gupta, V. Tuli and A. Goyal, Astrophys. J. \textbf{579} (2002) 374; A. Brillante and I. N. Mishustin,  EPL \textbf{105} (2014) 39001.


%%%%%%%%%%%%%%%%%%%%%%%%%%%%%%%%%%%%%%%%%%%%%%%%%%%%%%%%%%%%%%%%%%%%%%%%%%%%%%%%%%%%%%%%%%%%%%
%%%%%%%%%%%%%%%%%%%%%%%%%%%%%%%%%%%%%%%%%%%%%%%%%%%%%%%%%%%%%%%%%%%%%%%%%%%%%%%%%%%%%%%%%%%
%%%%%%%%%%%%%%%%%%%%%%%%%%%%%%%%%%%%%%%%%%%%%%%%%%%%%%%%%%%%%%%%%%%%%%%%%%%%%%%%%%%%%%%%%%

%\bibitem{TMT98} T. Tamaki, K. Maeda and T. Torii, Phys. Rev. D \textbf{57}, 4870 (1998).


%\bibitem{D38}P.A.M. Dirac, {\it Proc. R. Soc. A} \textbf{165}, 199 (1938).
%\bibitem{BR93}A. Barros and C. Romero, {\it Phys. Lett. A } \textbf{173}, 243 (1993).
%\bibitem{BS97}N. Banerjee and S. Sen, {\it Phys. Rev. D} \textbf{56}, 1334 (1997).
%\bibitem{Faraoni98}V. Faraoni, {\it Phys. Lett. A } \textbf{245}, 26                   (1998).
%\bibitem{Faraoni99}V. Faraoni, {\it Phys. Rev. D } \textbf{59}, 084021 (1999).
%\bibitem{R.book04}C. Rovelli, \textit{Quantum Gravity}, (Cambridge University Press, Cambridge, 2004).
%\bibitem{A02}T. Thiemann, \textit{Modern Canonical Quantum General Relativity}, (Cambridge             University Press, Cambridge, 2007).
%\bibitem{ASS04}M.B. Green, J.H. Schwarz and E. Witten, \textit{Superstring Theory}, (Cambridge               University Press, Cambridge, 1988).
%\bibitem{ZachosFairlie05}C.K. Zachos, D.B. Fairlie and T.L. Curtright (editors),                         {\it Quantum Mechanics in Phase Space}, (World Scientific, Singapore,                         2005).
%\bibitem{COR02}H. Garcia--Compean, O. Obregon and C. Ramirez, {\it Phys.               Rev. Lett.} \textbf{88}, 161301 (2002).
%\bibitem{KHS09}N. Khosravi and H.R. Sepangi, {\it Phys. Lett. B} \textbf{673}, 279 (2009).
%\bibitem{KHSV10}N. Khosravi, H.R. Sepangi and B. Vakili, {\it Gen. Rel. Gravit.} \textbf{42}, 1081 (2010).
%\bibitem{BS07}K.A. Bronnikov and A.A. Starobinsky, {\it JETP Lett.} \textbf{85}, 1 (2007).
%\bibitem{B09}P. Bonifacio, ``Spacetime Conformal Fluctuations and Quantum Dephasing", (Ph.D. Thesis,             University of Aberdeen, U.K., 2009), {\it gr-qc/0906.0463}.
%\bibitem{o'hanlon-tupper-72}J. O'Hanlon and B.O.J. Tupper, {\it Nuovo Cimento} \textbf{7B}, 305 (1972).
%\bibitem{FBOOK04}V. Faraoni, \textit{Cosmology in Scalar--Tensor Gravity}, (Kluwer, Dordrech, 2004).
%\bibitem{DjeSma04}A.E.F. Djema\"{\i} and H. Smail, {\it Commun. Theor. Phys.\/} {\bf 41}, 837 (2004).
%\bibitem{MFinprogress}B. Malekolkalami and M. Farhoudi, ``Noncommutative geometry and gravitomagnetism'', \textit{work in progress}.
%\bibitem{kss}N. Khosravi, H.R. Sepangi and M.M. Sheikh--Jabbari, {\it Phys. Lett. B} \textbf{647}, 219 (2007).
%\bibitem{kappa}J. Kowalski-Glikman and S. Nowak, \textit{Phys. Lett. B} \textbf{539}, 126 (2002).
%\bibitem{kappa1}J. Kowalski-Glikman, \textit{Lect. Notes Phys.} \textbf{669}, 131 (2005).
%\bibitem{gup1}A. Kempf, G. Mangano and R.B. Mann, \textit{Phys. Rev. D} \textbf{52}, 1108 (1995).
%\bibitem{gup2}A. Kempf and G. Mangano, \textit{Phys. Rev. D} \textbf{55} (1997) 7909.
%\bibitem{KHS08}N. Khosravi and H.R. Sepangi, {\it J. Cosmol. Astropart. Phys.} \textbf{011}, 0804 (2008).
%\bibitem{CST01}M. Chaichian, M.M. Sheikh--Jabbari and A. Tureanu, {\it Phys. Rev. Lett.} \textbf{86}, 2716 (2001).
%\bibitem{M.book05}V. Mukhanov, \textit{Physical Foundations of Cosmology}, (Cambridge University Press, Cambridge, 2005).
%\bibitem{W.book08}S. Weinberg, \textit{Cosmology}, (Oxford University Press, Oxford, 2008).
%\bibitem{NBV06}A. Nayeri, R.H. Brandenberger and C. Vafa, {\it Phys. Rev. Lett.} \textbf{97}, 021302 (2006).
%\bibitem{vsl}S. Alexander and J. Magueijo, ``Non--commutative geometry as a realization of varying             speed of light cosmology", \textit{hep-th/0104093}.
%\bibitem{brandenberger}R. Brandenberger and P.-M. Ho, {\it Phys. Rev. D} \textbf{66}, 023517 (2002).
%\bibitem{brandenberger1}S. Alexander, R. Brandenberger and J. Magueijo, {\it Phys. Rev. D} \textbf{67},                        081301 (2003).
%\bibitem{Faraoni.book}V. Faraoni, \textit{Cosmology in Scalar Tensor Gravity}        (Kluiwer Academic Publishers, Netherlands, 2004).
%\bibitem{KE95}S.J. Kolitch, and D.M. Eardley, {\it Ann. Phys. (N.Y.) } \textbf{241}, 128 (1995).
%\bibitem{MW95}J.P. Mimoso and D. Wands {\it Phys. Rev. D} \textbf{51}, 477 (1995).
%\bibitem{L95}J. E. Lidsey {\it Phys. Rev. D} \textbf{52}, 5407 (1995).
%\bibitem{L96}J. E. Lidsey {\it Class. Quant. Grav.} \textbf{13}, 2449 (1996).
%%%%%%%%%%%%%%%%%%%%%%%%%%%%%%%%%%%%%%%%%%%%%%%%%%%%%%%%%%%%%%%%%%%%%%%%%%%%%%%%%%%%%%%%%%%%%5
%%%%%%%%%%%%%%%%%%%%%%%%%%%%%%%%%%%%%%%%%%%%%%%%%%%%%%%%%%%%%%%%%%%%%%%%%%%%%%%%%%%%%%%%%%%%555
%%%%%%%%%%%%%%%%%%%%%%%%%%%%%%%%%%%%%%%%%%%%%%%%%%%%%%%%%%%%%%%%%%%%%%%%%%%%%%%%%%%%%%%%%%%%%%%%%%%%%%%%%%5
%%%%%%%%%%%%%%%%%%%%%%%%%%%%%%%%%%%%%%%%%%%%%%%%%%%%%%%%%%%%%%%%%%%%%%%%%%%%%%%%%%%%%%%%%%%%%%%%%%%%%%%
%\bibitem{DDB07}M. P. Dabrowski, T. Denkiewicz and D. Blaschke, {\it Ann. Phys.} \textbf{16}, 237 (2007).



%\bibitem{HN96-BFG96-M09-CDG13} Jai-chan Hwang and Hyerim Noh, Phys. Rev. D \textbf{54}, 1460 (1996);
%J. P. Baptista, J. C. Fabris and S. V. B. Gon�alves, Astrophys. Space Sci. \textbf{246}, 315 (1996);
%T. Matsuda, Classical Quantum Gravity \textbf{26}, 145016 (2009);
%J. A. R. Cembranos, A. de la Cruz Dombriz, and L. O. Garc\'{i}a, Phys. Rev. D \textbf{88}, 123507 (2013).
%\bibitem{10113896} O, Obregon and I. Quiros, Phys. Rev. D {\bf 84} 044005 (2011).
\end{thebibliography}
\end{document}